\documentclass[showpacs,preprintnumbers,amsmath,amssymb]{revtex4}

\usepackage{graphicx}
\usepackage{dcolumn}
\usepackage{bm}

\newcommand{\bq}{\begin{equation}}
\newcommand{\eq}{\end{equation}}
\newcommand{\bqa}{\begin{eqnarray}}
\newcommand{\eqa}{\end{eqnarray}}
\newcommand{\nn}{\nonumber \\}

\def\be     {\begin{equation}}
\def\ee     {\end{equation}}
\def\bea        {\begin{eqnarray}}
\def\eea        {\end{eqnarray}}
\def\bnn    {\begin{eqnarray*}}
\def\enn    {\end{eqnarray*}}
\def\dag    {\dagger}
\def\f      {\frac}

\def\br         {{\bf r}}

\def\non        {\nonumber \\}
\def\no     {\nonumber}
\def\f      {\frac}
\def\gam      {\gamma^}
\def\gamd      {\gamma_}
\def\i      {\imath}
\def\bl      {\biggl}
\def\br     {\biggr}
\def\bc      {\begin{center}}
\def\ec      {\end{center}}

\def\i         {\imath}
\def\limR    {\lim_{R \to 0}}
\def\Gam       {\Gamma}
\def\l          {\left}
\def\r          {\right}

\begin{document}

\title{Dilute magnetic topological semiconductors: What's new beyond the physics of dilute magnetic semiconductors?}
\author{Kyoung-Min Kim$^{1}$, Yong-Soo Jho$^{1}$, and Ki-Seok Kim$^{1,2}$}
\affiliation{ $^{1}$Department of Physics, POSTECH, Pohang, Gyeongbuk 790-784, Korea \\
$^{2}$Institute of Edge of Theoretical Science (IES), Hogil Kim Memorial building 5th floor,
POSTECH, Pohang, Gyeongbuk 790-784, Korea }
\date{\today}

\begin{abstract}
Role of localized magnetic moments in metal-insulator transitions \cite{DMFT_RMP} lies at the heart of modern condensed matter physics, for example, the mechanism of high T$_{c}$ superconductivity \cite{Lee_Nagaosa_Wen_RMP}, the nature of non-Fermi liquid physics near heavy fermion quantum criticality \cite{HFQCP_RMP}, the problem of metal-insulator transitions in doped semiconductors \cite{Doped_Si_Review,Disorder_Review,Disorder_Interaction_Review}, and etc. Dilute magnetic semiconductors have been studied for more than twenty years, achieving spin polarized electric currents in spite of low Curie temperatures \cite{DMS_Review}. Replacing semiconductors with topological insulators \cite{TI_Review_I,TI_Review_II}, we propose the problem of dilute magnetic topological semiconductors. Increasing disorder strength which corresponds to the size distribution of ferromagnetic clusters, we suggest a novel disordered metallic state, where Weyl metallic \cite{Weyl_Metal_I,Weyl_Metal_II,Weyl_Metal_III,Weyl_Metal_IV} islands appear to form inhomogeneous mixtures with topological insulating phases. Performing the renormalization group analysis combined with experimental results \cite{Kim_Kim_Sasaki_FMTI}, we propose a phase diagram in $(\lambda_{so},\Gamma,T)$, where the spin-orbit coupling $\lambda_{so}$ controls a topological phase transition from a topological semiconductor to a semiconductor with temperature $T$ and the distribution for ferromagnetic clusters $\Gamma$ gives rise to a novel insulator-metal transition from either a topological insulating or band insulating phase to an inhomogeneously distributed Weyl metallic state with such insulating islands. Since electromagnetic properties in Weyl metal are described by axion electrodynamics \cite{TI_Review_II,Axion_EM1,Axion_EM2,Axion_EM3}, the role of random axion electrodynamics in transport phenomena casts an interesting problem beyond the physics of percolation in conventional disorder-driven metal-insulator transitions \cite{Disorder_Review,Disorder_Interaction_Review}. We also discuss how to verify such inhomogeneous mixtures based on atomic force microscopy.
%
%
\end{abstract}

\maketitle

An endless effort has been performed to achieve spin polarized electric currents in semiconductors. Dilute magnetic semiconductors had been investigated for more than twenty years \cite{DMS_Review}, where such spin polarized electric currents have been realized but at low temperatures much below the room temperature, prohibiting us from device applications. However, interactions between doped magnetic ions and small number of charge carriers raised interesting and fundamental physics problems, for example, the nature of the RKKY (Ruderman-Kittel-Kasuya-Yosida) interaction \cite{RKKY} away from good metals, the mechanism of ferromagnetic ordering in randomly distributed magnetic ions, anomalous transport properties in the presence of scattering with random magnetic impurities, and etc.

In this study, we propose the problem of dilute magnetic topological semiconductors, replacing non-topological semiconductors with topological semiconductors. Recently, it has been reported that the evolution of average magnetic correlations from ferromagnetic- to antiferromagnetic- in Fe$_{x}$Bi$_{2}$Te$_{3}$ gives rise to changes in transport properties of magnetoresistivity and Hall effect, identified with topological ``phase transitions" driven by dynamics of doped magnetic impurities, where the paramagnetic topological ``semiconductor" of Bi$_{2}$Te$_{3}$ turns into a normal semiconductor with ferromagnetic-cluster glassy-like behaviors around $x \sim 0.025$, and it further evolves into a topological ``semiconductor" with valence-bond glassy-like behaviors, which spans over the region between $x \sim 0.03$ up to $x \sim 0.1$ \cite{Kim_Kim_Sasaki_FMTI}. Although these experiments could not reach the semiconducting regime, interactions between randomly distributed magnetic ions and itinerant electrons with topological properties cast a novel physics problem beyond the problem of dilute magnetic semiconductors, that is, interplay between the evolution of magnetic correlations in localized magnetic moments and anomalous transport phenomena in itinerant electrons with topological properties.

Performing the renormalization group analysis for an effective field theory to describe the first occurring ``phase transition" within the ``ferromagnetic" regime, we find that the variance of the distribution for randomly quenched effective magnetic fields due to ferromagnetic clusters goes toward an infinite fixed point as the concentration of magnetic ions increases. Recalling that time reversal symmetry breaking in this strong spin-orbit coupled system gives rise to the Weyl metallic state \cite{Weyl_Metal_I,Weyl_Metal_II,Weyl_Metal_III,Weyl_Metal_IV}, the infinite variance fixed point implies the emergence of randomly distributed Weyl metallic islands which coexist with topological semiconducting phases inhomogeneously, where local breaking of time reversal symmetry due to ferromagnetic clusters with large effective magnetic fields is responsible. See Fig. 1. Based on this physical picture, we propose a schematic phase diagram of Fig. 2 in $(\lambda_{so},\Gamma,T)$, where $\lambda_{so}$ is the spin-orbit coupling constant, $\Gamma$ is the variance of the distribution for randomly quenched effective magnetic fields given by ferromagnetic clusters, and $T$ is temperature. First of all, we find an unstable fixed point with $(\lambda_{so}^{c},\Gamma_{c})$ at $T = 0$, where $\lambda_{so}^{c}$ corresponds to the quantum critical point of a topological phase transition between a topological semiconductor $(\lambda_{so} \rightarrow \infty,\Gamma = 0)$ and a normal semiconductor $(\lambda_{so} = 0,\Gamma = 0)$ \cite{TI_BI_QCP}, and $\Gamma_{c}$ identifies a novel disordered quantum critical point between one fixed point of $(\lambda_{so},\Gamma = 0)$ and the other of $(\lambda_{so}^{c},\Gamma \rightarrow \infty)$ at $T = 0$. Although the nature of this infinite variance fixed point is not fully clarified within our perturbative renormalization group analysis, we conjecture to identify such a fixed point with the inhomogeneously distributed Weyl metallic state which coexists with insulating islands, as discussed before. The appearance of inhomogeneously distributed Weyl metallic islands suggests a novel disorder-driven insulator-metal transition, regarded to be counter intuitive since the metallic state results from increasing the strength of magnetic disorders. Frankly speaking, it is not clear at all whether or not the finite variance fixed point corresponds to this insulator-metal transition exactly because a percolation-type transition must be involved in order to have a genuine metallic state, where Weyl metallic islands should be connected to each other. However, this metal-insulator transition is beyond the percolation physics \cite{Percolation} since electromagnetic properties in Weyl metal are described not by conventional Maxwell dynamics but by axion electrodynamics \cite{TI_Review_II,Axion_EM1,Axion_EM2,Axion_EM3}. See the supplementary material for axion electrodynamics in a Weyl metallic phase. The role of random axion electrodynamics in transport phenomena of the disordered metallic state implies that the present metal-insulator transition does not fall into the class of either Anderson-type \cite{Disorder_Review} or Mott-type \cite{DMFT_RMP} metal-insulator transitions \cite{Disorder_Interaction_Review}, regarded as a novel class of metal-insulator transitions.

%
%
%

\begin{figure}[t]
\includegraphics[width=0.2\textwidth]{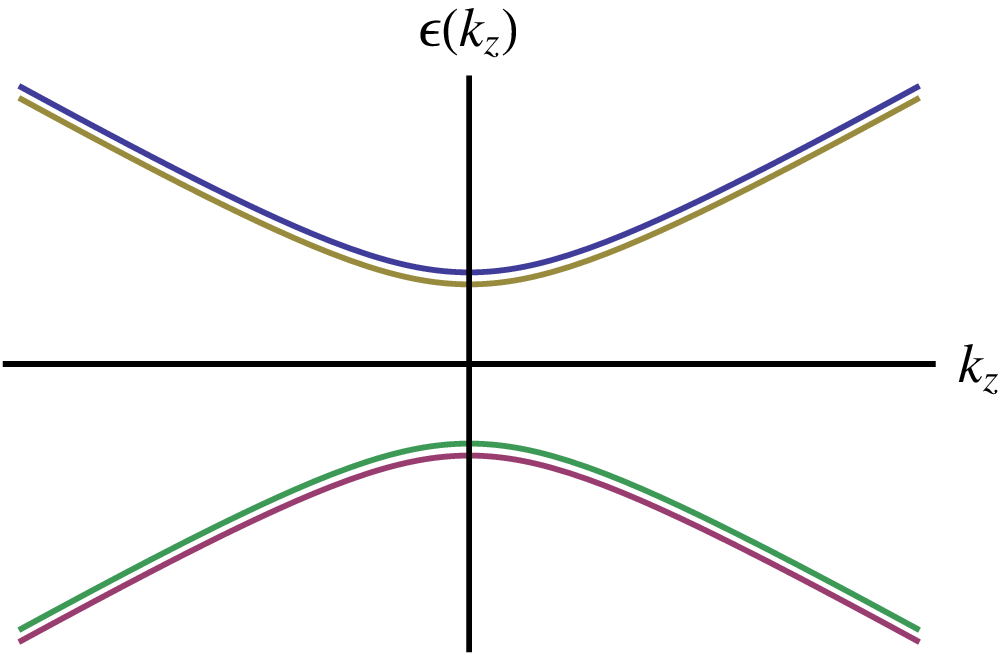}
\includegraphics[width=0.2\textwidth]{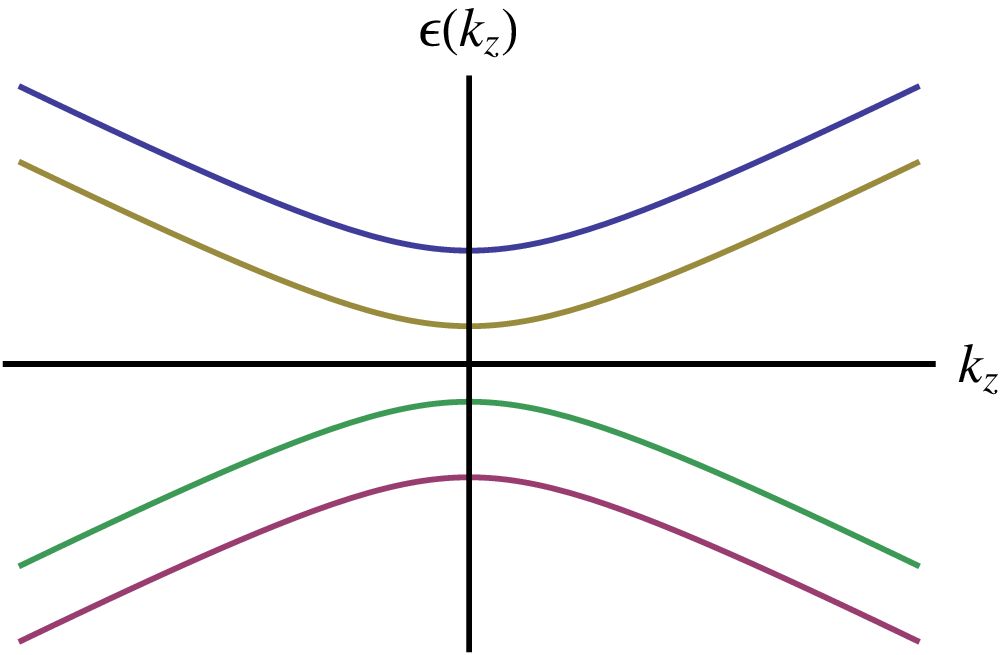}
\includegraphics[width=0.2\textwidth]{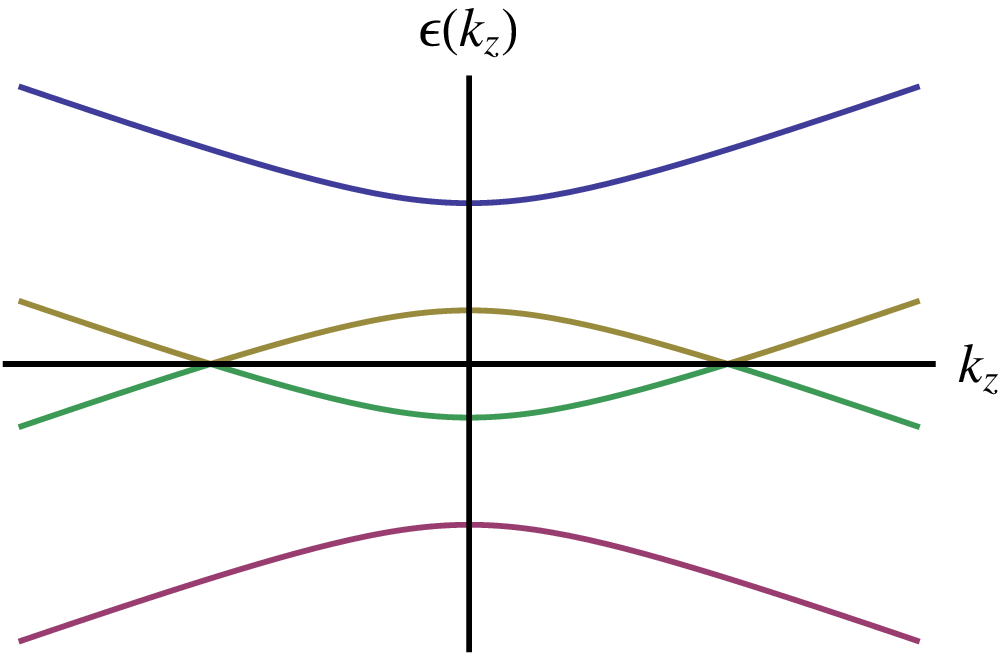}
\includegraphics[width=0.2\textwidth]{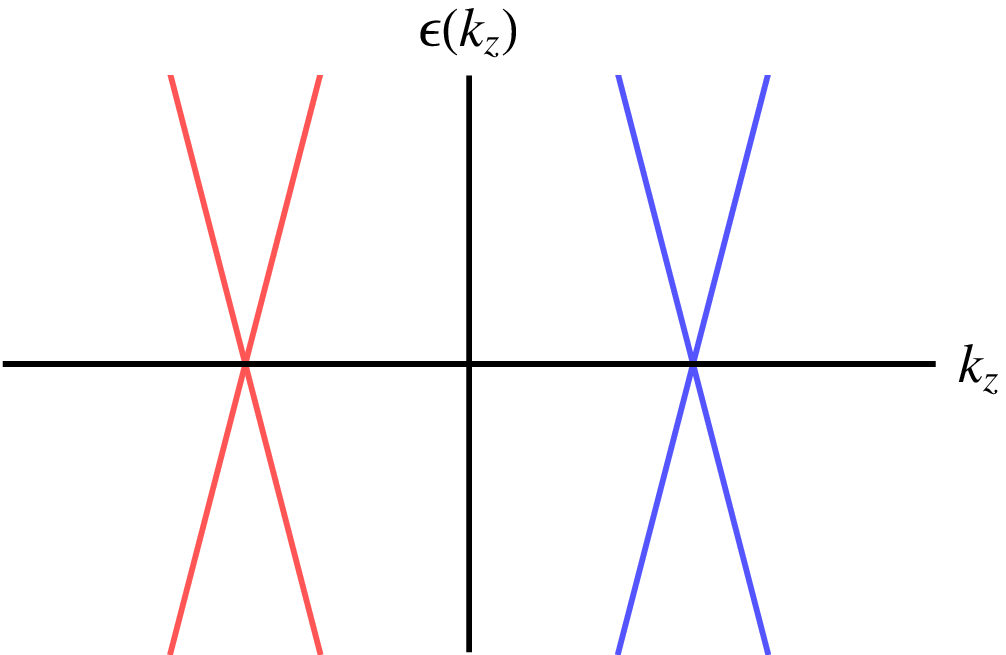}
\caption{Band structure of the Weyl metallic state. The presence of both time reversal and inversion symmetries gives rise to a degeneracy at each momentum point (First). Applying magnetic fields ($\bm{H} = H \bm{\hat{z}}$) larger than the semiconducting gap ($m(|\bm{k}|)$) in the strong spin-orbit coupled system, the gap closes to split into a pair of Weyl points with definite chirality, identified with a Weyl metallic phase. The second corresponds to $g H \ll m(|\bm{k}|)$ while the third corresponds to $g H \gg m(|\bm{k}|)$, where $g$ is Land$\acute{e}$ $g-$factor. Here, such magnetic fields are given by ferromagnetic clusters ($J \boldsymbol{\Phi}_{\bm{r}}$), conjectured to appear from RKKY interactions. As a result, randomly distributed ferromagnetic clusters give rise to inhomogeneously distributed Weyl metallic islands in places where effective magnetic fields generated by the ferromagnetic cluster exceed an insulating gap locally. The last corresponds to the case of a gapless semiconductor, i.e., Dirac semimetal, under magnetic fields.} \label{Band_Structure}
\end{figure}

\begin{figure}[t]
\includegraphics[width=0.8\textwidth]{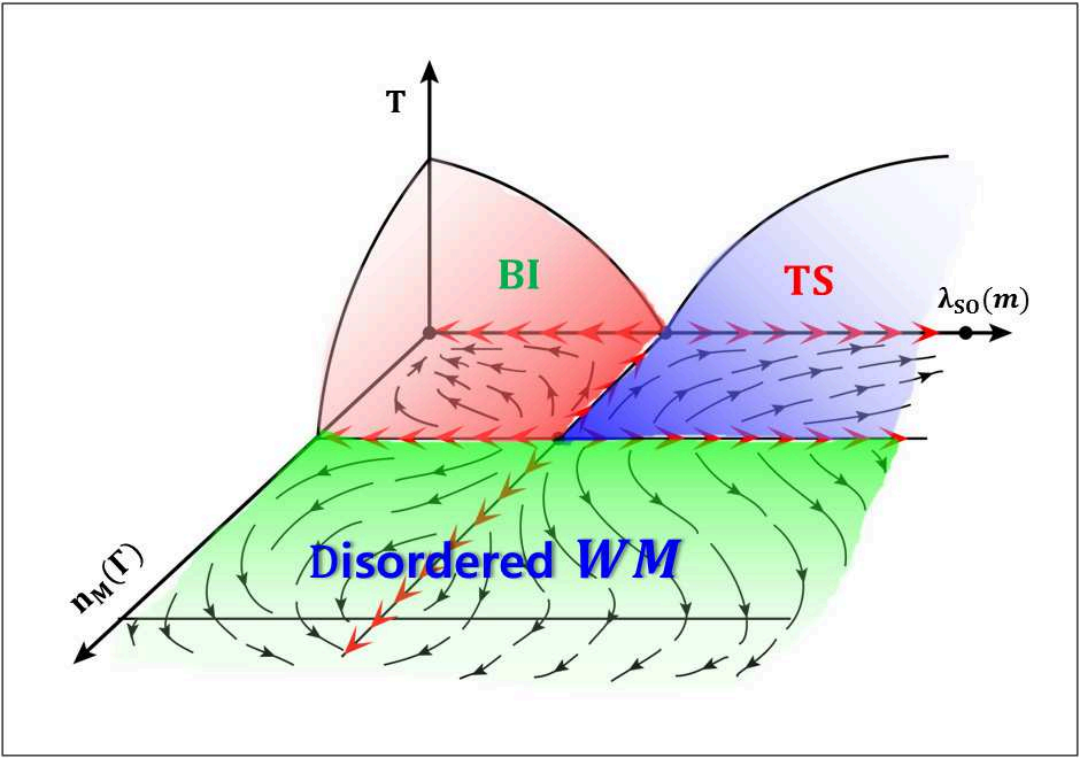}
\caption{A schematic phase diagram based on the renormalization group analysis for an effective field theory Eq. (3). The $\bm{y}-$axis represents a spin-orbit coupling constant $\lambda_{so}$ or a mass parameter $m$, and the $\bm{x}-$axis denotes a variance of the distribution for randomly quenched effective magnetic fields $\Gamma$, given by ferromagnetic clusters. $T$ is temperature. Arrows mean renormalization group flows. There exist three stable fixed points, where $(\lambda_{so} = 0,\Gamma = 0)$ and $(\lambda_{so} \rightarrow \infty,\Gamma = 0)$ correspond to a band insulating phase (BI) and a topological semiconducting state (TS), respectively, while the infinite variance fixed point of $(\lambda_{so}^{c},\Gamma \rightarrow \infty)$ is interpreted to be an inhomogeneously distributed Weyl metallic phase (Disordered WM) which coexists with randomly distributed insulating states. Two unstable fixed points imply two kinds of phase transitions. $(\lambda_{so}^{c},\Gamma = 0)$ is a quantum critical point between the topological insulating and normal semiconducting phases in the absence of magnetic impurities. $(\lambda_{so}^{c},\Gamma_{c})$ identifies a novel quantum critical point, conjectured to be associated with an insulator-metal transition, where the metallic state is not a conventional diffusive Fermi-liquid phase but quite an unconventional inhomogeneous Weyl metallic state. However, it is not clear at all whether or not this critical point coincides with this novel metal-insulator transition since a percolation-type transition must be involved in order to have a genuine metallic state, where Weyl metallic islands should be connected to each other. An important point is that electromagnetic properties of Weyl metal are described by axion electrodynamics, where they are unknown transport properties in this disordered metallic state and at the disordered quantum critical point.} \label{Phase_Diagram}
\end{figure}

The problem of dilute magnetic topological semiconductors differs from that of randomly doped magnetic impurities on the surface state of a topological insulator. One may speculate that half-quantized Hall conductance appears with Anderson localization if doped magnetic ions exhibit ferromagnetic ordering. On the other hand, an anomalous metallic phase can emerge to fall into the universality class of the quantum Hall plateau-plateau transition in the paramagnetic phase although an actual transition occurs between the quantum Hall plateau and the gapless surface state, where Anderson localization may not exist due to the presence of time reversal symmetry in average \cite{Fu_Kane_NLsM}. Although self-consistency must be incorporated to determine both the magnetic structure and Anderson localization at the same time, this surface-state problem should be distinguished from the problem of dilute magnetic topological semiconductors in the respect that axion electrodynamics does not appear. Randomly arising axion electrodynamics is the characteristic feature of dilute magnetic topological semiconductors.

We start from an effective-model free energy \cite{Axion_EM3}
\bqa && F = - T \int_{-\infty}^{\infty} d I_{\bm{r}\bm{r}'} P(I_{\bm{r}\bm{r}'}) \ln \int D \psi_{\bm{k}\tau} D \boldsymbol{S}_{\bm{r}\tau} \exp \Bigl[ - \int_{0}^{1/T} d \tau \int \frac{d^{3} \boldsymbol{k}}{(2\pi)^{3}} \psi_{\sigma\alpha}^{\dagger}(\boldsymbol{k},\tau) \Bigl\{ (\partial_{\tau} - \mu) \boldsymbol{I}_{\sigma\sigma'} \otimes \boldsymbol{I}_{\alpha\alpha'} \nn && + v \boldsymbol{k} \cdot \boldsymbol{\sigma}_{\sigma\sigma'} \otimes \boldsymbol{\tau}_{\alpha\alpha'}^{z} + m(|\boldsymbol{k}|) \boldsymbol{I}_{\sigma\sigma'} \otimes \boldsymbol{\tau}_{\alpha\alpha'}^{x} \Bigr\} \psi_{\sigma'\alpha'}(\boldsymbol{k},\tau) - \int_{0}^{1/T} d \tau \int
d^{3} \boldsymbol{r} J \psi_{\sigma\alpha}^{\dagger}(\boldsymbol{r},\tau) (\boldsymbol{\sigma}_{\sigma\sigma'} \otimes \boldsymbol{I}_{\alpha\alpha'}) \psi_{\sigma'\alpha'}(\boldsymbol{r},\tau) \cdot \boldsymbol{S}(\bm{r},\tau) \nn && - \int_{0}^{1/T} d \tau \int d^{3} \boldsymbol{r} \int d^{3} \boldsymbol{r}' I_{\boldsymbol{r}\boldsymbol{r}'} \boldsymbol{S}(\bm{r},\tau) \cdot \boldsymbol{S}(\bm{r}',\tau) - \mathcal{S}_{B} \Bigr] .
\eqa
Here, $\psi_{\sigma\alpha}(\boldsymbol{k},\tau)$ represents a four-component Dirac spinor, where $\sigma$ and $\alpha$ are spin and chiral indexes, respectively. $\bm{\sigma}_{\sigma\sigma'}$ and $\bm{\tau}_{\alpha\alpha'}$ are Pauli matrices acting on spin and ``orbital" spaces. The relativistic dispersion is represented in the chiral basis, where each eigen value of $\bm{\tau}_{\alpha\alpha'}^{z}$ expresses either $+$ or $-$ chirality, respectively. The mass term can be formulated as $m(|\boldsymbol{k}|) = m - \rho |\bm{k}|^{2}$, where $\mbox{sgn}(m) \mbox{sgn}(\rho) > 0$ corresponds to a topological insulating state while $\mbox{sgn}(m) \mbox{sgn}(\rho) < 0$ corresponds to a normal band insulating phase. $\mu$ is a chemical potential, controlled by doping.
%
%
%
%
Magnetic impurities $\boldsymbol{S}(\boldsymbol{r},\tau)$ experience RKKY interactions, denoted by $I_{\boldsymbol{r}\boldsymbol{r}'}$. Since they are doped at random positions, we take the coupling constant as a random variable, described by the gaussian distribution function $P(I_{\bm{r}\bm{r}'})$. Such magnetic impurities also interact with conduction electrons, described by the Kondo-type interaction $J$.
%
%
$\mathcal{S}_{B}$ is a Berry phase term in the spin coherent-state representation \cite{Spin_Textbook}.

Hinted from the recent experiment \cite{Kim_Kim_Sasaki_FMTI}, we introduce both ferromagnetic and valence-bond order parameters. Introducing a neutral fermion field to describe an impurity spin as $\bm{S}(\bm{r},\tau) = \frac{1}{2} f_{\sigma}^{\dagger}(\bm{r},\tau) \bm{\sigma}_{\sigma\sigma'} f_{\sigma'}(\bm{r},\tau)$ with a single occupancy constraint $f_{\sigma}^{\dagger}(\bm{r},\tau) f_{\sigma}(\bm{r},\tau) = 1$, and performing the standard decomposition \cite{Lee_Nagaosa_Wen_RMP} in this effective model, we construct
\bqa &&
\mathcal{F}_{MF}[\boldsymbol{\Phi}_{\bm{r}}, \chi_{\bm{r}\bm{r}'}, \lambda_{\bm{r}}; \mu, T] = - T \int_{-\infty}^{\infty} d I_{\bm{r}\bm{r}'} P(I_{\bm{r}\bm{r}'}) \ln \int D \psi_{\sigma\alpha}(\boldsymbol{r},\tau) D f_{\sigma}(\boldsymbol{r},\tau) \nn && \exp \Bigl[ - \int_{0}^{1/T} d \tau \int \frac{d^{3} \boldsymbol{k}}{(2\pi)^{3}} \psi_{\sigma\alpha}^{\dagger}(\boldsymbol{k},\tau) \Bigl\{ (\partial_{\tau} - \mu) \boldsymbol{I}_{\sigma\sigma'} \otimes \boldsymbol{I}_{\alpha\alpha'} + v \boldsymbol{k} \cdot \boldsymbol{\sigma}_{\sigma\sigma'} \otimes \boldsymbol{\tau}_{\alpha\alpha'}^{z} + m(|\boldsymbol{k}|) \boldsymbol{I}_{\sigma\sigma'} \otimes \boldsymbol{\tau}_{\alpha\alpha'}^{x} \Bigr\} \psi_{\sigma'\alpha'}(\boldsymbol{k},\tau) \nn && - \int_{0}^{1/T} d \tau \int d^{3} \boldsymbol{r} J \psi_{\sigma\alpha}^{\dagger}(\boldsymbol{r},\tau) (\boldsymbol{\sigma}_{\sigma\sigma'} \otimes \boldsymbol{I}_{\alpha\alpha'}) \psi_{\sigma'\alpha'}(\boldsymbol{r},\tau) \cdot \boldsymbol{\Phi}_{\bm{r}} - \int_{0}^{1/T} d \tau \int
d^{3} \boldsymbol{r} \int d^{3} \boldsymbol{r}' \Bigl\{ f_{\sigma}^{\dagger}(\boldsymbol{r},\tau) \Bigl( (\partial_{\tau} + \lambda_{\bm{r}}) \delta^{(3)}(\bm{r} - \bm{r}') \nn && - I_{\bm{r}\bm{r}'} \chi_{\bm{r}\bm{r}'} \Bigr) f_{\sigma}(\boldsymbol{r}',\tau) + I_{\bm{r}\bm{r}'} f_{\sigma}^{\dagger}(\boldsymbol{r},\tau) (\boldsymbol{\Phi}_{\bm{r}'} \cdot
\boldsymbol{\sigma})_{\sigma\sigma'} f_{\sigma'}(\boldsymbol{r},\tau) \Bigr\} - \frac{1}{T} \int d^{3} \boldsymbol{r} \int d^{3} \boldsymbol{r}' \Bigl\{ \lambda_{\bm{r}} \delta^{(3)}(\bm{r} - \bm{r}') + I_{\bm{r}\bm{r}'} (\chi_{\bm{r}\bm{r}'} \chi_{\bm{r}'\bm{r}} \nn && - \boldsymbol{\Phi}_{\bm{r}} \cdot \boldsymbol{\Phi}_{\bm{r}'}) \Bigr\} \Bigr] , \eqa
where $\boldsymbol{\Phi}_{\bm{r}}$ is a ferromagnetic order parameter, $\chi_{\bm{r}\bm{r}'}$ is a valence-bond singlet order parameter, and $\lambda_{\bm{r}}$ is a Lagrange multiplier field to impose the single occupancy constraint, respectively, given by a functional of $I_{\bm{r}\bm{r}'}$ in the self-consistent mean-field analysis.

Our scenario is as follows. A control parameter is $\langle I_{\bm{r}\bm{r}'} \rangle = \int_{-\infty}^{\infty} d I_{\bm{r}\bm{r}'} P(I_{\bm{r}\bm{r}'}) I_{\bm{r}\bm{r}'}$, which evolves from ferromagnetic- to antiferromagnetic-, increasing the concentration of magnetic impurities. Within the ferromagnetic-in-average region of $\mbox{sgn} \langle I_{\bm{r}\bm{r}'} \rangle < 0$, we expect $\langle \chi_{\bm{r}\bm{r}'} \rangle = \int_{-\infty}^{\infty} d I_{\bm{r}\bm{r}'} P(I_{\bm{r}\bm{r}'}) \chi_{\bm{r}\bm{r}'} = 0$ and $\langle \boldsymbol{\Phi}_{\bm{r}} \rangle = L^{-3} \int d^{3} \boldsymbol{r}' \int_{-\infty}^{\infty} d I_{\bm{r}\bm{r}'} P(I_{\bm{r}\bm{r}'}) \boldsymbol{\Phi}_{\bm{r}'} = 0$ but an anomalous power-law dependence of the spin susceptibility, given by $L^{-3} \int d^{3} \boldsymbol{r} L^{-3} \int d^{3} \boldsymbol{r}' \langle \boldsymbol{\Phi}_{\bm{r}} \cdot \boldsymbol{\Phi}_{\bm{r}'} \rangle = L^{-3} \int d^{3} \boldsymbol{r} L^{-3} \int d^{3} \boldsymbol{r}' \int_{-\infty}^{\infty} d I_{\bm{r}\bm{r}'} P(I_{\bm{r}\bm{r}'}) \boldsymbol{\Phi}_{\bm{r}} \cdot \boldsymbol{\Phi}_{\bm{r}'} \propto T^{-\eta_{\bm{\Phi}}}$ as observed by the experiment. Within the antiferromagnetic-in-average region of $\mbox{sgn} \langle I_{\bm{r}\bm{r}'} \rangle > 0$, we expect $\langle \boldsymbol{\Phi}_{\bm{r}} \rangle = L^{-3} \int d^{3} \boldsymbol{r}' \int_{-\infty}^{\infty} d I_{\bm{r}\bm{r}'} P(I_{\bm{r}\bm{r}'}) \boldsymbol{\Phi}_{\bm{r}'} = 0$ and $\langle \chi_{\bm{r}\bm{r}'} \rangle = \int_{-\infty}^{\infty} d I_{\bm{r}\bm{r}'} P(I_{\bm{r}\bm{r}'}) \chi_{\bm{r}\bm{r}'} = 0$ but an anomalous power-law dependence of the spin-singlet susceptibility, given by $L^{-3} \int d^{3} \boldsymbol{r} L^{-3} \int d^{3} \boldsymbol{r}' L^{-3} \int d^{3} \boldsymbol{r}_{1} L^{-3} \int d^{3} \boldsymbol{r}_{1}' \langle \chi_{\bm{r}\bm{r}'} \chi_{\bm{r}_{1}\bm{r}_{1}'} \rangle \propto T^{-\eta_{\chi}}$, a liquid-like behavior of singlets. Here, $L$ is the size of our system.

This magnetic evolution gives rise to the variation in transport properties as discussed in the introduction. In particular, normal metallic transport properties in magnetoresistivity and Hall effect appear from topological semiconducting transport behaviors in the ferromagnetic-in-average region before the antiferromagnetic-in-average region. According to the above physical picture, randomly frozen magnetic clusters described by $\bm{\Phi}_{\bm{r}}$ generate effective magnetic fields to itinerant electrons. As a result, the ``local" spectrum of itinerant electrons becomes modified into $E_{\bm{r}}(\bm{k}) = - \mu \pm \sqrt{v^{2}(k_{x}^{2} + k_{y}^{2}) + (J |\bm{\Phi}_{\bm{r}}| \pm \sqrt{m^{2}(|\bm{k}|) + v^{2} k_{z}^{2}})^{2}}$. This local spectrum implies that the gap of a topological semiconductor vanishes at position $\bm{r}$ in the case of $J |\bm{\Phi}_{\bm{r}}| > |m(|\bm{k}|)|$, splitting the Dirac spectrum into a pair of Weyl points locally. See Fig. 1. Then, a Weyl metallic island arises from a topological semiconducting island at position $\bm{r}$, regarded to be an insulator-metal ``transition" driven by random magnetic moments. Inhomogeneously distributed topological semiconductor and Weyl metal islands are the characteristic feature of the dilute magnetic topological semiconductor in the ferromagnetic regime.


In order to understand the nature of such inhomogeneous mixtures, we construct an effective field theory based on the above physical picture. It is straightforward to show that randomly quenched effective magnetic fields ($J \bm{\Phi}_{\bm{r}}$) correspond to random chiral gauge fields ($\bm{c}$), rewriting the effective Hamiltonian of the ferromagnetic regime into the standard representation of the Dirac theory with the introduction of Dirac gamma matrices. An approximation is that the distribution function of random chiral gauge fields is gaussian, which corresponds to the fact that the distribution of effective magnetic moments of ferromagnetic clusters is gaussian. We point out that this distribution function should be determined self-consistently. For example, one can transform the distribution function of $P(I_{\bm{r}\bm{r}'})$ into $P(\bm{\Phi}_{\bm{r}})$, resorting to the solution $\bm{\Phi}_{\bm{r}}[I_{\bm{r}\bm{r}'}]$ of the saddle-point analysis in Eq. (2). If one starts from a log-normal distribution function for the RKKY interaction, he/she may get a power-law distribution function for the effective magnetic moment \cite{Kettemann}, which gives rise to quantum Griffiths phenomena \cite{Quantum_Griffiths}. Unfortunately, we do not touch this difficult issue in the present problem. Instead, we consider a gaussian distribution function for random chiral gauge fields, where ferromagnetic clusters are assumed to be independent with each other.

We start from an effective Dirac theory with random chiral gauge fields, $S = \int d^{4} x \bar{\psi} (i \gamma^{\mu} \partial_{\mu} - m + \gamma^{\mu} \gamma^{5} c_{\mu}) \psi$, where the distribution function of $c_{\mu}$ is assumed to be gaussian with its variance $\Gamma$. Applying the replica trick and performing the gaussian integral for random chiral gauge fields, we find that effective nonlocal-in-time ``interactions" of chiral currents arise between different replicas \cite{Disorder_Review}. Rewriting the bare action in terms of renormalized fields and renormalized coupling constants with the introduction of counter terms, i.e., $\mathcal{S}_{B} = \mathcal{S}_{R} + \mathcal{S}_{CT}$, we construct an effective field theory for renormalization group analysis
\bqa \mathcal{S}_{R} &=& \int_{0}^{\beta} d \tau \int d^{d-1} \bm{r} \Bigl\{ \bar{\psi}_{R}^{a}(\bm{r},\tau) (i \gamma^{\tau} \partial_{\tau} + i v_{R} \bm{\gamma} \cdot \bm{\nabla} - m_{R}) \psi_{R}^{a}(\bm{r},\tau) \nn &+& \frac{\Gamma_{R}}{2} \int_{0}^{\beta} d \tau' \bar{\psi}_{R}^{a}(\bm{r},\tau) \gamma^{\mu} \gamma^{5} \psi_{R}^{a}(\bm{r},\tau) \bar{\psi}_{R}^{a'}(\bm{r},\tau') \gamma_{\mu} \gamma^{5} \psi_{R}^{a'}(\bm{r},\tau') \Bigr\} , \nn \mathcal{S}_{CT} &=& \int_{0}^{\beta} d \tau \int d^{d-1} \bm{r} \Bigl\{ \bar{\psi}_{R}^{a}(\bm{r},\tau) ( \delta_{\psi}^{\omega} i \gamma^{\tau} \partial_{\tau} + \delta_{\psi}^{\bm{k}} i v_{R} \bm{\gamma} \cdot \bm{\nabla} - \delta_{m} m_{R}) \psi_{R}^{a}(\bm{r},\tau) \nn &+& \delta_{\Gamma} \frac{\Gamma_{R}}{2} \int_{0}^{\beta} d \tau' \bar{\psi}_{R}^{a}(\bm{r},\tau) \gamma^{\mu} \gamma^{5} \psi_{R}^{a}(\bm{r},\tau) \bar{\psi}_{R}^{a'}(\bm{r},\tau') \gamma_{\mu} \gamma^{5} \psi_{R}^{a'}(\bm{r},\tau') \Bigr\} \eqa
Here, ${\psi}_{R}^{a}(\bm{r},\tau)$ is a renormalized electron field with a replica index $a = 1, ..., R$, and $v_{R}$, $m_{R}$, $\Gamma_{R}$ are renormalized velocity, renormalized mass, renormalized variance, respectively. $\delta_{\psi}^{\omega}$, $\delta_{\psi}^{\bm{k}}$, $\delta_{m}$, and $\delta_{\Gamma}$ are introduced to absorb infinities resulting from quantum corrections. These renormalized field and parameters are related with the bare field and parameters as follows \bqa && \psi_{B}^{a}(\bm{r},\tau) = {Z_{\psi}^{\omega}}^{\frac{1}{2}} \psi_{R}^{a}(\bm{r},\tau) , ~~~ v_{B} = Z_{\psi}^{\bm{k}} {Z_{\psi}^{\omega}}^{-1} v_{R} , \nn && m_{B} = Z_{m} {Z_{\psi}^{\omega}}^{-1} m_{R} , ~~~ \Gamma_{B} = \mu^{\varepsilon} Z_{\Gamma} {Z_{\psi}^{\omega}}^{-2} \Gamma_{R} , \eqa where renormalization constants are given by \bqa && Z_{\psi}^{\omega} = 1 + \delta_{\psi}^{\omega} , ~~~ Z_{\psi}^{\bm{k}} = 1 + \delta_{\psi}^{\bm{k}} , \nn && Z_{m} = 1 + \delta_{m} , ~~~ Z_{\Gamma} = 1 + \delta_{\Gamma}  . \eqa Here, $\mu$ is a scale of momentum, distinguished from the chemical potential before and $\varepsilon = d - 3$.


Performing the standard procedure for the renormalization group analysis, we find renormalization group equations, where both vertex and self-energy corrections are introduced self-consistently. See Fig. 3, where all quantum corrections are shown as Feynman's diagrams up to the one-loop order for vertex corrections and the two-loop order for self-energy corrections. All details are shown in our supplementary material. Here, we point out that the renormalization constant of the ``interaction" vertex remains to be $Z_{\Gamma} = 1$, where the divergence of the particle-hole ladder diagram is canceled by that of the particle-particle channel while other vertex corrections do not give rise to divergences. On the other hand, the Fock diagram in the one-loop order and both the rainbow diagram and the crossed diagram with a vertex correction in the two-loop order for self-energy corrections contribute to the wave-function renormalization constant while others do not cause divergences. In particular, the role of the rainbow diagram turns out to be crucial in the emergence of a novel metallic fixed point of $\Gamma_{R} \rightarrow \infty$ and $m_{R} \rightarrow 0$, identified with a disordered Weyl metallic phase. As a result, we find \bqa && \frac{d \ln \Gamma_{R}}{d \ln \mu} = 1 - a_{\Gamma} \Gamma_{R} - b_{\Gamma} \Gamma_{R}^{2} , \nn && \frac{d \ln m_{R}}{d \ln \mu} = - 1 - a_{m} \Gamma_{R} + b_{m} \Gamma_{R}^{2} , \eqa where positive numerical constants are given by $a_{\Gamma} =\frac{2}{\pi}$, $b_{\Gamma} = \frac{29}{4\pi^{2}}$, $a_{m} = \frac{3}{\pi}$, and $b_{m} = \frac{3}{2\pi^{2}}$. We emphasize that the chemical potential lies between the band gap. The renormalization group flow of these equations is shown in Fig. 4, which confirms our proposed phase diagram (Fig. 2). First, we focus on the quantum critical point of the topological phase transition, identified with $m_{R} = 0$. Then, it is easy to see that there exists an unstable disorder fixed point $\Gamma_{R} = \Gamma_{c}$, which separates two stable fixed points of $\Gamma_{R} = 0$ and $\Gamma_{R} \rightarrow \infty$. This means that the Dirac semimetallic state, arising at the critical point, remains to be stable in the case of weak randomness, expected since the density of states vanishes. However, it is quite interesting that anti-screening appears for random  fluctuations in chiral currents in contrast with those in charge currents. Recall that the electric charge is screened to decrease at low energies \cite{Lee_Nagaosa_Wen_RMP}. Second, we start from an insulating phase, increasing the variance of random chiral gauge fields. Then, we reach a novel stable fixed point of $(m_{R} \rightarrow 0, \Gamma_{R} \rightarrow \infty)$, separated from two insulating fixed points of $(m_{R} \rightarrow \infty, \Gamma_{R} \rightarrow 0)$ (topological insulator) and $(m_{R} \rightarrow - \infty, \Gamma_{R} \rightarrow \infty)$ (band insulator). The appearance of this fixed point is quite surprising since the mass parameter renormalizes to vanish, which originates from random fluctuations of chiral currents. Although metallicity can be enhanced by the interplay between disorders and interactions \cite{Anderson_Mott}, the present metallicity results from the interplay between randomness and topology of a band structure in the approach of an effective field theory.

\begin{figure}[t]
\includegraphics[width=0.3\textwidth]{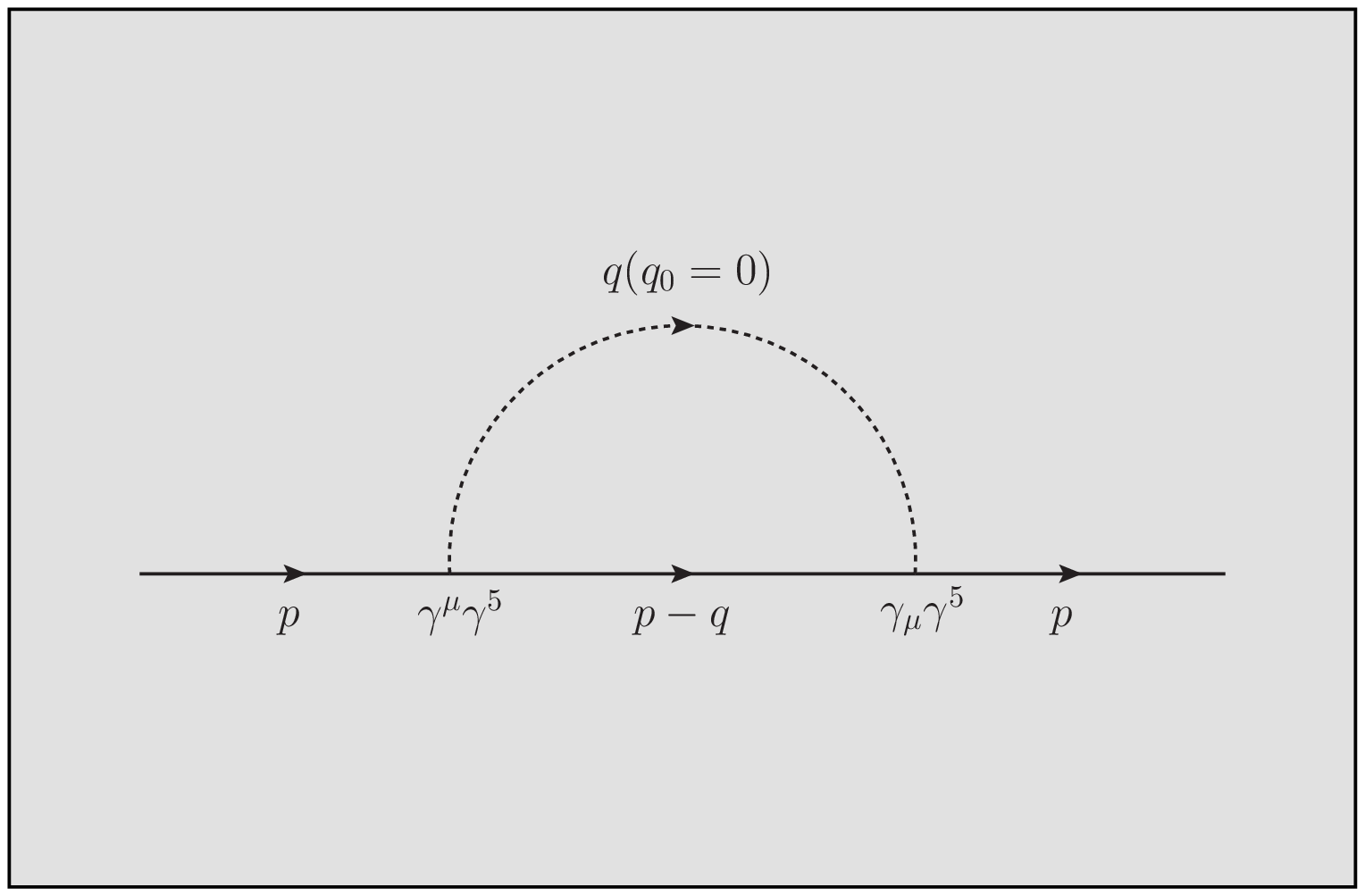}
\includegraphics[width=0.3\textwidth]{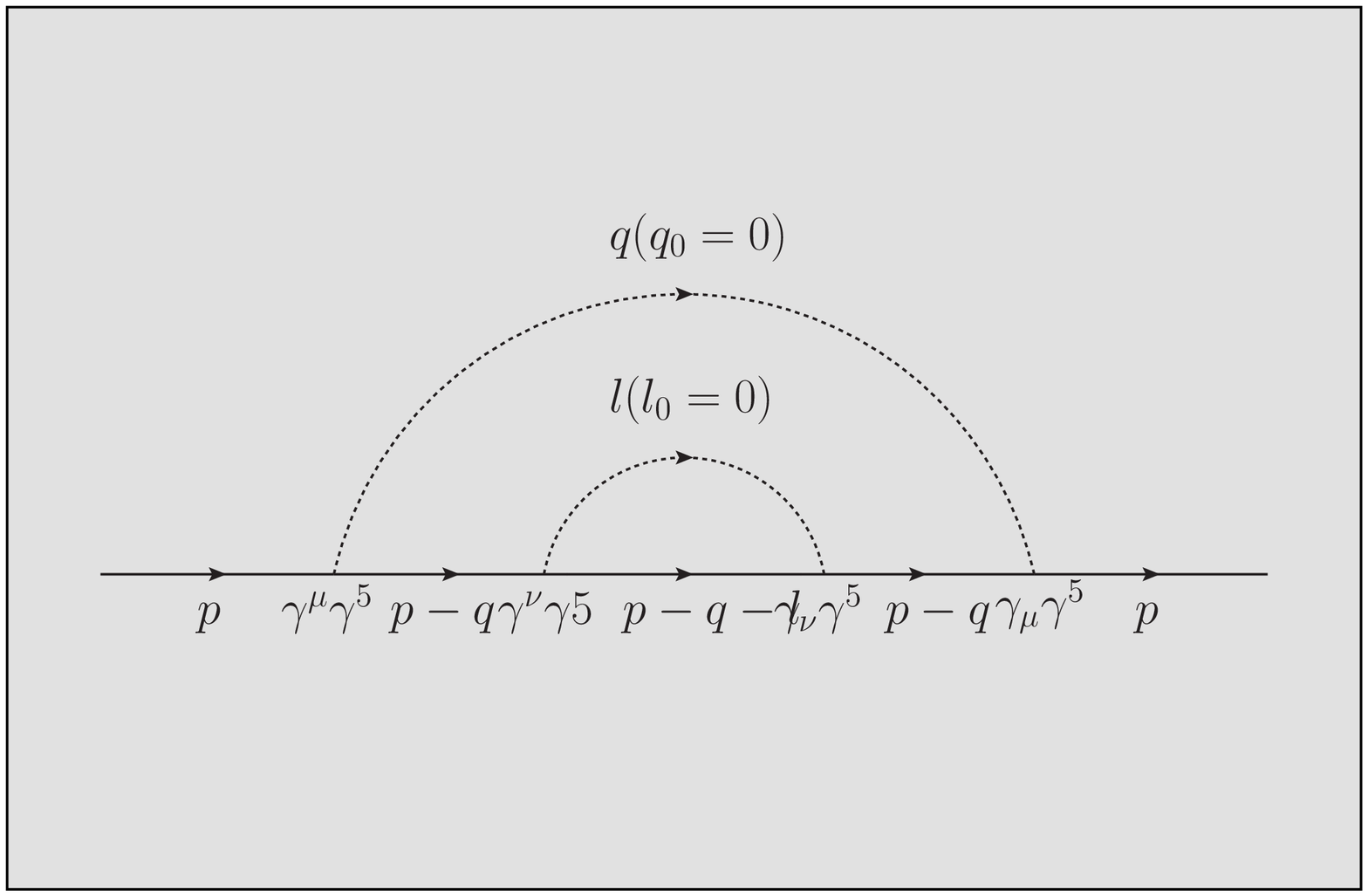}
\includegraphics[width=0.3\textwidth]{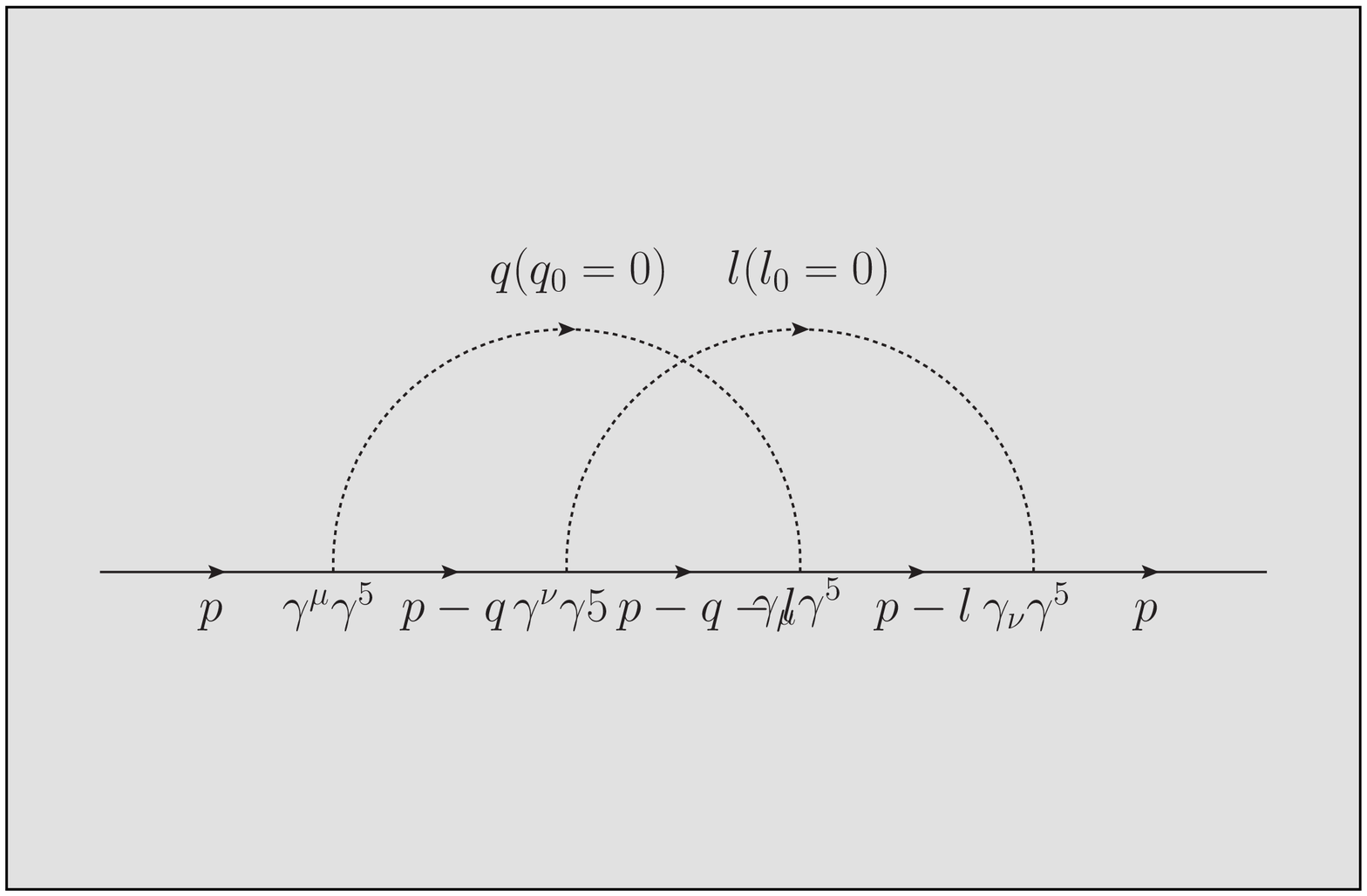}
\\
\includegraphics[width=0.3\textwidth]{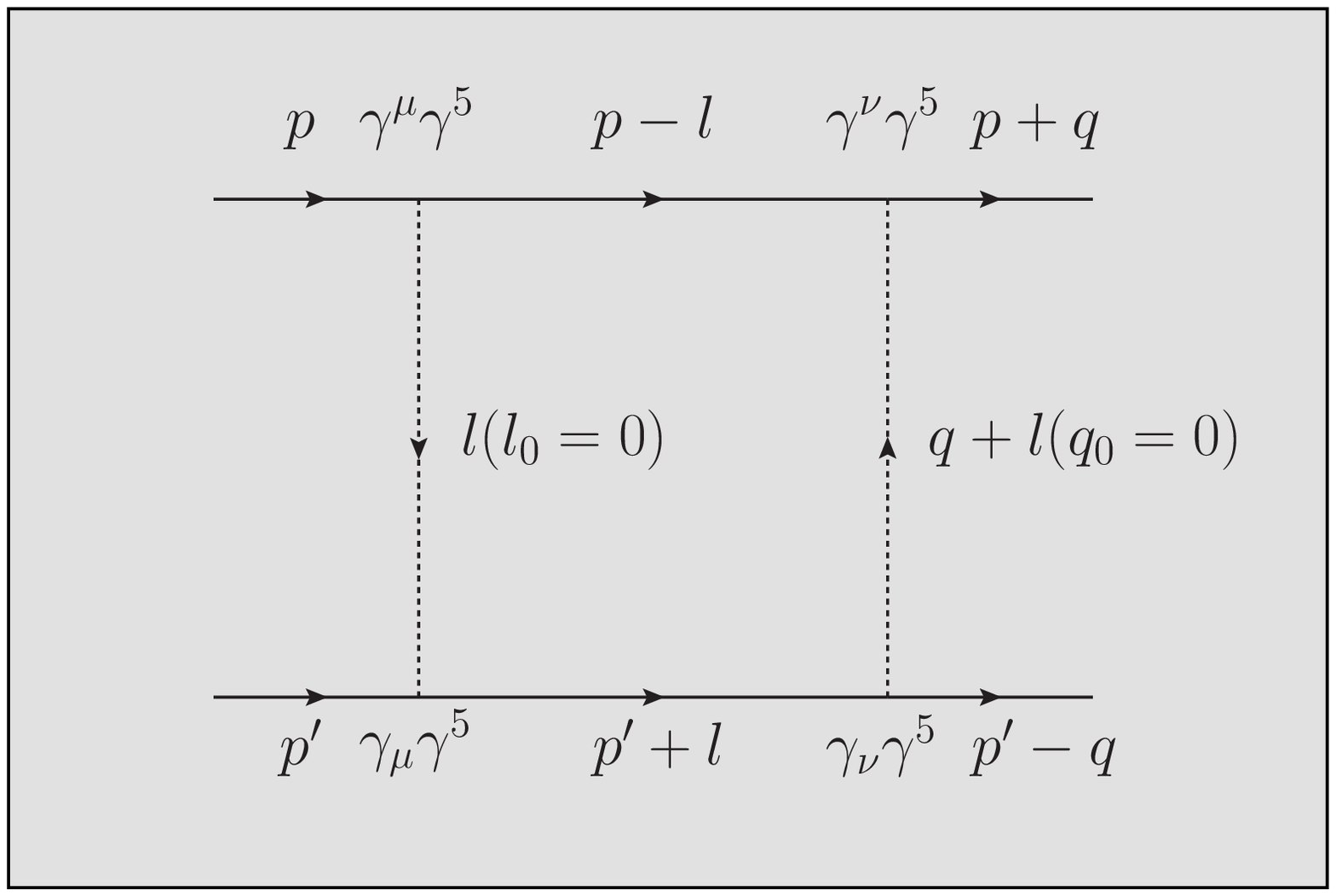}
\includegraphics[width=0.3\textwidth]{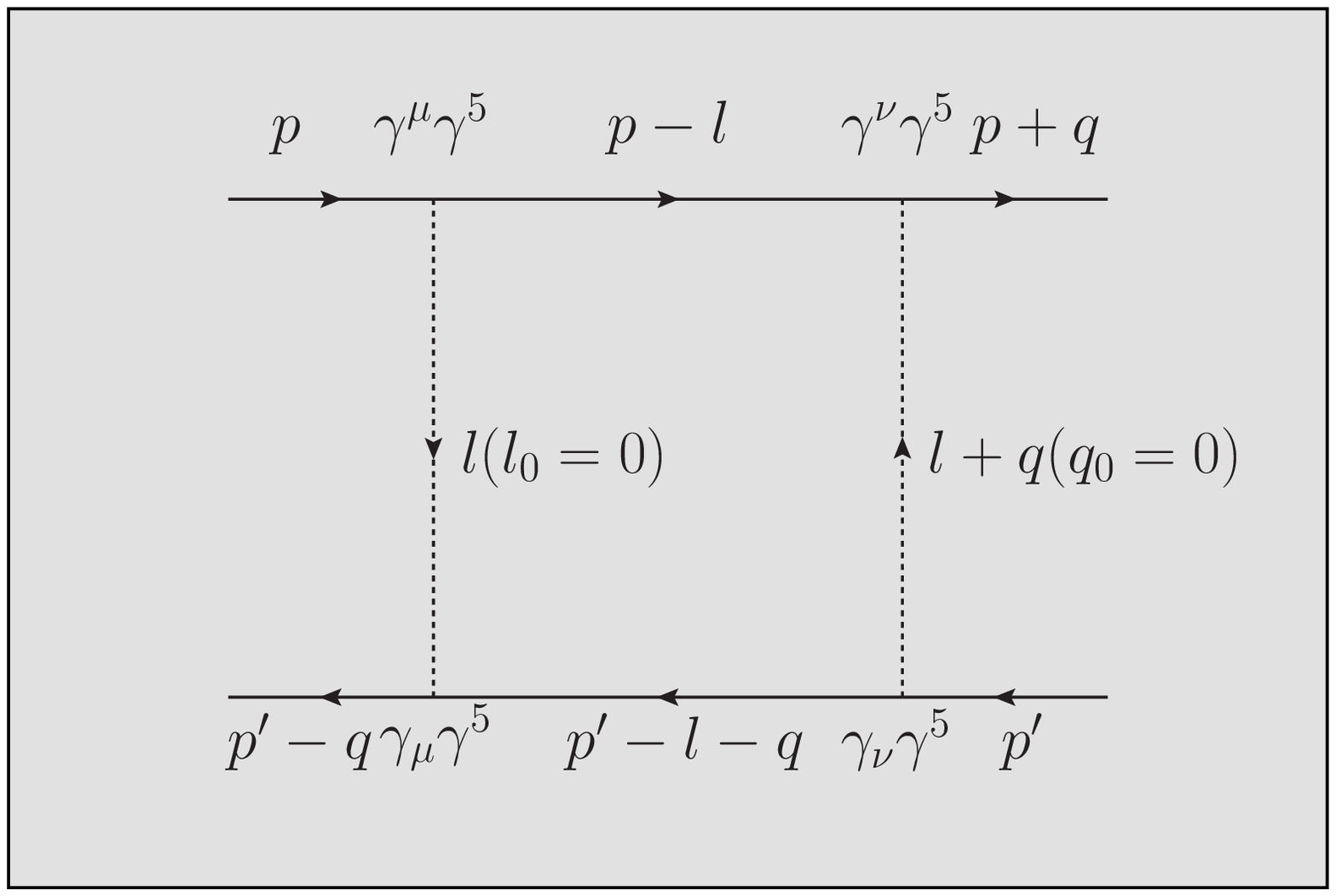}
\includegraphics[width=0.3\textwidth]{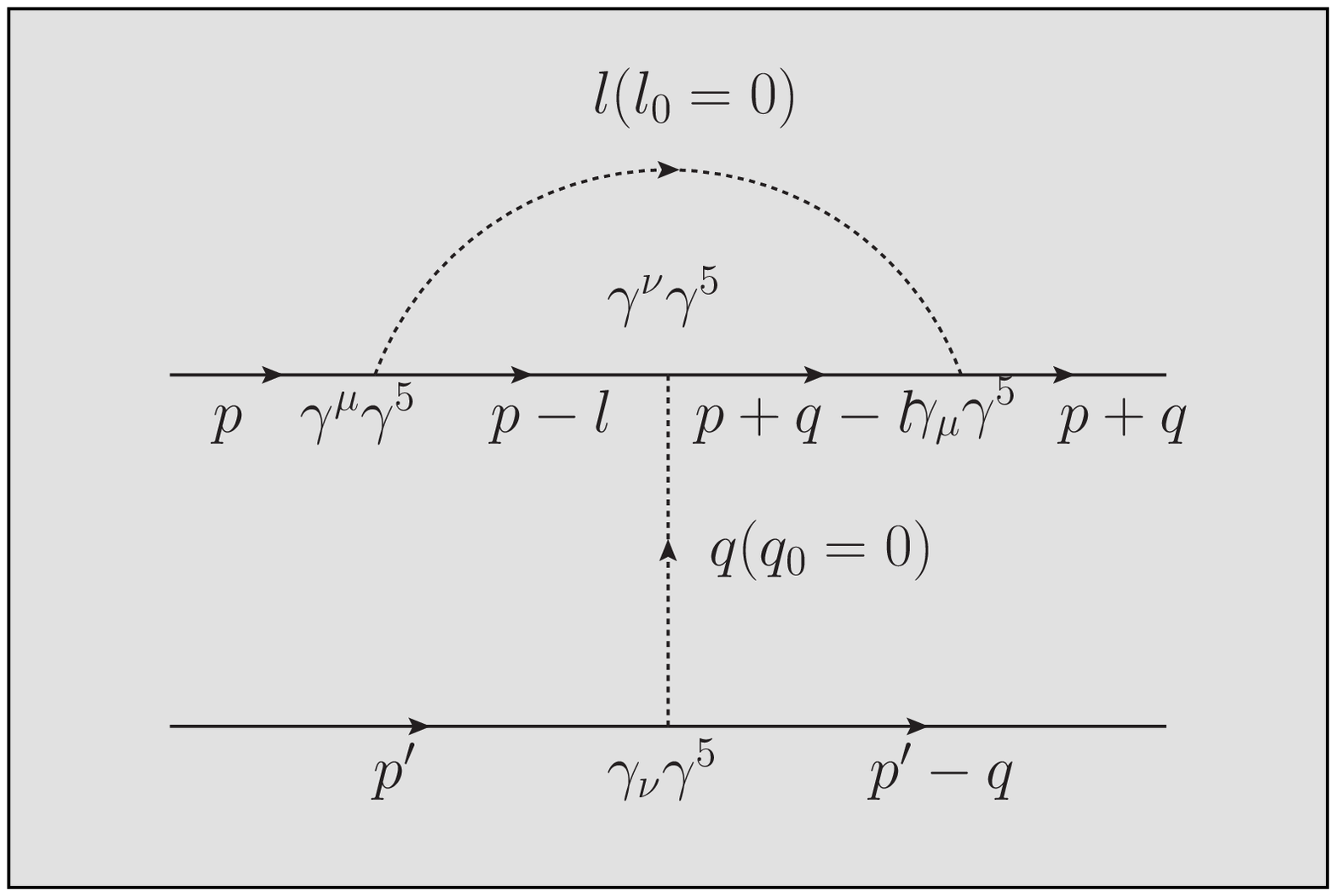}
\caption{Feynman's diagrams up to the one-loop order for vertex corrections and the two-loop order for self-energy corrections. First of all, we point out that quantum corrections including fermion loops vanish in the replica limit of $N_{R} \rightarrow 0$. Three types of quantum corrections contribute to the vertex renormalization. It turns out that the $1/\varepsilon$ divergence in the ladder diagram of the particle-hole channel is canceled by that of the particle-particle channel, where the diagram with a vertex correction does not cause the divergence. As a result, the vertex renormalization constant remains to be $Z_{\Gamma} = 1$. The Fock diagram results in the $1/\varepsilon$ divergence for the fermion self-energy in the one-loop order, and both the rainbow diagram and the crossed diagram with a vertex correction also causes that in the two-loop order. The wave-function renormalization constant $Z_{\psi}^{\omega}$ is given by these three contributions. In particular, the role of the rainbow diagram turns out to be essential, meaning that we cannot reach the disorder driven novel metallic fixed point without it, where the sign of $b_{m}$ changes from negative to positive in the renormalization group equation for the mass parameter.} \label{Feynman_Diagram}
\end{figure}

One may understand the emergence of this novel metallic fixed point as follows. First of all, $\Gamma_{R} \rightarrow \infty$ is difficult to be compatible with $m_{R} \rightarrow \pm \infty$ since $\Gamma_{R} \rightarrow \infty$ implies that most regions become Weyl metallic with $\Gamma_{R} \gg m_{R}$, giving rise to gap closing inevitably. The only consistent way for the existence of $(m_{R} \rightarrow \pm \infty, \Gamma_{R} \rightarrow \infty)$ is that the mass parameter increases faster than the variance of effective magnetic fields. Actually, we find that the mass gap increases faster than the variance if we neglect the renormalization given by the rainbow diagram. This means that the Weyl metallic island does not occur and the insulating phase survives although the variance goes toward an infinite fixed point. This does not make any sense because we fail to figure out the nature of a quantum phase transition between two insulating phases given by $(m_{R} \rightarrow \infty, \Gamma_{R} = 0)$ and $(m_{R} \rightarrow \infty, \Gamma_{R} \rightarrow \infty)$, focusing on the topological semiconducting side. Incorporating the contribution of the rainbow diagram, we observe that the sign of $b_{m}$ in the renormalization group equation for the mass parameter changes from negative to positive, giving rise to the disorder driven metallic fixed point. We interpret this infinite variance fixed point with zero mass gap as an inhomogeneously distributed Weyl metallic state, where transport properties are described by axion electrodynamics, which should be distinguished from the diffusive Fermi-liquid state \cite{Disorder_Review}. The nature of the quantum phase transition between $(m_{R} \rightarrow \pm \infty, \Gamma_{R} = 0)$ and $(m_{R} = 0, \Gamma_{R} \rightarrow \infty)$ is expected to be the first order, where an insulating phase persists just before the disordered Weyl metallic state. However, we cannot exclude the possibility of an additional phase transition associated with percolation, which may be responsible for a genuine insulator-metal transition beyond the present description.

\begin{figure}[t]
\includegraphics[width=0.4\textwidth]{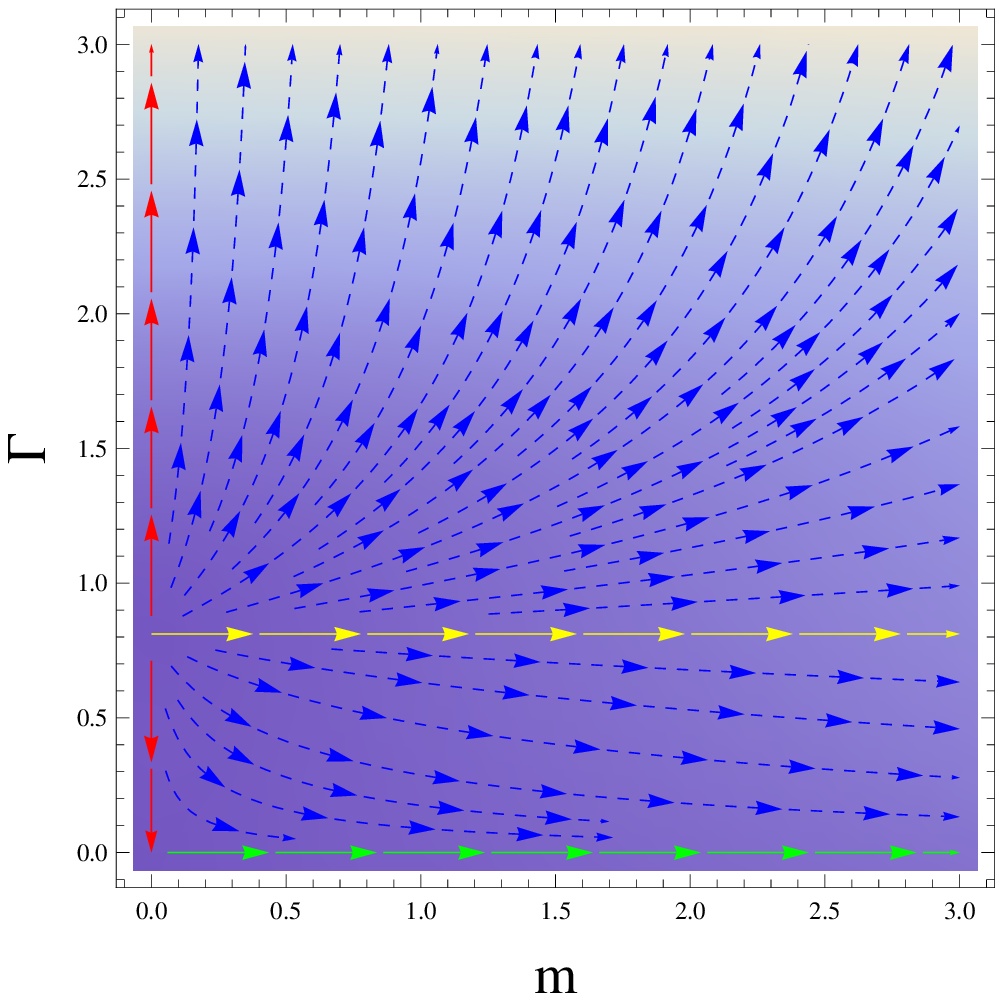}
\includegraphics[width=0.4\textwidth]{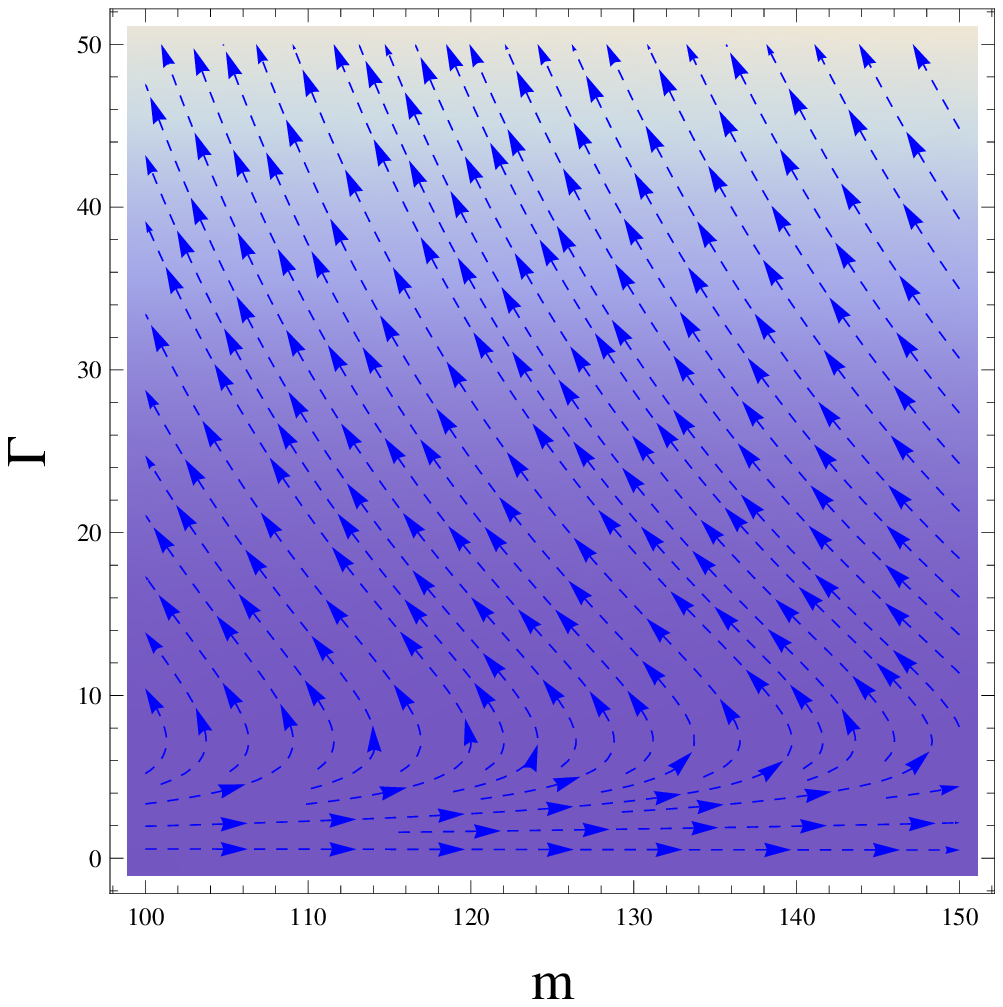}
\caption{Renormalization group flow as the solution of the coupled renormalization group equations (6). A characteristic feature is the emergence of a novel stable fixed point $(m_{R} = 0, \Gamma_{R} \rightarrow \infty)$, identified with an inhomogeneously distributed Weyl metallic phase which coexists with insulating islands (right). This metallic fixed point originates from random fluctuations of chiral currents due to effective random magnetic fields of ferromagnetic clusters. There exists a quantum phase transition of the second order between the Dirac semimetallic state $(m_{R} = 0, \Gamma_{R} = 0)$ and the disordered Weyl metallic phase $(m_{R} = 0, \Gamma_{R} \rightarrow \infty)$, identified with a disorder-driven quantum critical point $(m_{R} = 0, \Gamma_{R} = \Gamma_{c})$ (left). On the other hand, it would be the first order quantum phase transition between an insulating phase of either $(m_{R} \rightarrow \infty, \Gamma_{R} = 0)$ (topological insulator) or $(m_{R} \rightarrow - \infty, \Gamma_{R} = 0)$ (normal semiconductor) and the disordered Weyl metal state. See the text for more details.} \label{Renormalization_Group_Flow}
\end{figure}

We speculate what would happen when the chemical potential lies above the band gap, resulting in a Fermi surface. First of all, the presence of the Fermi surface changes the engineering dimension of the variance $\Gamma_{R}$ from $+1$ to $-1$, making it relevant. Keeping the physics of anti-screening, we write down the renormalization group equation for the ``interaction" vertex \bqa && \frac{d \ln \Gamma_{R}}{d \ln \mu} = - 1 - c \Gamma_{R} , \nonumber \eqa where $c$ is a positive numerical constant. This allows the $\Gamma_{R} \rightarrow \infty$ fixed point only in the low energy limit. Considering the emergence of randomly distributed Weyl metallic islands in the case of zero chemical potential, we expect that the Dirac point is separated into a pair of Weyl points locally due to local time reversal symmetry breaking if the critical point of $m_{R} = 0$ is taken into account for example, and thus the single Fermi surface with degeneracy in the Dirac spectrum splits into a pair of chiral Fermi surfaces locally, which encloses each Weyl point with definite chirality. The nature of a pair of chiral Fermi surfaces turns out to differ from a normal Fermi surface in the respect that both the Berry curvature, which originates from the Weyl point identified with a magnetic monopole in momentum space, and chiral anomaly, which means that this pair of Weyl points are not independent but connected to each other, change electromagnetic properties seriously, described by the axion electrodynamics \cite{Weyl_Metal_I,Weyl_Metal_IV,Axion_EM2,Axion_EM3}, as discussed before . The emergence of a randomly distributed pair of chiral Fermi surfaces is an essential feature when the chemical potential lies above the band gap, regarded to be an extended physical picture of the case of zero chemical potential.

Although we expect that this infinite variance fixed point should exhibit the strong inhomogeneity, its thermodynamic nature looks much complicated, where randomly distributed ferromagnetic clusters would interact with each other beyond our present description. Then, quantum Griffiths phenomena \cite{Quantum_Griffiths} are expected to appear, implying that the nontrivial power-law exponent of the spin susceptibility at low temperatures should show non-universal continuous evolutions within the ferromagnetic-in-average region. Such quantum Griffiths effects may be incorporated, resorting to a power-law distribution function instead of the gaussian distribution function for random chiral gauge fluctuations.

We believe that this infinite randomness fixed point can be verified by atomic force microscopy. Although the local electronic spectrum will not show strong inhomogeneity around zero bias in the metallic regime, it should be observed deep inside the spectrum around $- \mu$, where $\mu$ is the chemical potential. In a certain region a gap feature appears while such a gap does not exist in the vicinity of the same energy scale at a different position, where a Weyl metallic island exists. See Fig. 5.

\begin{figure}[t]
\includegraphics[width=0.8\textwidth]{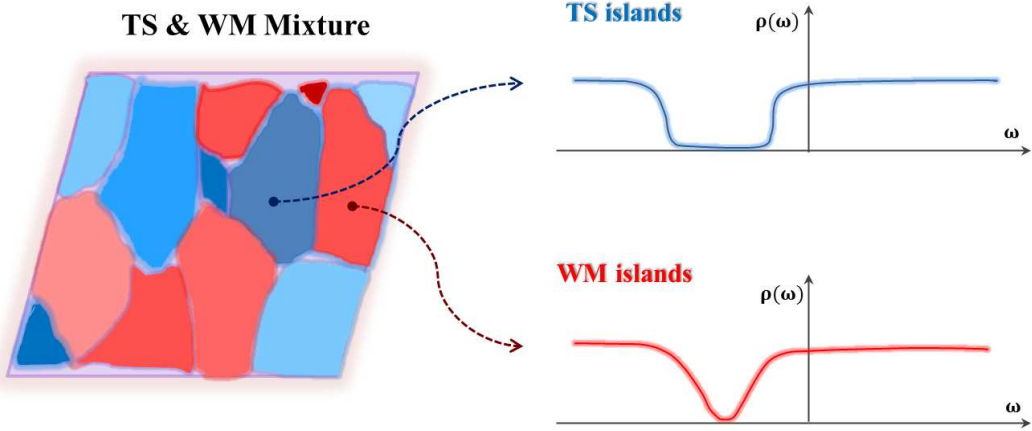}
\caption{A schematic picture for the local density of states probed by atomic force microscopy. Consider the case when the chemical potential lies above the band gap, which corresponds to a metallic state. Recalling that the infinite variance fixed point is identified with inhomogeneous mixtures between a normal Fermi surface with degeneracy in a Dirac spectrum and a pair of chiral Fermi surfaces without degeneracy in a pair of Weyl spectrum, we predict that a gap feature appears in a certain region which corresponds to the region of small ferromagnetic clusters while a v-shaped pseudogap feature results at a different position which coincides with the region of large ferromagnetic clusters exceeding the band gap.} \label{AFM}
\end{figure}


In summary, we proposed the problem of dilute magnetic topological semiconductors, novel physics of which beyond that of dilute magnetic semiconductors is the emergence of randomly distributed Weyl metallic islands. Performing the renormalization group analysis for an effective Dirac theory with random chiral gauge fluctuations, expected to encode the information of randomly quenched magnetic moments, we find that the variance of random chiral gauge fields reaches an infinite fixed point as long as average magnetic correlations remain to be ferromagnetic, which enforces the mass gap to vanish. As a result, we find a disorder driven novel metallic phase and an associated insulator-metal phase transition beyond either the Anderson or the Mott metal-insulator transition, where this metallic state appears to be identified with the infinite variance fixed point. Recalling that quantum Griffiths phenomena may arise in the vicinity of this infinite variance fixed point, we predicted continuous nonuniversal changes in the temperature exponent of the uniform spin susceptibility. In addition, we claimed that this picture of inhomogeneous mixtures can be verified by atomic force microscopy. However, a difficult fundamental problem remains, that is, how to understand transport coefficients near this infinite variance fixed point, where random axion electrodynamics arises to govern electromagnetic properties, identified with the problem of dilute magnetic topological semiconductors.

\section*{Acknowledgement}

This study was supported by the Ministry of Education, Science, and Technology (No. 2012R1A1B3000550 and No. 2011-0030785) of the National Research Foundation of Korea (NRF) and by TJ Park Science Fellowship of the POSCO TJ Park Foundation.

\appendix

\section{From effective magnetic fields to chiral gauge fields}

The kinetic-energy sector for dynamics of bulk electrons can be rewritten as the standard representation of the Dirac theory in the following way
\bnn
S[\psi^\dag,\psi]
&=&\int_{0}^{\beta} d \tau \int \frac{d^{3} \boldsymbol{k}}{(2\pi)^{3}} \psi_{\sigma\alpha}^{\dagger}(\boldsymbol{k},\tau) \l\{\partial_{\tau} \boldsymbol{I}_{\sigma\sigma'} \otimes \boldsymbol{I}_{\alpha\alpha'} + v_F\boldsymbol{k} \cdot \boldsymbol{\sigma}_{\sigma\sigma'} \otimes \boldsymbol{\tau}_{\alpha\alpha'}^{z} + m(|\boldsymbol{k}|) \boldsymbol{I}_{\sigma\sigma'} \otimes \boldsymbol{\tau}_{\alpha\alpha'}^{x} \r\} \psi_{\sigma'\alpha'}(\boldsymbol{k},\tau)\\&=&\int_{0}^{\beta} d \tau \int d^{3} \boldsymbol{x}\psi^{\dag}(\boldsymbol{x},\tau) \br\{ \partial_{\tau}\begin{pmatrix} I_{2\times2}&0\\0&I_{2\times2}\end{pmatrix} + v_F (-\i\boldsymbol{\triangledown})\cdot \begin{pmatrix} \boldsymbol{\sigma}&0\\0&-\boldsymbol{\sigma}\end{pmatrix} + m \begin{pmatrix} 0& I_{2\times2}\\ I_{2\times2}&0\end{pmatrix} \bl\} \psi(\boldsymbol{x},\tau)\\
&=&\int_{0}^{\beta} d \tau  \int d^{3} \boldsymbol{x} \psi^{\dag}(\boldsymbol{x},\tau) \begin{pmatrix} 0& I_{2\times2}\\ I_{2\times2}&0\end{pmatrix}\br\{ \partial_{\tau}\begin{pmatrix} I_{2\times2}&0\\0&I_{2\times2}\end{pmatrix} + v_F \i\boldsymbol{\triangledown}\cdot \begin{pmatrix}0& \boldsymbol{\sigma}\\-\boldsymbol{\sigma}&0\end{pmatrix} + m I_{4\times4}\bl\} \psi(\boldsymbol{x},\tau)\\
&=&\int d^{4}x \bar{\psi}(x)\l\{\i\gam0\partial_{\tau} + v_F \i\boldsymbol{\triangledown}\cdot \boldsymbol{\gamma} + m \r\} \psi(x) , \\
\enn
where Dirac gamma matrices are given by $\gam 0=\begin{pmatrix} 0& -\imath\\-\imath&0\end{pmatrix}$ and $\gam k=\begin{pmatrix}0& -\sigma^k\\ \sigma^k&0\end{pmatrix}$.
%
%
%
%

Next, we consider an effective Zeeman coupling term, $H_{eff} = H_{0} - J\boldsymbol{\Phi}\cdot\boldsymbol{S} \equiv H_0+ H_{int}$, where $H_{0}$ is a free Dirac Hamiltonian and $\boldsymbol{\Phi}$ is an effective magnetic moment given by a ferromagnetic cluster with the Kondo coupling $J$. $\boldsymbol{S}=\f{1}{2}\psi^{\dag}(I\otimes\boldsymbol{\sigma})\psi$ represents a spin of itinerant electrons of the bulk sample. Then, it is easy to show that an effective magnetic field is equal to a chiral gauge field in this Dirac theory, given by \be H_{int}=\psi^\dag\l(-\f{1}{2}J\boldsymbol{\Phi}\cdot(I\otimes\boldsymbol{\sigma})\r)\psi=\psi^\dag\beta\l(-\f{1}{2}J\boldsymbol{\Phi}\cdot\beta I\otimes\boldsymbol{\sigma}\r)\psi=\bar{\psi}\l(\f{1}{2}J\boldsymbol{\Phi}\cdot\boldsymbol{\gamma}\gam5\r)\psi\equiv\bar{\psi}\l(\boldsymbol{C}\cdot\boldsymbol{\gamma}\gam5\r)\psi . \no\ee
In the last equality we used the identity of $\beta I\otimes\boldsymbol{\sigma}=\begin{pmatrix}0&1\\1&0\end{pmatrix}\begin{pmatrix}\boldsymbol{\sigma}&0\\ 0&\boldsymbol{\sigma}\end{pmatrix} =\begin{pmatrix}0&\boldsymbol{\sigma}\\\boldsymbol{\sigma}&0\end{pmatrix}=-\boldsymbol{\gamma}\gam5$.
Generally, we introduce a time-component of the chiral gauge field and represent the Zeeman coupling term as $ H_{int}=\bar{\psi}(C_\mu\gam\mu\gam5)\psi $.

We reach the following expression for an effective field theory in the ferromagnetic-in-average regime
\bea
S[\bar{\psi},\psi]
&=&\int d^{4}x\bar{\psi}(x)(\imath\gam \mu\partial_{\mu}+m)\psi(x)+\int d^{4}x\bar{\psi}(x)C_\mu\gam\mu\gam5\psi(x)\\
&\equiv&S_{0}[\bar{\psi},\psi]+S_{dis}[\bar{\psi},\psi; C_\mu] .\no
\eea
Then, an effective free energy becomes \bqa && \mathcal{F} = - T \int D C_{\mu}(\boldsymbol{x})P[C_{\mu}(\boldsymbol{x})] \ln \int D(\bar{\psi}(x),\psi(x)) \exp\bigl(- S_{0}[\bar{\psi},\psi] - S_{dis}[\bar{\psi},\psi; C_\mu]\bigr) , \eqa
where $P[C_{\mu}(\boldsymbol{x})] = \mathcal{N} e^{-\int d^3\boldsymbol{x}\f{[C_{\mu}(\boldsymbol{x})]^2}{2\Gam}}$ is the distribution function for chiral gauge fields with their variance $\Gam$, originating from randomly quenched ferromagnetic clusters. The coefficient $\mathcal{N}$ is determined from the normalization condition of $\mathcal{N} \int D C_{\mu}e^{-\int d^3\boldsymbol{x}\f{[C_{\mu}(\boldsymbol{x})]^2}{2\Gam}}=1$.

\section{Axion electrodynamics in the Weyl metallic phase}

We start from QED$_{4}$ (quantum electrodynamics in one time and three spatial dimensions) with the topological-in-origin $\bm{E}\cdot\bm{B}$ term, \bqa && Z_{QED_4} = \int D \psi(x) \exp\Bigl[ - \int_{0}^{\beta} d \tau \int d^{3} \bm{r} \Bigl\{ \bar{\psi}(x) \Bigl( i \gamma^{\mu} [\partial_{\mu} + i e A_{\mu}] + m \Bigr) \psi(x) - \frac{1}{4} F_{\mu\nu} F^{\mu\nu} + \theta(\bm{r}) \frac{e^{2}}{16 \pi^{2}} \epsilon^{\mu\nu\rho\delta} F_{\mu\nu} F_{\rho\delta} \Bigr\} \Bigr] , \label{QED4} \nn \eqa where $\psi(x)$ with $x = (\bm{r},\tau)$ is a four-component Dirac spinor and the coefficient $\theta({\bm{r}})$ is spatially modulated. Resorting to the anomaly equation \bqa && \partial_{\mu} (\bar{\psi} \gamma^{\mu} \gamma^{5} \psi) = - \frac{e^{2}}{16 \pi^{2}} \epsilon^{\mu\nu\rho\delta} F_{\mu\nu} F_{\rho\delta} , \eqa one may rewrite the above expression as follows \bqa && Z_{WM} = \int D \psi(x) \exp\Bigl[ - \int_{0}^{\beta} d \tau \int d^{3} \bm{r} \Bigl\{ \bar{\psi}(x) \Bigl( i \gamma^{\mu} [\partial_{\mu} + i e A_{\mu}] + m + c_{\mu} \gamma^{\mu} \gamma^{5} \Bigr) \psi(x) - \frac{1}{4} F_{\mu\nu} F^{\mu\nu} \Bigr\} \Bigr] , \nn \eqa where the chiral gauge field $c_{\mu} = (c_{\tau}, \bm{c})$ is given by $c_{\tau} = 0 $ and $\bm{c} = \bm{\nabla}_{\bm{r}} \theta(\bm{r})$. In the previous section we have shown that topological insulators under magnetic fields can be described by Eq. (A1), identical to Eq. (B3). Effective magnetic fields are identified with $\bm{\nabla}_{\bm{r}} \theta(\bm{r})$.

It is straightforward to integrate over gapped fermion excitations, resulting in an effective field theory for electromagnetic fields \bqa && \mathcal{L}_{axion} = - \frac{1}{4} F_{\mu\nu} F^{\mu\nu} + \theta(\bm{r},t) \frac{e^{2}}{16 \pi^{2}} \epsilon^{\mu\nu\rho\delta} F_{\mu\nu} F_{\rho\delta} , \eqa where time dependence in $\theta(\bm{r},t)$ has been introduced for generality. Applying the least action principle to Eq. (B4), we reach Maxwell equations to describe the axion electrodynamics \bqa && \bm{\nabla} \cdot \bm{D} = 4 \pi \rho + 2 \alpha (\bm{\nabla} P_{3} \cdot \bm{B})  , \nn && \bm{\nabla} \times \bm{H} - \frac{1}{c} \frac{\partial \bm{D}}{\partial t} = \frac{4\pi}{c} \bm{j} - 2 \alpha \Bigl( (\bm{\nabla} P_{3} \times \bm{E}) + \frac{1}{c} (\partial_{t} P_{3}) \bm{B} \Bigr) , \nn && \bm{\nabla} \times \bm{E} + \frac{1}{c} \frac{\partial \bm{B}}{\partial t} = 0 , ~~~~~ \bm{\nabla} \cdot \bm{B} = 0 , \eqa where we follow the standard cgs notation with $P_{3}(\bm{r},t) \propto \theta(\bm{r},t)$ and the fine structure constant $\alpha$ \cite{Axion_EM1}.

\section{Effective field theory for renormalization group analysis in the replica trick}

A physical observable is defined as follows \be <O(\bar{\psi},\psi)> = \int DC_\mu P[C_\mu]\f{\int D(\bar{\psi},\psi)O(\bar{\psi},\psi)e^{-S_{0}[\bar{\psi},\psi]}e^{-S_{int}[\bar{\psi},\psi;C_\mu]}}{\int D(\bar{\psi},\psi)e^{-S_{0}[\bar{\psi},\psi]}e^{-S_{int}[\bar{\psi},\psi;C_\mu]}} , \ee which can be formulated from \be <O(\bar{\psi},\psi)>=\int DC_\mu P[C_\mu]\f{\delta}{\delta J}\br|_{J=0}\log{Z[C_\mu,J]}, ~~~~~ Z[C_\mu,J]=\int D(\bar{\psi},\psi)e^{-S_{0}[\bar{\psi},\psi]}e^{-S_{int}[\bar{\psi},\psi;C_\mu] + \int d^{4} x J O(\bar{\psi},\psi)} , \label{averaged observable}\ee where $J$ is a source coupled to an operator $O(\bar{\psi},\psi)$ locally. Since the averaging procedure for disorder is not straightforward within this formulation, we take the replica trick of $\log{Z}=\limR\f{Z^R-1}{R}$, where the replicated partition function is given by $Z^R=\int D(\bar{\psi}^a,\psi^a)\exp{\l[-\sum_{a=1}^RS[\bar{\psi}^a,\psi^a;C_\mu]+\int d^{4} x J \sum_{a=1}^R O(\bar{\psi}^{a},\psi^{a})\r]}$ with a replica index ``a". Then, the above expression is reformulated as follows
\bea
&&<O(\bar{\psi},\psi; C_\mu)>\non
&=&\limR\f{1}{R}\int DC_\mu P[C_\mu]\f{\delta}{\delta J}\br|_{J=0}\l(Z^R-1\r)\non
&=&\limR\f{1}{R}\int DC_\mu P[C_\mu] \int D(\bar{\psi}^a,\psi^a)\sum_{a=1}^{R}O(\bar{\psi^a},\psi^a)e^{-\sum_{a=1}^{R}S[\bar{\psi^a},\psi^a;C_\mu]}\non
&=&\limR\f{1}{R}\sum_{a=1}^{R}\int DC_\mu e^{-\int d^3\boldsymbol{x}\f{[C_\mu(\boldsymbol{x})]^2}{2\Gam}} \int D(\bar{\psi}^a,\psi^a)O(\bar{\psi^a},\psi^a)e^{-\sum_{a=1}^{R}S_0[\bar{\psi^a},\psi^a]}e^{-\sum_{a=1}^R\int^\beta_0\tau\int d^{3}\boldsymbol{x}\bar{\psi}^a(x)C_\mu\gam\mu\gam5\psi^a(x)}\non
&=& \limR\f{1}{R}\sum_{a=1}^{R}\int D(\bar{\psi}^a,\psi^a) O(\bar{\psi^a},\psi^a)e^{-\sum_{a=1}^{R}S_{0}[\bar{\psi^a}\psi^a]}e^{\sum_{b,c=1}^{R}\int^\beta_0d\tau\int^\beta_0 d\tau'\int d^3\boldsymbol{x}\f{\Gam}{2}(\bar{\psi^b}_\tau\gam\mu\gam5\psi^b_{\tau})(\bar{\psi^c}_{\tau'}\gam\mu\gam 5\psi^c_{\tau'})}\non
&=&\limR\f{1}{R}\sum_{a=1}^{R}\int D(\bar{\psi},\psi)O(\bar{\psi^a},\psi^a)e^{-\sum_{a=1}^{R}S_0[\bar{\psi^a},\psi^a]-\sum_{a,b=1}^{R}S_{dis}[\bar{\psi^a},\psi^a,\bar{\psi^b},\psi^b]}\label{e.effective observable}
\eea
where the average for disorder has been performed first to result in $S_{dis}[\bar{\psi^b},\psi^b,\bar{\psi^c},\psi^c]\equiv \int^\beta_0
d\tau\int^\beta_0 d\tau'\int d^3\boldsymbol{x}\f{\Gam}{2}\l(\bar{\psi^b}_\tau(\boldsymbol{x})\gam\mu\gam 5\psi^b_{\tau}(\boldsymbol{x})\r)\l(\bar{\psi^c}_{\tau'}(\boldsymbol{x})\gamd\mu\gam 5\psi^c_{\tau'}(\boldsymbol{x})\r)$. We point out the positive sign, arising from lowering the index from $\gamma^{\mu}$ to $\gamma_{\mu}$. Averaging for random chiral gauge fluctuations gives rise to effective interactions between chiral currents with all replicas, where effective all-time interactions allow momentum exchange only (not energy exchange).

An effective field theory is given by \be S_{B}= \int d^{d}x\bar{\psi}_{B}^{a}(\imath\gam0\partial_0+v_B\imath\gam k\partial_{k}+m_{B})\psi_{B}^{a}+\int d\tau\int d\tau' \int d^{d-1}\boldsymbol{x}\f{\Gam_{B}}{2} (\bar{\psi_{B}}^{b}(\gam \mu\gam 5)\psi_{B}^{b})_{\tau}(\bar{\psi_{B}}^{c}\gamd{\mu}\gam 5\psi_{B}^{c})_{\tau'} \label{Blagrangian}\ee in the replica trick, where Einstein convention has been used. B (R) stands for ``bare" (``renormalized"). Performing the dimensional analysis, where space and time coordinates have $-1$ in mass dimension, we observe $dim[\psi]=\f{d-1}{2}$, $dim[m]=1$, and $dim[\Gam]=3-d$. In this respect we perform the renormalization group analysis in $d=3+\epsilon$ dimensions, where $\epsilon$ is a small parameter. In the end of the calculation the dimension is analytically continued to the physical dimension ($d=4$), setting $\epsilon = 1$.

Taking into account quantum corrections, divergences are generated, which can be absorbed by renormalization constants, redefining fields and parameters. Rewriting the effective field theory in terms of renormalized fields and parameters, we obtain
\bea S_{B} &=&\int d^{d} x \l(Z_{\psi}^\omega\bar{\psi}_{R}^{a}\imath\gam0\partial_0\psi_{R}^{a}+ Z_{\psi}^{\boldsymbol{k}} v_R\bar{\psi}_{R}^{a}\imath\gam k\partial_{k}\psi_{R}^{a}+Z_mm_{R}\bar{\psi}_{R}^{a}\psi_{R}^{a}\r)+\int d\tau \int d\tau' \int d^{d-1}\boldsymbol{x}Z_{\Gam}\f{\Gam_{R}}{2}(\bar{\psi}_{R}^{b}\gam \mu\gam 5\psi_{R}^{b})_{\tau}(\bar{\psi}_{R}^{c}\gamd{\mu}\gam 5\psi_{R}^{c})_{\tau'} \no
\eea
with $\psi_{B}^{a}=(Z_\psi^\omega)^{1/2}\psi_{R}^{a}$, $m_{B}=Z_{m}(Z_\psi^\omega)^{-1} m_{R}$, $v_{B}=Z_\psi^{\boldsymbol{k}}(Z_\psi^\omega)^{-1} v_{R}$, and $\Gam_{B}=Z_{\Gam}(Z_\psi^\omega)^{-2}\Gam_{R}$, where $Z_\psi^\omega$ is a wave-function renormalization constant, $Z_{m}$, mass renormalization, $Z_\psi^{\boldsymbol{k}}$, velocity renormalization, and $Z_{\Gam}$, vertex renormalization.

It is more cultural to rewrite this field theory, separating the renormalized part from counter terms that absorb divergences in the following way
\bea
S_B&=&S_R+S_{CT}\non
S_{R}&=&\int d^{d} x \bar{\psi}_{R}^{a}\l(\imath\gam0\partial_0+ v_R\imath\gam k\partial_{k}+m_{R}\r)\psi_{R}^{a}+\int d\tau
\int d\tau' \int d^{d-1}\boldsymbol{x}\f{\Gam_{R}}{2}(\bar{\psi}_{R}^{b}\gam \mu\gam 5\psi_{R}^{b})_{\tau}(\bar{\psi}_{R}^{c}\gamd{\mu}\gam 5\psi_{R}^{c})_{\tau'}\label{Rlagrangian}\\
S_{CT}&=&\int d^{d}x\delta_{\psi}\bar{\psi}_{R}^{a}\l(\delta_\psi^\omega\imath\gam0\partial_0+ \delta_\psi^{\boldsymbol{k}}v_R\imath\gam k\partial_{k}+\delta_mm_{R}\r)\psi_{R}^{a}+\int d\tau \int d\tau' \int d^{d-1}\boldsymbol{x}\delta_{\Gam}\f{\Gam_{R}}{2}(\bar{\psi}_{R}^{b}\gam \mu\gam 5\psi_{R}^{b})_{\tau}(\bar{\psi}_{R}^{c}\gamd{\mu}\gam 5\psi_{R}^{c})_{\tau'}\label{CTlagrangian}
\eea
where $Z_\psi^\omega=1+\delta_\psi^\omega$, $Z_\psi^{\boldsymbol{k}}=1+\delta_\psi^{\boldsymbol{k}}$, $Z_{m}=1+\delta_{m}$, and $Z_{\Gam}=1+\delta_{\Gam}$.

\section{Evaluation of Feynman's diagrams}

\subsection{Self-energy corrections}

\subsubsection{Feynman's diagrams}

Within the replica trick, we are allowed to perform the perturbative analysis. The Green's function of $G(x,y)=\f{T}{L^d}\sum_{p,q}e^{-\i p\cdot x+\i q\cdot y}G(p,q)$ with $G(p,q)=<\psi(p),\bar{\psi}(q)>$ is evaluated as follows
\bnn
&&G(p,q)\\
&=&\limR\f{1}{R}\sum_{a=1}^{R}\int D(\bar{\psi},\psi)\psi^a(p)\bar{\psi}^a(q)e^{-\sum_{\alpha=1}^{R}S_{0}[\bar{\psi^\alpha} \psi^\alpha]}e^{\sum_{b,c=1}^{R}\int^\beta_0d\tau\int^\beta_0 d\tau'\int d^d\boldsymbol{x}\f{\Gam_R}{2}(\bar{\psi^b}_\tau\gam\mu\gam 5\psi^b_{\tau})(\bar{\psi^c}_\tau'\gamd\mu\gam 5\psi^c_{\tau'})}\\
&=&\limR\f{1}{R}\sum_{a=1}^{R}\int D(\bar{\psi},\psi)e^{-\sum_{\alpha=1}^{R}S_{0}[\bar{\psi^\alpha},\psi^\alpha]}\bl[\psi^a(p)\bar{\psi}^a(q)-\f{\Gam}{2}\sum_{b,c=1}^{R}\sum_{p_i}\psi^a(p)\bar{\psi}^a(q)\bar{\psi^b}(p_1)\gam\mu\gam 5\psi^b(p_2)\bar{\psi^c}(p_3)\gamd\mu\gam 5\psi^c(p_4)\\
&&\times\delta^{(3)}(\boldsymbol{p}_1-\boldsymbol{p}_2+\boldsymbol{p}_3-\boldsymbol{p}_4)\delta_{p^0_1,p^0_2}\delta_{p^0_3,p^0_4}+\l(-\f{\Gam_R}{2}\r)^2\sum_{b,c,d,e}\sum_{p_i,q_i}\psi^a(p)\bar{\psi}^a(q)\bar{\psi^b}(p_1)\gam\mu\gam 5\psi^b(p_2)\bar{\psi^c}(p_3)\gamd\mu\gam 5\psi^c(p_4)\\
&&\times\bar{\psi^d}(q_1)\gam\nu\gam 5\psi^d(q_2)\bar{\psi^e}(q_3)\gamd\nu\gam 5\psi^e(q_4)\delta^{(3)}(\boldsymbol{p}_1-\boldsymbol{p}_2+\boldsymbol{p}_3-\boldsymbol{p}_4)\delta_{p^0_1,p^0_2}\delta_{p^0_3, p^0_4}\delta^{(3)}(\boldsymbol{q}_1-\boldsymbol{q}_2+\boldsymbol{q}_3-\boldsymbol{q}_4)\delta_{q^0_1,
q^0_2}\delta_{q^0_3, q^0_4}+O(\Gam_R^3)\br]\\
&=&\limR\f{1}{R}\sum_{a=1}^{R}<\psi^a(p)\bar{\psi}^a(q)>_0+\limR\f{1}{R}\sum_{a,b,c=1}^{R}\l(-\f{\Gam_R}{2}\r)\sum_{p_i}<\psi^a(p)\bar{\psi}^a(q)\bar{\psi^b}(p_1)\gam\mu\gam 5\psi^b(p_2)\bar{\psi^c}(p_3)\gamd\mu\gam 5\psi^c(p_4)>_0\delta^{(4)}(p_i)\\
&&+\limR\f{1}{R}\sum_{\substack{a,b,c\\d,e}}\l(-\f{\Gam_R}{2}\r)^2\sum_{p_i,q_i}<\psi^a(p)\bar{\psi}^a(q)\bar{\psi^b}(p_1)\gam\mu\gam 5\psi^b(p_2)\bar{\psi^c}(p_3)\gamd\mu\gam 5\psi^c(p_4)\bar{\psi^d}(q_1)\gam\nu\gam 5\psi^d(q_2)\bar{\psi^e}(q_3)\gamd\nu\gam 5\psi^e(q_4)>_0\\&&
\times\delta^{(4)}(p_i)\delta^{(4)}(q_i)+O(\Gam_R^3)
\enn
where we introduced a short-hand-notation of $\delta^{(4)}(p_i)=\delta^{(3)}(\boldsymbol{p}_1-\boldsymbol{p}_2+\boldsymbol{p}_3-\boldsymbol{p}_4)\delta_{p^0_1, p^0_2}\delta_{p^0_3, p^0_4}$ with the four-vector notation of $x = (\tau,\bm{x})$ and $p = (\omega_{n},\bm{p})$. The first term is just the bare propagator. From now on, we omit momentum arguments and summations for the moment in order to focus on replica indices.

\begin{figure}[t]
\includegraphics[width=0.8\textwidth]{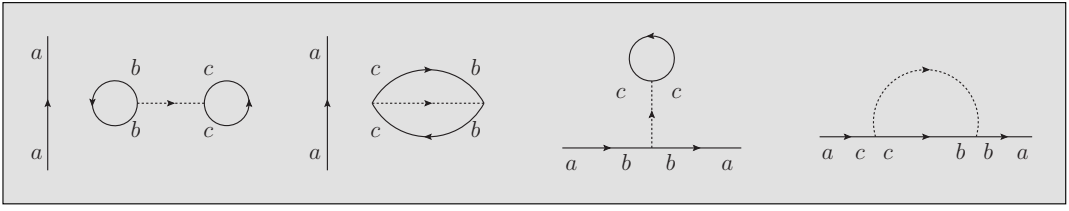}
\caption{All possible quantum corrections in the first-order without the replica limit.} \label{p.self-1all}
\end{figure}

First-order corrections are given by (Fig. \ref{p.self-1all})
\bnn
&&\limR\f{1}{R}\sum_{a,b,c=1}^{R}<\psi^a_i\bar{\psi}^a_j\bar{\psi^b}_k(\gam\mu\gam5)_{kl}\psi^b_l\bar{\psi^c}_m(\gamd\mu\gam5)_{mn}\psi^c_n>_0\\
&=&\limR\f{1}{R}\sum_{a,b,c=1}^{R}\br[<\psi^a_i\bar{\psi}^a_j>_0<\psi^b_l\bar{\psi^b}_k>_0<\psi^c_n\bar{\psi^c}_m>_0(\gam\mu\gam 5)_{kl}(\gamd\mu\gam5)_{mn}+<\psi^a_i\bar{\psi}^a_j>_0<\psi^c_n\bar{\psi^b}_k>_0<\psi^b_l\bar{\psi^c}_m>_0(\gam\mu\gam5)_{kl}(\gamd\mu\gam5)_{mn}\\
&&-2<\psi^a_i\bar{\psi}^b_k>_0<\psi^b_l\bar{\psi}^a_j>_0<\psi^c_n\bar{\psi}^c_m>_0(\gam\mu\gam5)_{kl}(\gamd\mu\gam5)_{mn}+2<\psi^b_l\bar{\psi}^a_j>_0<\psi^c_n\bar{\psi}^b_k>_0<\psi^a_i\bar{\psi}^c_m>_0(\gam\mu\gam5)_{kl}(\gamd\mu\gam5)_{mn}\bl]\\
&=&\limR\f{1}{R}\sum_{a,b,c=1}^{R}\br[G^a_{ij}G^b_{lk}G^c_{nm}(\gam\mu\gam 5)_{kl}(\gamd\mu\gam 5)_{mn}\delta_{aa}\delta_{bb}\delta_{cc}+G^a_{ij}G^c_{nk}G^b_{lm}(\gam\mu\gam 5)_{kl}(\gamd\mu\gam 5)_{mn}\delta_{aa}\delta_{cb}\delta_{bc}\\
&&-2G^a_{ik}G^b_{lj}G^c_{nm}(\gam\mu\gam 5)_{kl}(\gamd\mu\gam 5)_{mn}\delta_{ab}\delta_{ba}\delta_{cc}+2G^b_{lj}G^c_{nk}G^a_{im}(\gam\mu\gam
5)_{kl}(\gamd\mu\gam 5)_{mn}\delta_{ba}\delta_{cb}\delta_{ac}\bl]\\
&=&\limR\f{1}{R}\br[\sum_{a,b,c=1}^RG^atr[G^b\gam\mu\gam 5]tr[G^c\gamd\mu\gam 5]\delta_{aa}\delta_{bb}\delta_{cc}+\sum_{a,b,c=1}^RG^atr[G^c(\gam\mu\gam 5)G^b(\gamd\mu\gam 5)]\delta_{aa}\delta_{cb}\delta_{bc}\\
&&-2\sum_{a,b,c=1}^RG^a(\gam\mu\gam 5)G^btr[G^c(\gamd\mu\gam 5)]\delta_{ab}\delta_{ba}\delta_{cc}+2\sum_{a,b,c=1}^RG^a(\gamd\mu\gam 5)G^c(\gam\mu\gam 5)G^b\delta_{ba}\delta_{cb}\delta_{ac}\bl]
\enn
where ``2" results from identical contributions and $-$ comes from the odd number of fermion loops (one loop). Since all Green's functions with different replica indices are identical, the
first term is proportional to $R^3$, the second, $R^2$, the third, $R^2$, and the forth, $R$. Taking the replica limit of $\limR\f{1}{R}$, only the forth term survives. As a result, we find
\begin{figure}[t]
\includegraphics[width=0.5\textwidth]{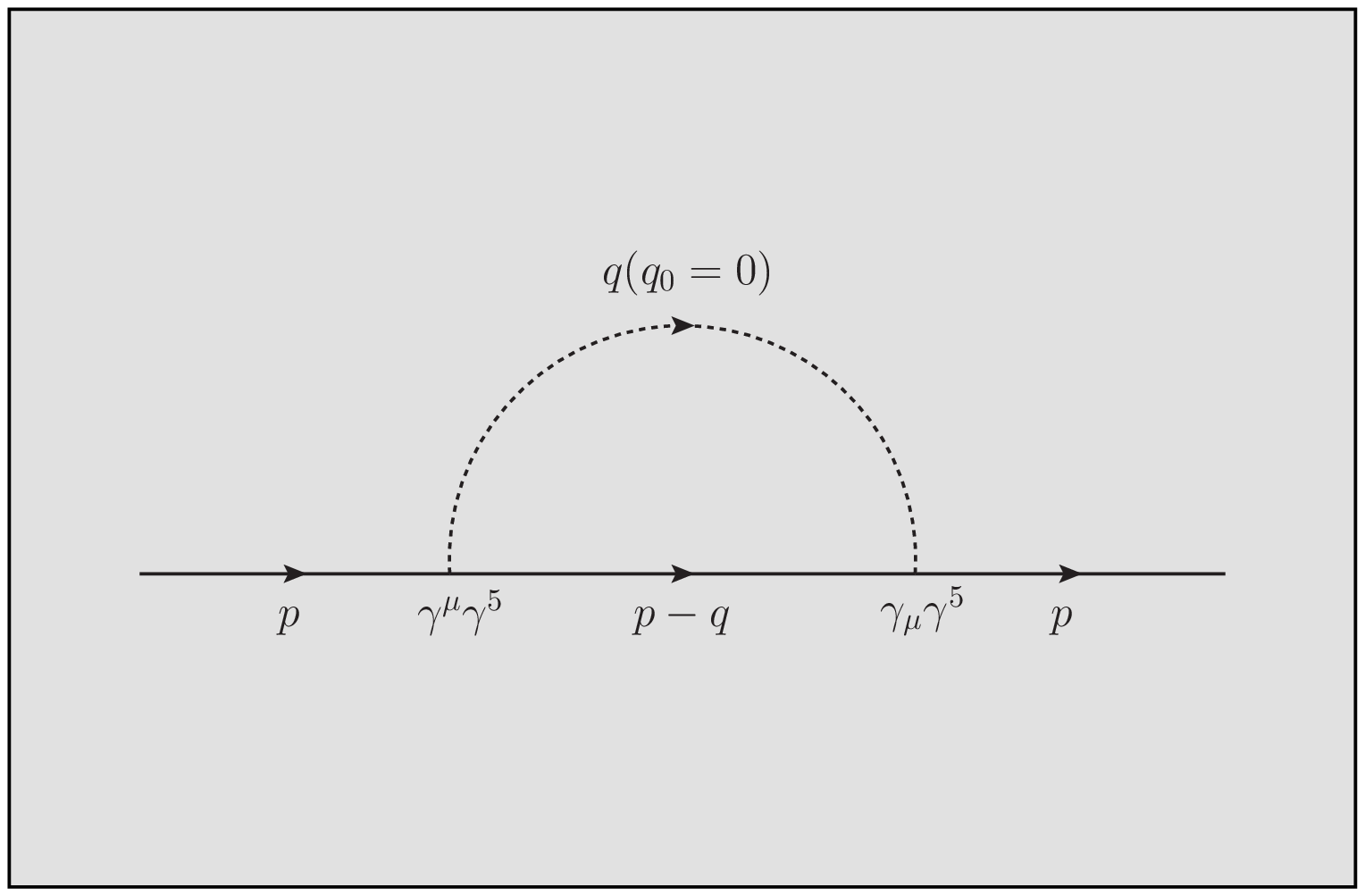}
\caption{The Fock correction, which contributes to the wave-function renormalization constant only in the first order.} \label{p.self-1st}
\end{figure}
\be
G^{(1)}=G(p)\l(-\f{\Gam}{2}\r)\sum_{q}\gam\mu\gam 5G(p-q)\gamd\mu\gam 5G(p)G(p)\Sigma^{(1)}G(p) \label{e.self-1st}
\ee
in the one-loop order (Fig. \ref{p.self-1st}). Here, we point out that Feynman diagrams whose internal propagators are not connected to external lines (the third diagram in Fig. \ref{p.self-1all}) always vanish in the replica limit. In other words, contributions with fermion loops vanish identically in the replica limit.

Omitting vacuum and one-particle reducible diagrams, we have self-energy corrections in the second order
\begin{figure}[t]
\includegraphics[width=1\textwidth]{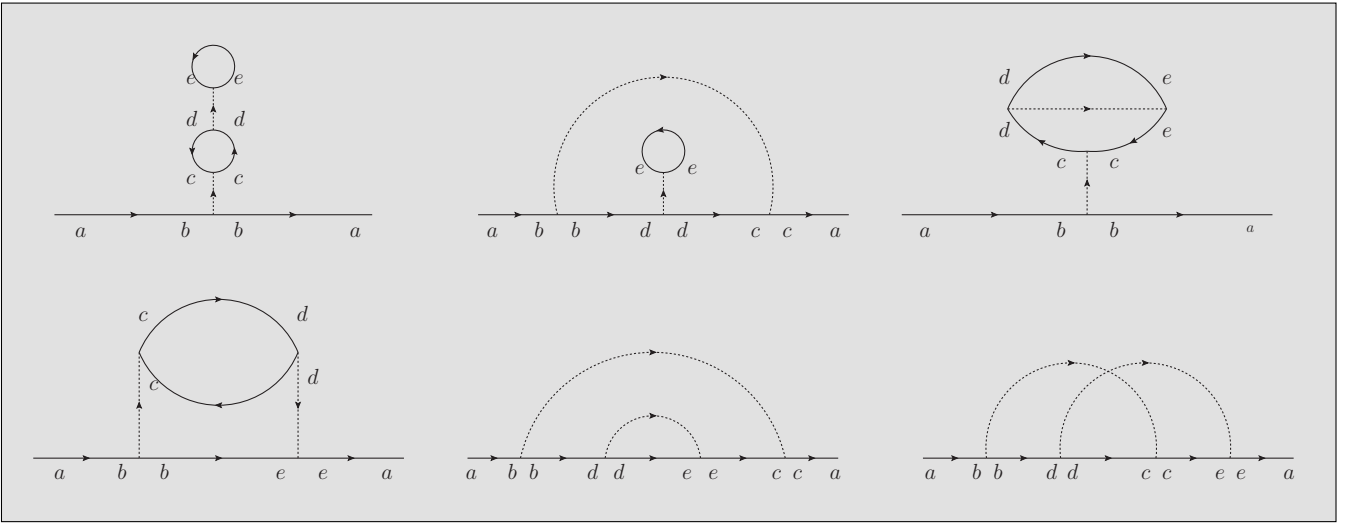}
\caption{All possible second-order self-energy diagrams without the replica limit, omitting vacuum  and one-particle reducible diagrams.} \label{p.self-2all}
\end{figure}
\bnn
&&\limR\f{1}{R}\sum_{a,b,c,d,e}\br[<\psi^a\bar{\psi}^a\bar{\psi^b}\gam\mu\gam5\psi^b\bar{\psi^c}\gamd\mu\gam 5\psi^c\bar{\psi^d}\gam\nu\gam 5\psi^b\bar{\psi^c}\gamd\nu\gam5\psi^c>_0\bl]_{1PI}\\
&=&\limR\f{1}{R}\sum_{a,b,c,d,e}\br[8G^aG^b\gam\mu\gam5G^bG^atr[G^c\gam\nu\gam 5G^d\gamd\mu\gam 5]tr[G^e\gamd\nu\gam5]\delta_{ab}\delta_{ba}\delta_{cd}\delta_{dc}\delta_{ee}-4G^a\gam\mu\gam5G^b\gam\nu\gam 5G^d\gamd\mu\gam 5\\
&&\times G^ctr[G^e\gamd\nu\gam5]\delta_{ab}\delta_{bd}\delta_{dc}\delta_{ca}\delta_{ee}-4G^a\gam\mu\gam5G^btr[G^c\gam\nu\gam 5G^d\gamd\nu\gam 5G^e\gamd\mu\gam5]\delta_{ab}\delta_{ba}\delta_{cd}\delta_{de}\delta_{ec}+(-1)16G^a\gam\mu\gam
5G^b\gamd\nu\gam 5\\
&&\times G^etr[G^c\gam\nu\gam5G^d\gamd\mu\gam5]\delta_{ab}\delta_{be}\delta_{ea}\delta_{cd}\delta_{dc}+8G^a\gam\mu\gam5G^b\gam\nu\gam 5G^d\gamd\nu\gam5G^e\gamd\mu\gam5G^c\delta_{ab}\delta_{bd}\delta_{de}\delta_{ec}\delta_{ca}+8G^a\gam\mu\gam 5G^b\gam\nu\gam 5G^d\gamd\mu\gam 5\\
&&\times G^c\gamd\nu\gam 5G^e\delta_{ab}\delta_{bd}\delta_{dc}\delta_{ce}\delta_{ea}\bl] .
\enn See Fig. \ref{p.self-2all}. The first term is proportional to $R^3$, the second, third, and forth, $R^2$, the fifth and last, $R$. As a result, only the fifth and last terms survive in the replica limit. Therefore, the relevant self-energy correction is given by (Fig. \ref{p.self-2nd})
\begin{figure}[t]
\includegraphics[width=0.4\textwidth]{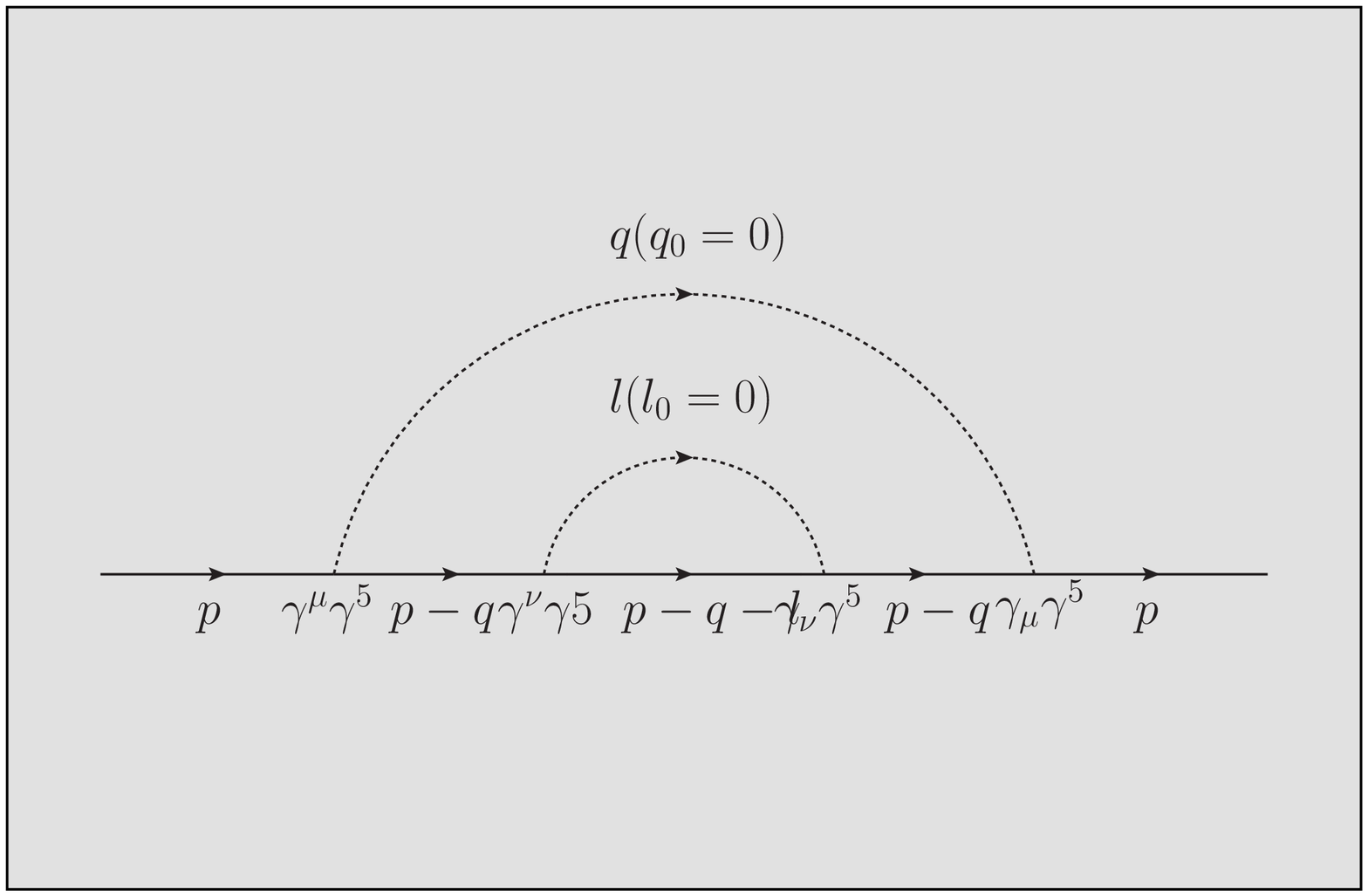}
\includegraphics[width=0.4\textwidth]{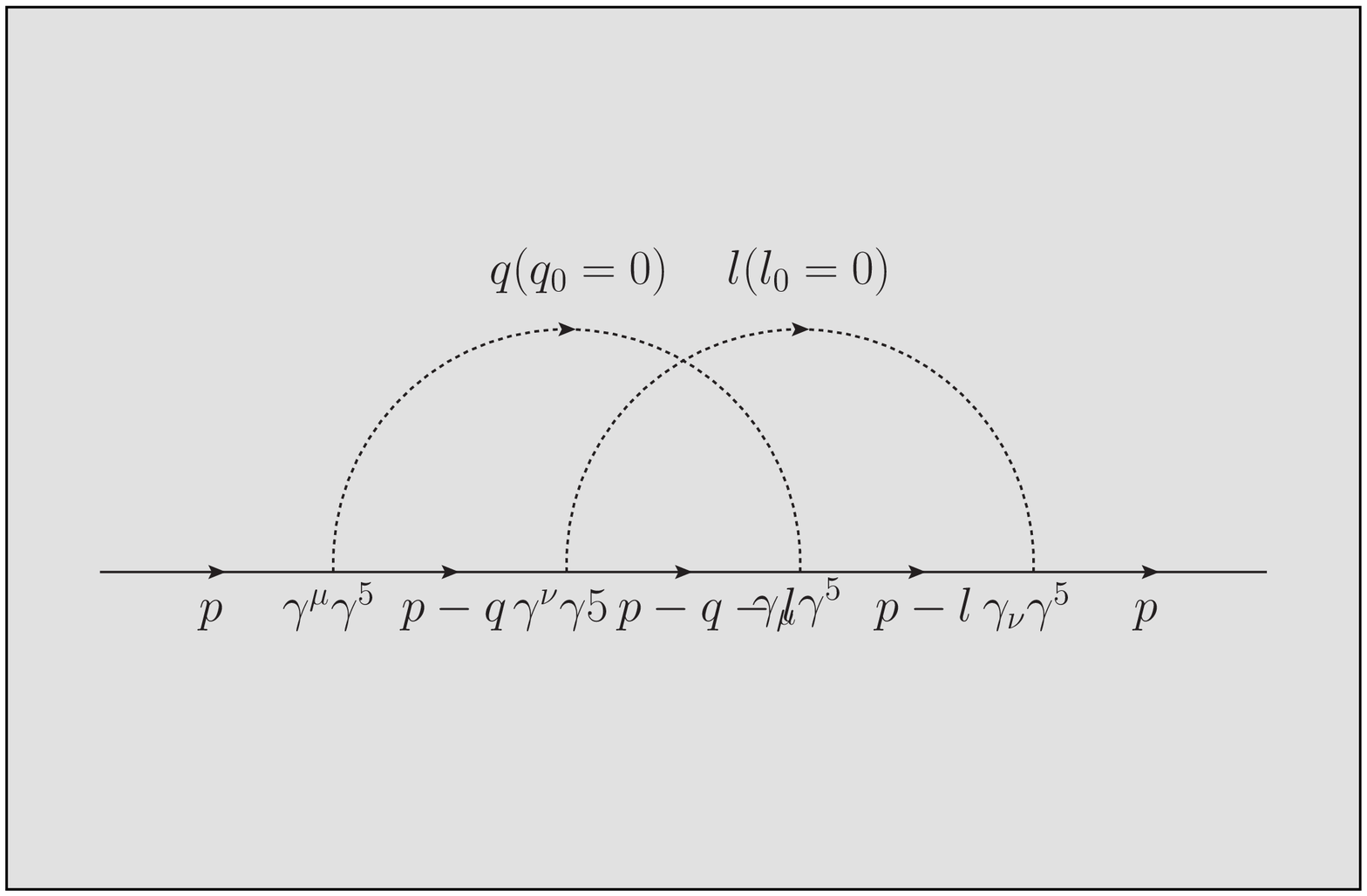}
\caption{Relevant second-order self-energy corrections in the replica limit.} \label{p.self-2nd}
\end{figure}
\bea
G^{(2),r}(p)&=&G(p)\l(-\f{\Gam_R}{2}\r)^2\sum_{q,l}\gam\mu\gam 5G(p-q)\gam\nu\gam5G(p-q-l)\gamd\nu\gam 5G(p-q)\gamd\mu\gam 5 G(p)=G(p) \Sigma^{(2),r}G(p)\non
G^{(2),c}(p)&=&G(p)\l(-\f{\Gam_R}{2}\r)^2\sum_{q,l}\gam\mu\gam 5G(p-q)\gam\nu\gam5G(p-q-l)\gamd\mu\gam 5G(p-l)\gamd\nu\gam 5 G(p)=G(p) \Sigma^{(2),c}G(p)\non
\Sigma^{(2),r}&=&\l(-\f{\Gam_R}{2}\r)^2\sum_{q,l}\gam\mu\gam 5G(p-q)\gam\nu\gam5G(p-q-l)\gamd\nu\gam 5G(p-q)\gamd\mu\gam 5\label{e.self-2nd-rainbow}\\
\Sigma^{(2),c}&=&\l(-\f{\Gam_R}{2}\r)^2\sum_{q,l}\gam\mu\gam 5G(p-q)\gam\nu\gam5G(p-q-l)\gamd\mu\gam 5G(p-l)\gamd\nu\gam5\label{e.self-2nd-cross} .
\eea

\subsubsection{Evaluation of relevant Feynman's diagrams}

The first-order Fock diagram (Fig. \ref{p.self-1st}) is
\bnn
\Sigma^{(1)}
&=&-\f{\Gam_R}{2}\int \f{d^dq}{(2\pi)^d}2\pi\delta (q_0)\gam\mu\gam
5\f{\not{p}-\not{q}-m}{(p-q)^2-m^2}\gam\mu\gam 5\\
&=&-\f{\Gam_R}{2}\int \f{d^{d-1}\boldsymbol{q}}{(2\pi)^{d-1}}\f{\gam\mu (p_0\gam 0+p_k\gam k-q_k\gam
k+m)\gamd\mu}{-p_0^2-(\boldsymbol{p}-\boldsymbol{q})^2-m^2}\\
&=&\f{\Gam_R}{2}\int \f{d^{d-1}\boldsymbol{q}}{(2\pi)^{d-1}}\f{(2-d)q_k\gam k-2p_0\gam
0+4m}{\boldsymbol{q}^2+p_0^2+m^2}\\
&=&\f{\Gam_R}{2}\l[\f{-2p_0\gam
0+4m}{(4\pi)^{\f{d-1}{2}}}\f{\Gam(1-\f{d-1}{2})}{\Gam(1)}\f{1}{(p_0^2+m^2)^{1=\f{d-1}{2}}}\r]\\
&=&\f{\Gam_R}{2(4\pi)}(-2p_0\gam
0+4m)\br(-\f{2}{\epsilon}-\gamma-\log{(p_0^2+m^2)}+\log{4\pi}+O(\epsilon)\bl) .
\enn Then, a relevant part for renormalization is \be \Sigma^{(1)}=-\f{\Gam_R}{4\pi}\f{1}{\epsilon}(-2p_0\gam 0+4m)+O(1) \label{r.self-1st}\ee

The second-order rainbow diagram (the first diagram in Fig. \ref{p.self-2nd}) is
\bnn
\Sigma^{(2),r}
&=&\l(-\f{\Gam_R}{2}\r)^2\int \f{d^dq}{(2\pi)^d}2\pi\delta (q_0)\int \f{d^dl}{(2\pi)^d}2\pi\delta(l_0)\gam\mu\gam 5\f{\not{p}-\not{q}-m}{(p-q)^2-m^2}\gam\nu\gam 5\f{\not{p}-\not{q}-\not{l}-m}{(p-q-l)^2-m^2}\gamd\nu\gam5\\
&&\times\f{\not{p}-\not{q}-m}{(p-q)^2-m^2}\gamd\mu\gam 5\\
&=&\f{\Gam_R^2}{4}\int \f{d^{d-1}\boldsymbol{q}}{(2\pi)^{d-1}}\int\f{d^{d-1}\boldsymbol{l}}{(2\pi)^{d-1}}\gam\mu\f{p_0\gam 0+p_k\gam k-q_k\gam k+m}{-p_0^2-(\boldsymbol{p}-\boldsymbol{q})^2-m^2}\gam\nu\f{p_0\gam 0+p_l\gam l-q_l\gam l-l_l\gam l-m}{-p_0^2-(\boldsymbol{p}-\boldsymbol{q}-\boldsymbol{l} )^2-m^2}\gamd\nu\\
&&\times\f{p_0\gam 0+p_m\gam m-q_m\gam m+m}{-p_0^2-(\boldsymbol{p}-\boldsymbol{q})^2-m^2}\gamd\mu\\
&=&-\f{\Gam_R^2}{4}\int \f{d^{d-1}\boldsymbol{q}}{(2\pi)^{d-1}}\gam\mu\f{p_0\gam 0+p_k\gam
k-q_k\gam k+m}{(\boldsymbol{p}-\boldsymbol{q})^2+p_0^2+m^2}\gam\nu\br[\int\f{d^{d-1}\boldsymbol{l}}{(2\pi)^{d-1}}\f{p_0\gam 0+p_l\gam l-q_l\gam l-l_l\gam l-m}{(\boldsymbol{p}-\boldsymbol{q}-\boldsymbol{l})^2+p_0^2+m^2}\bl]\gamd\nu\\
&&\times\f{p_0\gam 0+p_m\gam m-q_m\gam m+m}{(\boldsymbol{p}-\boldsymbol{q})^2+p_0^2+m^2}\gamd\mu\\
&=&-\f{\Gam_R^2}{4}\int \f{d^{d-1}\boldsymbol{q}}{(2\pi)^{d-1}}\gam\mu\f{-q_k\gam k+p_0\gam 0+m}{\boldsymbol{q}^2+p_0^2+m^2}\gam\nu\br[\int\f{d^{d-1}\boldsymbol{l}}{(2\pi)^{d-1}}\f{-l_l\gam l+p_0\gam 0-m}{\boldsymbol{l}^2+p_0^2+m^2}\bl]\gamd\nu\f{-q_m\gam m+p_0\gam 0+m}{\boldsymbol{q}^2+p_0^2+m^2}\gamd\mu\\
&=&-\f{\Gam_R^2}{4}\int \f{d^{d-1}\boldsymbol{q}}{(2\pi)^{d-1}}\gam\mu\f{-q_k\gam k+p_0\gam 0+m}{\boldsymbol{q}^2+p_0^2+m^2}\br[\f{1}{(4\pi)^{\f{d-1}{2}}}\f{\Gam(1-\f{d-1}{2})}{\Gam(1)}\f{\gam\nu(p_0\gam 0-m)\gamd\nu}{(p_0^2+m^2)^{1-\f{d-1}{2}}}\bl]\f{-q_m\gam m+p_0\gam 0+m}{\boldsymbol{q}^2+p_0^2+m^2}\gamd\mu\\
&=&-\f{\Gam_R^2}{4}\f{\Gam(\f{3-d}{2})}{(4\pi)^{\f{d-1}{2}}(p_0^2+m^2)^{\f{3-d}{2}}}\int\f{d^{d-1}\boldsymbol{q}}{(2\pi)^{d-1}}\f{\gam\mu(-q_k\gam k+p_0\gam 0+m)(-2p_0\gam 0-4m)(-q_m\gam m+p_0\gam 0+m)\gamd\mu}{(\boldsymbol{q}^2+p_0^2+m^2)^2} .
\enn
Rearranging the numerator as follows
\bnn
N&=&\gam\mu(-q_k\gam k+p_0\gam 0+m)(-2p_0\gam 0-4m)(-q_m\gam m+p_0\gam 0+m)\gamd\mu\\
&=&-q_kq_m\gam\mu\gam k(2p_0\gam 0+4m)\gam m\gamd\mu-\gam\mu(p_0\gam 0+m)(2p_0\gam 0+4m)(p_0\gam 0+m)\gamd\mu\\
&=&-q_kq_m\gam\mu\gam k(2p_0\gam 0+4m)\gam m\gamd\mu+(-4p_0^3\gam 0+32mp_0^2+20m^2p_0\gam 0-16m^3)\\
&=&-q_kq_m\gam\mu\gam k(2p_0\gam 0+4m)\gam m\gamd\mu+f(p)
\enn
with $f(p)=-4p_0^3\gam 0+32mp_0^2+20m^2p_0\gam 0-16m^3$, we obtain
\bnn
\Sigma^{(2),r}
&=&-\f{\Gam_R^2}{4}\f{\Gam(\f{3-d}{2})}{(4\pi)^{\f{d-1}{2}}(p_0^2+m^2)^{\f{3-d}{2}}}\int
\f{d^{d-1}\boldsymbol{q}}{(2\pi)^{d-1}}\f{-q_kq_m\gam\mu\gam k(2p_0\gam 0+4m)\gam
m\gamd\mu+f(p)}{(\boldsymbol{q}^2+p_0^2+m^2)^2}\\
&=&-\f{\Gam_R^2}{4}\f{\Gam(\f{3-d}{2})}{(4\pi)^{\f{d-1}{2}}(p_0^2+m^2)^{\f{3-d}{2}}}\br[-\f{\gam\mu\gam
k(2p_0\gam 0+4m)\gamd
k\gamd\mu}{2(4\pi)^{\f{d-1}{2}}}\f{\Gam(2-\f{d-1}{2}-1)}{\Gam(2)(p_0^2+m^2)^{2-\f{d-1}{2}-1}}+\f{f(p)}{(4\pi)^{\f{d-1}{2}}}\f{\Gam(2-\f{d-1}{2})}{\Gam(2)(p_0^2+m^2)^{2-\f{d-1}{2}}}\bl]\\
&=&-\f{\Gam_R^2}{4}\f{\Gam(\f{3-d}{2})}{(4\pi)^{\f{d-1}{2}}(p_0^2+m^2)^{\f{3-d}{2}}}\br[-\f{2(d-2)(d-1)p_0\gam
0+4d(d-1)m}{2(4\pi)^{\f{d-1}{2}}}\f{\Gam(\f{3-d}{2})}{(p_0^2+m^2)^{\f{3-d}{2}}}+\f{f(p)}{(4\pi)^{\f{d-1}{2}}}\f{\Gam(\f{5-d}{2})}{(p_0^2+m^2)^{\f{5-d}{2}}}\bl]\\
&=&\f{\Gam_R^2}{4(4\pi)^2}\l(-\f{2}{\epsilon}-\gamma+\log{(4\pi)}-\log{(p_0^2+m^2)}+O(\epsilon)\r)^2\l((1+\epsilon)(2+\epsilon)p_0\gam
0+2(3+\epsilon)(2+\epsilon)m\r)\\
&&-\f{\Gam^2}{4(4\pi)^2}\l(-\f{2}{\epsilon}-\gamma+\log{(4\pi)}-\log{(p_0^2+m^2)}+O(\epsilon)\r)^2\l(-\f{\epsilon}{2}\r)\f{-4p_0^3\gam
0+32mp_0^2+20m^2p_0\gam 0-16m^3}{p_0^2+m^2} .
\enn
We note that the second term vanishes when we use the on-shell condition, given by $p_0\gam 0+m=0$ and $p_0^2=m^2$. As a result, we obtain
\be \Sigma^{(2),r}= \f{\Gam_R^2}{16\pi^2}\f{1}{\epsilon}(3p_0\gam 0+10m)+O(1)+O\l(\epsilon^{-2}\r) . \label{r.self-2nd-rainbow}\ee

The second-order crossed diagram (the second diagram in Fig. \ref{p.self-2nd}) is
\bnn
\Sigma^{(2),c}
&=&\l(-\f{\Gam_R}{2}\r)^2\int \f{d^dq}{(2\pi)^d}2\pi\delta (q_0)\int \f{d^dl}{(2\pi)^d}2\pi\delta(l_0)\gam\mu\gam 5\f{\not{p}-\not{q}-m}{(p-q)^2-m^2}\gam\nu\gam 5\f{\not{p}-\not{q}-\not{l}-m}{(p-q-l)^2-m^2}\gamd\mu\gam 5\f{\not{p}-\not{l}-m}{(p-l)^2-m^2}\gam\nu\gam 5\\
&=&\f{\Gam_R^2}{4}\int \f{d^{d-1}\boldsymbol{q}}{(2\pi)^{d-1}}\int\f{d^{d-1}\boldsymbol{l}}{(2\pi)^{d-1}}\gam\mu\f{p_0\gam 0+p_k\gam k-q_k\gam k+m}{-p_0^2-(\boldsymbol{p}-\boldsymbol{q})^2-m^2}\gam\nu\f{p_0\gam 0+p_l\gam l-q_l\gam l-l_l\gam l-m}{-p_0^2-(\boldsymbol{p}-\boldsymbol{q}-\boldsymbol{l})^2-m^2}\gamd\mu\\
&&\times\f{p_0\gam 0+p_m\gam m-l_m\gam m+m}{-p_0^2-(\boldsymbol{p}-\boldsymbol{l})^2-m^2}\gamd\nu\\
&=&-\f{\Gam_R^2}{4}\int \f{d^{d-1}\boldsymbol{l}}{(2\pi)^{d-1}}\int\f{d^{d-1}\boldsymbol{q}}{(2\pi)^{d-1}}\gam\mu\f{-q_k\gam k+p_0\gam 0+m}{\boldsymbol{q}^2+p_0^2+m^2}\gam\nu\f{-q_l\gam l-l_l\gam l-p_l\gam l+p_0\gam 0-m}{(\boldsymbol{q}+\boldsymbol{l}+\boldsymbol{p})^2+p_0^2+m^2}\gamd\mu\f{-l_m\gam m+p_0\gam
0+m}{\boldsymbol{l}^2+p_0^2+m^2}\gamd\nu\\
&=&-\f{\Gam_R^2}{4}\int \f{d^{d-1}\boldsymbol{l}}{(2\pi)^{d-1}}\l[\int\f{d^{d-1}\boldsymbol{q}}{(2\pi)^{d-1}}\f{N}{D}\r]\f{-l_m\gam m+p_0\gam 0+m}{\boldsymbol{l}^2+p_0^2+m^2}\gamd\nu ,
\enn
where the denominator $D^{-1}$ and the numerator $N$ are given by
\bnn
D^{-1}&=&\int^1_0dx[\boldsymbol{q}^2+x(1-x)(\boldsymbol{l}+\boldsymbol{p})^2+p_0^2+m^2]^{-2}=\int^1_0dx[\boldsymbol{q}^2+\Delta(l,p)]^{-2} \no \enn and \bnn
N&=&\gam\mu(-q_k\gam k+x(l_k+p_k)\gam k+p_0\gam 0+m)\gam\nu(-q_l\gam l-(1-x)(l_l+p_l)\gam l+p_0\gam 0-m)\gamd\mu\\
&=&q_kq_l\gam\mu\gam k\gam\nu\gam l\gamd\mu+\gam\mu(x(l_k+p_k)\gam k+p_0\gam
0+m)\gam\nu(-(1-x)(l_l+p_l)\gam l+p_0\gam0+m)\gamd\mu\\
&=&q_kq_l(-2\gam l\gam\nu\gam k+(4-d)\gam k\gam\nu\gam l)+f(l,p) ,
\enn respectively. In order to obtain these equations, we have used the following identity
\bnn
\f{1}{(p-q)^2-m^2}\f{1}{(p-l)^2-m^2}
&=&\int^1_0dx\l[x\l((p-q)^2-m^2\r)+(1-x)\l((p-l)^2-m^2\r)\r]^{-2}\\
&=&\int^1_0dx\l[p^2-2\l((1-x)l+xq\r)\cdot p+xq^2+(1-x)l^2-m^2\r]^{-2}\\
&& p\rightarrow p+(1-x)l+xq\\
&=&\int^1_0dx\l[p^2-\l((1-x)l+xq\r)^2+xq^2+(1-x)l^2-m^2\r]^{-2}\\
&=&\int^1_0dx\l[p^2+x(1-x)(l-q)^2-m^2\r]^{-2} , \\
\enn which will be also used in the evaluation of vertex corrections. Performing the momentum integral with these expressions, we obtain
\bnn
\int \f{d^{d-1}\boldsymbol{q}}{(2\pi)^{d-1}}\f{N}{D}
&=&\int^1_0dx\int \f{d^{d-1}\boldsymbol{q}}{(2\pi)^{d-1}}\f{q_kq_l(-2\gam l\gam\nu\gam k+(4-d)\gam k\gam\nu\gam l)+f(l,p)}{[\boldsymbol{q}^2+\Delta(l,p)]^2}\\
&=&\int^1_0dx\l[\f{1}{(4\pi)^{\f{d-1}{2}}}\f{(2-d)\gam l\gam\nu\gamd l}{2}\f{\Gam(1-\f{d-1}{2})}{\Gam(2)}\f{1}{\Delta(l,p)^{1-\f{d-1}{2}}}+\f{f(l,p)}{(4\pi)^{\f{d-1}{2}}}\f{\Gam(2-\f{d-1}{2})}{\Gam(2)}\f{1}{\Delta(l,p)^{2-\f{d-1}{2}}}\r]\\
&=&\int^1_0dx\l[\f{(2-d)(3-d)\gam\nu}{2(4\pi)^{\f{d-1}{2}}}\f{\Gam(\f{3-d}{2})}{\Delta(l,p)^{\f{3-d}{2}}}+\f{f(l,p)}{(4\pi)^{\f{d-1}{2}}}\f{\Gam(\f{5-d}{2})}{\Delta(l,p)^{\f{5-d}{2}}}\r] .
\enn Then, we have
\bnn
\Sigma^{(2),c}
&=&-\f{\Gam_R^2}{4}\int\f{d^{d-1}\boldsymbol{l}}{(2\pi)^{d-1}}\int^1_0dx\l[\f{(2-d)(3-d)\gam\nu}{2(4\pi)^{\f{d-1}{2}}}\f{\Gam(\f{3-d}{2})}{\Delta^{\f{3-d}{2}}}+\f{f(l,p)}{(4\pi)^{\f{d-1}{2}}}\f{\Gam(\f{5-d}{2})}{\Delta^{\f{5-d}{2}}}\r]\f{-l_m\gam m+p_0\gam 0+m}{\boldsymbol{l}^2+p_0^2+m^2}\gamd\nu\\
&=&-\f{\Gam_R^2}{4}\int^1_0dx\br[\f{(2-d)(3-d)}{2(4\pi)^{\f{d-1}{2}}}\f{\Gam(\f{3-d}{2})}{\Delta^{\f{3-d}{2}}}\int
\f{d^{d-1}\boldsymbol{l}}{(2\pi)^{d-1}}\f{-(2-d)l_m\gam m-2p_0\gam 0+4m}{\boldsymbol{l}^2+p_0^2+m^2}\\
&&+\f{\Gam(\f{5-d}{2})}{(4\pi)^{\f{d-1}{2}}}\int\f{d^{d-1}\boldsymbol{l}}{(2\pi)^{d-1}}\f{f(l,p)}{\Delta^{\f{5-d}{2}}}\f{-l_m\gam m+p_0\gam 0+m}{\boldsymbol{l}^2+p_0^2+m^2}\gamd\nu\bl]\\
&=&-\f{\Gam_R^2}{4}\int^1_0dx\br[\f{(2-d)(3-d)}{2(4\pi)^{\f{d-1}{2}}}\f{\Gam(\f{3-d}{2})}{\Delta^{\f{3-d}{2}}}\int\f{d^{d-1}\boldsymbol{l}}{(2\pi)^{d-1}}\f{-(2-d)l_m\gam m-2p_0\gam 0+4m}{\boldsymbol{l}^2+p_0^2+m^2}+\f{\Gam(\f{5-d}{2})}{(4\pi)^{\f{d-1}{2}}}\int\f{d^{d-1}\boldsymbol{l}}{(2\pi)^{d-1}}\\
&&\times\f{\gam\mu(x(l_k+p_k)\gam k+p_0\gam
0+m)\gam\nu(-(1-x)(l_l+p_l)\gam l+p_0\gam 0+m)\gamd\mu}{\Delta^{\f{5-d}{2}}}\f{-l_m\gam m+p_0\gam 0+m}{\boldsymbol{l}^2+p_0^2+m^2}\gamd\nu\bl]\\
&=&-\f{\Gam_R^2}{4}\int^1_0dx\br[\f{(2-d)(3-d)}{2(4\pi)^{\f{d-1}{2}}}\f{\Gam(\f{3-d}{2})}{\Delta^{\f{3-d}{2}}}\int\f{d^{d-1}\boldsymbol{l}}{(2\pi)^{d-1}}\f{-(2-d)l_m\gam m-2p_0\gam 0+4m}{\boldsymbol{l}^2+p_0^2+m^2}+\f{\Gam(\f{5-d}{2})}{(4\pi)^{\f{d-1}{2}}}\int\f{d^{d-1}\boldsymbol{l}}{(2\pi)^{d-1}}\f{N}{D}\bl] ,
\enn
where
\bnn
D^{-1}
&=&\l[x(1-x)(\boldsymbol{l}+\boldsymbol{p})^2+p_0^2+m^2\r]^{\f{5-d}{2}}[\boldsymbol{l}^2+p_0^2+m^2]^{-1}\\
&=&\int^1_0dy\f{[x(1-x)]^{\f{5-d}{2}}y^{\f{5-d}{2}-1}}{[y((\boldsymbol{l}+\boldsymbol{p})^2+(x(1-x))^{-1}(p_0^2+m^2))+(1-y)(\boldsymbol{l}^2+p_0^2+m^2)]^{\f{7-d}{2}}}\f{\Gam(\f{5-d}{2}+1)}{\Gam(\f{5-d}{2})}\\
&=&\f{\Gam(\f{7-d}{2})}{\Gam(\f{5-d}{2})}\int^1_0dy\f{[x(1-x)]^{\f{5-d}{2}}y^{\f{3-d}{2}}}{[\boldsymbol{l}^2+2y\boldsymbol{p}\cdot\boldsymbol{l}+y\boldsymbol{p}^2+((x(1-x))^{-1}+1-y)(p_0^2+m^2)]^{\f{7-d}{2}}}\\
&=&\f{\Gam(\f{7-d}{2})}{\Gam(\f{5-d}{2})}\int^1_0dy\f{[x(1-x)]^{\f{5-d}{2}}y^{\f{3-d}{2}}}{[\boldsymbol{l}^2+y(1-y)\boldsymbol{p}^2-((x(1-x))^{-1}+1-y)\boldsymbol{p}^2]^{\f{7-d}{2}}}\\
&=&\f{\Gam(\f{7-d}{2})}{\Gam(\f{5-d}{2})}\int^1_0dy\f{[x(1-x)]^{\f{5-d}{2}}y^{\f{3-d}{2}}}{[\boldsymbol{l}^2+\Delta(p)]^{\f{7-d}{2}}}
\\
&&\textrm{for the forth equality: $\boldsymbol{l}\rightarrow \boldsymbol{l}-y\boldsymbol{p}$ and $-p_0^2=\boldsymbol{p}^2+m^2$,}
\\
N
&=&\gam\mu(x(l_k+(1-y)p_k)\gam k+p_0\gam 0+m)\gam\nu(-(1-x)(l_l+(1-y)p_l)\gam l+p_0\gam 0-m)\gamd\mu(-l_m\gam m+yp_m\gam m\\
&&+p_0\gam 0+m)\gamd\nu\\
&=&\br[-x(1-x)l_kl_l\gam\mu\gam k\gam\nu\gam l\gamd\mu+xl_k\gam\mu\gam
k\gam\nu(-(1-x)(1-y)p_l\gam l+p_0\gam
0-m)\gamd\mu+\gam\mu(x(1-y)+p_k\gam k\\
&&+p_0\gam 0+m)\gam\nu(-(1-x)l_l\gam l)\gamd\mu+\gam\mu(x(1-y)p_k\gam k+p_0\gam
0+m)\gam\nu(-(1-x)(1-y)p_l\gam l+p_0\gam
0-m)\gamd\mu\bl]\\
&&\times\l[-l_m\gam m+yp_m\gam m+p_0\gam 0+m]\gamd\nu\r]\\
&=&-x(1-x)l_kl_l\gam\mu\gam k\gam\nu\gam l\gamd\mu(yp_m\gam m+p_0\gam
0+m)\gamd\nu-xl_kl_m\gam\mu\gam
k\gam\nu(-(1-x)(1-y)p_l\gam l+p_0\gam 0-m)\gamd\mu\gam m\gamd\nu\\
&&+(1-x)l_ll_m\gam\mu(x(1-y)p_k\gam k+p_0\gam 0+m)\gam\nu\gam l\gamd\mu\gam
m\gamd\nu+\gam\mu(x(1-y)p_k\gam k+p_0\gam
0+m)\\
&&\times\gam\nu(-(1-x)(1-y)p_l\gam l+p_0\gam 0-m)\gamd\mu(yp_m\gam m+p_0\gam 0+m)\gamd\nu\\
&=&l_kl_l\br[ -x(1-x)\gam\mu\gam k\gam\nu\gam l\gamd\mu(yp_m\gam m+p_0\gam
0+m)\gamd\nu-x\gam\mu\gam
k\gam\nu\l(-(1-x)(1-y)p_m\gam m+p_0\gam 0-m\r)\gamd\mu\gam l\gamd\nu\\
&&+(1-x)\gam\mu\l(x(1-y)p_m\gam m+p_0\gam 0+m\r)\gam\nu\gam k\gamd\mu\gam
l\gamd\nu\bl]+\gam\mu(x(1-y)p_k\gam k+p_0\gam
0+m)\\
&&\times\gam\nu(-(1-x)(1-y)p_l\gam l+p_0\gam 0-m)\gamd\mu(yp_m\gam m+p_0\gam 0+m)\gamd\nu\\
&=&l_kl_l\cdot f(p)+g(p) .
\enn
As a result, the above expression becomes more simplified in the following way
\bnn
\Sigma^{(2),c}
&=&-\f{\Gam_R^2}{4}\int^1_0dx\f{(2-d)(3-d)}{2(4\pi)^{\f{d-1}{2}}}\f{\Gam(\f{3-d}{2})}{\Delta^{\f{3-d}{2}}}\int
\f{d^{d-1}\boldsymbol{l}}{(2\pi)^{d-1}}\f{-(2-d)l_m\gam m-2p_0\gam 0+4m}{\boldsymbol{l}^2+p_0^2+m^2}\\
&&-\f{\Gam_R^2}{4}\f{\Gam(\f{5-d}{2})}{(4\pi)^{\f{d-1}{2}}}\int^1_0dx\int
\f{d^{d-1}\boldsymbol{l}}{(2\pi)^{d-1}}\f{\Gam(\f{7-d}{2})}{\Gam(\f{5-d}{2})}\int^1_0dy\f{y^{\f{3-d}{2}}}{[x(1-x)]^{\f{5-d}{2}}}\f{l_kl_l\cdot
f(p)+g(p)}{[\boldsymbol{l}^2+\Delta(p)]^{\f{7-d}{2}}} .
\enn

The first line is easy to perform integrals. It seems to have a double pole, but the presence of $(2-d)(3-d)$ in the numerator gives only $\f{1}{\epsilon}$ as a leading term as follows
\bnn
&&-\f{\Gam_R^2}{4}\int^1_0dx\f{(2-d)(3-d)}{2(4\pi)^{\f{d-1}{2}}}\f{\Gam(\f{3-d}{2})}{\Delta^{\f{3-d}{2}}}\f{-2p_0\gam
0+4m}{(4\pi)^{\f{d-1}{2}}}\f{\Gam(\f{3-d}{2})}{(p_0^2+m^2)^{\f{3-d}{2}}}\\
&=&-\f{\Gam_R^2}{8}\f{(1+\epsilon)\epsilon}{(4\pi)^2}\l(-\f{2}{\epsilon}-\gamma-\log{\Delta}+\log{4\pi}+O(\epsilon)\r)\l(-\f{2}{\epsilon}-\gamma-\log{(p_0^2+m^2)}+\log{4\pi}+O(\epsilon)\r)(-2p_0\gam 0+4m)\\
&=&-\f{\Gam^2}{32\pi^2}\f{1}{\epsilon}(-2p_0\gam 0+4m)+O(1) .
\enn

The evaluation of the second line is quite complicated especially because of the contribution from Dirac algebra in the anomalous dimension and potential appearance of poles by the integration for Feynman parameters of $(x,y)$. We need to analyze both $f(p)$ and $g(p)$ carefully, separating them from each other
\be -\f{\Gam_R^2}{4}\int^1_0dxdy\f{y^{\f{3-d}{2}}}{[x(1-x)]^{\f{5-d}{2}}}\l[\f{\Gam(3-d)}{2(4\pi)^{d-1}}\f{g_{kl}\cdot f(p)}{\Delta(p)^{3-d}}+\f{\Gam(4-d)}{(4\pi)^{d-1}}\f{g(p)}{\Delta(p)^{4-d}}\r] . \no \ee

First, let's analyze the second term with $g(p)$. At a glance, it doesn't have a pole. However, it is possible to diverge if the $x$ integral contains the contribution of $\int^1_0dx[x^{\f{d-3}{2}}(1-x)^{\f{d-5}{2}}]=\f{\Gam(\f{d-1}{2})\Gam(\f{d-3}{2})}{\Gam(d-2)}$, for example. We rearrange $g(p)$ and $\f{1}{\Delta(p)}$ as follows
\bnn
g(p)
&=&\gam\mu(x(1-y)p_k\gam k+p_0\gam 0+m)\gam\nu(-(1-x)(1-y)p_l\gam l+p_0\gam 0-m)\gamd\mu(yp_m\gam m+p_0\gam 0+m)\gamd\nu\\
&=&x(1-x)y(1-y)^2\times(\cdots)+x(1-x)(1-y)^2\times(\cdots)+\cdots+1\times(\cdots)\\
\f{1}{\Delta(p)}&=&\f{1}{(\boldsymbol{p})^{4-d}\l(\f{1}{x(1-x)}+(1-y)^2\r)^{4-d}}\\
&=&(\boldsymbol{p})^{d-4}\br[x(1-x)-[x(1-x)]^2(1-y)^2+[x(1-x)]^3(1-y)^4+\cdots\bl] ,
\enn
where $(\cdots)$ indicates quantities which depend on $p$ but not on $x$ and $y$. We note that the expansion of $\Delta(p)$ is justified because of $0<x,y<1$, which can not be done in a reciprocal way, i.e, $\l[\f{1}{(1-y)^2}-\f{1}{x(1-x)(1-y)^4}+\cdots\r]$. Then, we obtain
\bnn
&&\int^1_0dxdy x^{\f{d-5}{2}}(1-x)^{\f{d-5}{2}}y^{\f{3-d}{2}}\f{g(p)}{\Delta(p)^{4-d}}\\
&&=(\boldsymbol{p})^{d-4}\int^1_0dxdy[x(1-x)]^{\f{d-5}{2}}y^{\f{3-d}{2}}\l[x(1-x)y(1t-y)^2\times(\cdots)+\cdots+1\times(\cdots)\r]\l[x(1-x)-[x(1-x)]^2(1-y)^2+\cdots\r]\\
&&=(\boldsymbol{p})^{d-4}\int^1_0dxdy\br[x^{\f{d-3}{2}}(1-x)^{\f{d-3}{2}}y^{\f{3-d}{2}}\times(\cdots)+\cdots
\bl]\\
&&=(\boldsymbol{p})^{d-4}\l[\f{\Gam(\f{d-1}{2})\Gam(\f{d-1}{2})}{\Gam(d-1)}\f{\Gam(\f{5-d}{2})\Gam(1)}{\Gam(\f{7-d}{2})}+\cdots\r]\\
&&\xrightarrow[]{d\rightarrow3}finite ,
\enn
where we used the formula of $B(x,y)=\int^1_0dt\l[t^{x-1}(1-t)^{y-1}\r]=\f{\Gam(x)\Gam(y)}{\Gam(x+y)}$. The leading term is already finite, which leads us to conclude that the second term gives only a finite value.

Next, let's focus on the first term. The first term contains $\Gam(3-d)$, which gives rise to a divergence as $d\rightarrow 3$. Also, it may cause a divergence in the $x$ integral by the same reason before. In this respect it is more plausible to show a divergent behavior. Performing the Dirac algebra repeatedly,
\bnn
&&g_{kl}\cdot f(p)\\
&=&-x(1-x)\gam\mu\gam k\gam\nu\gamd k\gamd\mu(yp_m\gam m+p_0\gam 0+m)\gamd\nu-x\gam\mu\gam
k\gam\nu(-(1-x)(1-y)p_m\gam
m+p_0\gam 0-m)\gamd\mu\gamd k\gamd \nu\\
&&+(1-x)\gam\mu(x(1-y)p_m\gam m+p_0\gam 0+m)\gam\nu\gam k\gamd\mu\gamd k\gamd\nu\\
&=&-x(1-x)[-2\gamd k\gam\nu\gam k+(4-d)\gam k\gam\nu\gamd k](yp_m\gam m+p_0\gam
0+m)\gamd\nu+\br[x(1-x)(1-y)p_m(-2\gam
m\gam\nu\gam k\\
&&+(4-d)\gam k\gam\nu\gam m)-xp_0(-2\gam 0\gam\nu\gam k+(4-d)\gam k\gam\nu\gam
0)+xm(4g^{k\nu}+(d-4)\gam k\gam\nu)\bl]\gamd
k\gamd\nu\\
&&+\br[x(1-x)(1-y)p_m(-2\gam k\gam\nu\gam m+(4-d)\gam m\gam\nu\gam k)+(1-x)p_0(-2\gam
k\gam\nu\gam 0+(4-d)\gam 0\gam\nu\gam
k)\\
&&+(1-x)m(4g^{\nu k}+(d-4)\gam\nu\gam k)\bl]\gamd k\gamd\nu\\
&=&-x(1-x)yp_m\gam k\l(4g_k^m+(d-4)\gamd k\gam m\r)(2-d)-x(1-x)p_0\gam k\l(4g_k^0+(d-4)\gamd
k\gam 0\r)(2-d)\\
&&-x(1-x)m\gam k\gamd k(2-d)^2-2x(1-x)(1-y)p_m\gam m(d-1)d+x(1-x)(1-y)p_m\gam
k(4g^m_k+(d-4)\gam m\gamd k)(4-d)\\
&&+2xp_0\gam 0(d-1)d-xp_0\gam k(4g^0_k+(d-4)\gam 0\gamd
k)(4-d)+4xm(d-1)+xm(d-4)(2-d)(d-1)\\&&-2x(1-x)(1-y)p_m\gam k(4g^m_k+(d-4)\gam m\gamd
k)+x(1-x)(1-y)p_m\gam m(4-d)(d-1)d\\
&&-2(1-x)p_0\gam k(4g^0_k+(d-4)\gam 0\gamd k)+(1-x)p_0\gam
0(4-d)(1-d)d+4(1-x)m(d-1)+(1-x)m(d-4)(d-1)d\\
&=&-4x(1-x)yp_m\gam m(2-d)-x(1-x)yp_m\gam m(2-d)(d-4)(d-1)-x(1-x)p_0\gam 0(d-4)(2-d)(d-1)\\
&&-x(1-x)m(2-d)^2(d-1)-2x(1-x)(1-y)p_m\gam m(d-1)d+4x(1-x)(1-y)p_m\gam m(4-d)\\
&&+x(1-x)(1-y)p_m\gam m(d-4)^2(d-3)+2xp_0\gam 0(d-1)d+xp_0\gam 0(4-d)(d-4)(d-1)+4xm(d-1)\\
&&+xm(d-4)(2-d)(d-1)-8x(1-x)(1-y)p_m\gam m-2x(1-x)(1-y)p_m\gam m(d-4)(3-d)\\
&&+x(1-x)(1-y)p_m\gam m(4-d)(d-1)d+2(1-x)p_0\gam 0(d-4)(d-1)+(1-x)p_0\gam 0(4-d)(d-1)d\\
&&+4(1-x)m(d-1)+(1-x)m(d-4)(d-1)d\\
&=&p_m\gam m\br[x(1-x)(1-y)[-2d(d-1)+4(4-d)+(d-4)^2(d-3)-8+2(d-4)(d-3)-(d-4)(d-1)d]\\
&&+x(1-x)y[4(d-2)+(d-4)(d-2)(d-1)] \bl]+p_0\gam
0\br[x(1-x)[(d-4)(d-2)(d-1)]+(1-x)[2(d-4)(d-1)\\
&&-(d-4)(d-1)d]+x[2d(d-1)-(d-4)^2(d-1)] \bl]+m\br[x(1-x)[-(d-2)^2(d-1)]\\
&&+(1-x)[4(d-1)+d(d-4)(d-1)]+x[4(d-1)-(d-4)(d-2)(d-1)] \bl]\\
&=&p_m\gam m\br[x(1-x)(1-y)[-6d^2+28d-16]+x(1-x)y[d^3-7d^2+18d-16] \bl]\\
&&+p_0\gam 0\br[x(1-x)[d^3-7d^2+14d-8]+(1-x)[-d^3+7d^2-14d+8]+x[-d^3+11d^2-26d+16]\bl]\\
&&+m\br[x(1-x)[-d^3+5d^2-8d+4]+(1-x)[d^3-5d^2+8d-4]+x[-d^3+7d^2-10d+4] \bl] , \\
\enn we simplify the above expression as follows
\bnn
&&-\f{\Gam_R^2}{4}\int^1_0dxdy\f{y^{\f{3-d}{2}}}{[x(1-x)]^{\f{5-d}{2}}}\l[\f{\Gam(3-d)}{2(4\pi)^{d-1}}\f{g_{kl}\cdot
f(p)}{\Delta(p)^{3-d}}\r]\\
&=&-\f{\Gam_R^2}{8(4\pi)^{d-1}}\int^1_0dxdy\f{y^{\f{3-d}{2}}}{[x(1-x)]^{\f{5-d}{2}}}\f{\Gam(3-d)}{\Delta(p)^{3-d}}\br(p_m\gam m\br[x(1-x)(1-y)[-6d^2+28d-16]+x(1-x)y[d^3-7d^2\\
&&+18d-16] \bl]+p_0\gam 0\br[x(1-x)[d^3-7d^2+14d-8]+(1-x)[-d^3+7d^2-14d+8]+x[-d^3+11d^2-26d+16]\bl]\\
&&+m\br[x(1-x)[-d^3+5d^2-8d+4]+(1-x)[d^3-5d^2+8d-4]+x[-d^3+7d^2-10d+4]\bl]\bl)
\enn
There appears a simple pole already due to $\Gam(3-d)$, but another pole can be made by the $x$ integration. The $y$ integration turns out not to cause a divergence. Since we are interested in contributions with a simple pole, we are allowed to take two kinds of terms only: One has a double pole with $\epsilon$ in a numerator and the other contains a simple pole only. Performing integrals carefully, we find
\bnn
&&-\f{\Gam_R^2}{8(4\pi)^{d-1}}\f{\Gam(3-d)}{\Delta(p)^{3-d}}\br(p_m\gam
m\br[\f{\Gam(\f{d-1}{2})^2}{\Gam(d-1)}\f{\Gam(\f{5-d}{2})\Gam(2)}{\Gam(\f{9-d}{2})}[-6d^2+28d-16]+\f{\Gam(\f{d-1}{2})^2}{\Gam(d-1)}\f{\Gam(\f{7-d}{2})}{\Gam(\f{9-d}{2})}[d^3-7d^2+18d-16]\bl]\\
&&+p_0\gam
0\br[\f{\Gam(\f{d-1}{2})^2}{\Gam(d-1)}\f{\Gam(\f{5-d}{2})}{\Gam(\f{7-d}{2})}[d^3-7d^2+14d-8]+\f{\Gam(\f{d-3}{2})\Gam(\f{d-1}{2})}{\Gam(d-2)}\f{\Gam(\f{5-d}{2})}{\Gam(\f{7-d}{2})}[-d^3+7d^2-14d+8]\\
&&+\f{\Gam(\f{d-3}{2})\Gam(\f{d-1}{2})}{\Gam(d-2)}\f{\Gam(\f{5-d}{2})}{\Gam(\f{7-d}{2})}[-d^3+11d^2-26d+16]\bl]+m\br[\f{\Gam(\f{d-1}{2})^2}{\Gam(d-1)}\f{\Gam(\f{5-d}{2})}{\Gam(\f{7-d}{2})}[-d^3+5d^2-8d+4]\\
&&+\f{\Gam(\f{d-3}{2})\Gam(\f{d-1}{2})}{\Gam(d-2)}\f{\Gam(\f{5-d}{2})}{\Gam(\f{7-d}{2})}[d^3-5d^2+8d-4]+\f{\Gam(\f{d-3}{2})\Gam(\f{d-1}{2})}{\Gam(d-2)}\f{\Gam(\f{5-d}{2})}{\Gam(\f{7-d}{2})}[-d^3+7d^2-10d+4]\bl]\bl)\\
&=&-\f{\Gam_R^2}{128\pi^2}\br(-\f{1}{\epsilon}+\cdots \bl)\br(p_m\gam
m\l[8+O(\epsilon)\r]+p_0\gam
0\l[-2+O(\epsilon)+\Gam(\f{\epsilon}{2})(12+14\epsilon)\r]+m\l[-2+O(\epsilon)+\Gam(\f{\epsilon}{2})(12+10\epsilon)\r]\bl)\\
&=&-\f{\Gam_R^2}{128\pi^2}\f{1}{\epsilon}\br(-8p_m\gam m-26p_0\gam
0-18m\bl)+O\l(\epsilon^{-2}\r)+O(1) .
\enn
As a result, the self-energy correction from the crossed diagram is given by
\bea
\Sigma^{(2),c}
&=&-\f{\Gam_R^2}{32\pi^2}\f{1}{\epsilon}(-2p_0\gam 0+4m)-\f{\Gam_R^2}{128\pi^2}\f{1}{\epsilon}\br(-8p_m\gam m-26p_0\gam 0-18m\bl)+O(\f{1}{\epsilon^2})+O(1)\non
&=&-\f{\Gam_R^2}{128\pi^2}\f{1}{\epsilon}(-34p_0\gam 0-8p_m\gam m-2m)+O(\f{1}{\epsilon^2})+O(1) . \label{r.self-2nd-cross}
\eea It is important to notice that the sign of the self-energy correction in the two-loop order differs from that of the Fock diagram in the one-loop order, which turns out to play a central role in the renormalization group equation for the mass parameter.

\subsection{Vertex corrections}

\subsubsection{Feynman's diagrams}

\begin{figure}[t]
\includegraphics[width=0.8\textwidth]{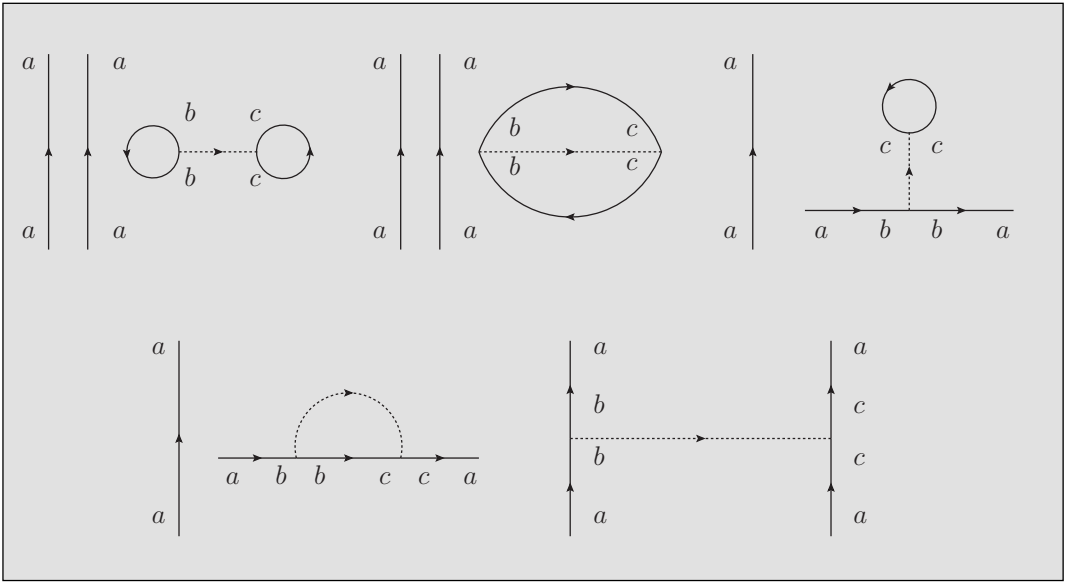}
\caption{All possible first-order corrections for the four-point function without the replica limit.} \label{vertex-1st}
\end{figure}

\begin{figure}[t]
\includegraphics[width=0.4\textwidth]{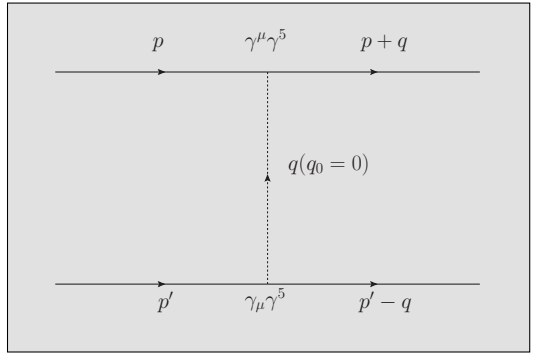}
\caption{Tree level vertex} \label{vertex-tree}
\end{figure}

The vertex renormalization can be found from the four-point function of $G(x,x',y,y')=\f{T^2}{L^{2d}}\sum_{p,p',q,q'}e^{-\i p\cdot x+\i p'\cdot x'-\i q\cdot y+\i q'\cdot y'}G(p,p',q,q')$ with $G(p,p',q,q')=<T\l[\psi(p)\bar{\psi(p')}\psi(q)\bar{\psi(q')}\r]>$. Performing the perturbative analysis up to the $\Gamma_{R}^{2}$ order, we obtain
\bnn
&&G(p,p',q,q')\\
&=&\limR\f{1}{R}\sum_{a=1}^{R}\int D(\bar{\psi},\psi)\psi^a(p)\bar{\psi}^a(p')\psi^a(q)\bar{\psi}^a(q')e^{-\sum_{\alpha=1}^{R}S_{0 [\bar{\psi^\alpha},\psi^\alpha]}}e^{\sum_{b,c=1}^{R}\int^\beta_0d\tau\int^\beta_0 d\tau'\int d^d\boldsymbol{x}\f{\Gam_R}{2}(\bar{\psi^b}_\tau\gam\mu\gam 5\psi^b_{\tau})(\bar{\psi^c}_\tau'\gam\mu\gam 5\psi^c_{\tau'})}\\
&=&\limR\f{1}{R}\sum_{a=1}^{R}\int D(\bar{\psi},\psi)e^{-\sum_{\alpha=1}^{R}S_{0}[\bar{\psi^\alpha} \psi^\alpha]}\br[\psi^a(p)\bar{\psi}^a(p')\psi^a(q)\bar{\psi}^a(q')+\sum_{b,c=1}^{R}\sum_{p_i}\l(-\f{\Gam_R}{2}\r)\psi^a(p)\bar{\psi}^a(p')\psi^a(q)\bar{\psi}^a(q')\\
&&\times(\bar{\psi^b}(p_1)\gam\mu\gam 5\psi^b(p_2))(\bar{\psi^c}(p_3)\gam\mu\gam 5\psi^c(p_4))\delta^{(3)}(\boldsymbol{p}_1-\boldsymbol{p}_2+\boldsymbol{p}_3-\boldsymbol{p}_4)\delta_{p^0_1,
p^0_2}\delta_{p^0_3, p^0_4}+\sum_{b,c,d,e=1}^{R}\sum_{p_i,q_i}\l(-\f{\Gam_R}{2}\r)^2\psi^a(p)\bar{\psi}^a(p')\\
&&\times \psi^a(q)\bar{\psi}^a(q')(\bar{\psi^b}(p_1)\gam\mu\gam 5\psi^b(p_2))(\bar{\psi^c}(p_3)\gam\mu\gam 5\psi^c(p_4))(\bar{\psi^d}(q_1)\gam\mu\gam 5\psi^d(q_2))(\bar{\psi^e}(q_3)\gam\mu\gam
5\psi^e(q_4))\delta^{(3)}(\boldsymbol{p}_1-\boldsymbol{p}_2+\boldsymbol{p}_3\\
&&-\boldsymbol{p}_4)\delta_{p^0_1,p^0_2}\delta_{p^0_3,p^0_4}\delta^{(3)}(\boldsymbol{q}_1-\boldsymbol{q}_2+\boldsymbol{q}_3-\boldsymbol{q}_4)\delta_{q^0_1,
q^0_2}\delta_{q^0_3, q^0_4}+O(\Gam_R^3) \bl]\\
&=&\limR\f{1}{R}\sum_{a=1}^{R}<\psi^a(p)\bar{\psi}^a(p')\psi^a(q)\bar{\psi}^a(q')>_0+\limR\f{1}{R}\sum_{a,b,c=1}^{R}\l(-\f{\Gam_R}{2}\r)\sum_{p_i}<\psi^a(p)\bar{\psi}^a(p')\psi^a(q)\bar{\psi}^a(q')\bar{\psi^b}(p_1)\gam\mu\gam 5\\
&&\times\psi^b(p_2)\bar{\psi^c}(p_3)\gam\mu\gam 5\psi^c(p_4)>_0\delta^{(4)}(p_i)+\limR\f{1}{R}\sum_{a,b,c,d,e=1}^{R}\l(-\f{\Gam_R}{2}\r)^2\sum_{p_i,q_i}<\psi^a(p)\bar{\psi}^a(p')\psi^a(q)\bar{\psi}^a(q')\bar{\psi^b}(p_1)\gam\mu\gam 5\\
&&\times\psi^b(p_2)\bar{\psi^c}(p_3)\gam\mu\gam 5\psi^c(p_4)\bar{\psi^d}(q_1)\gam\mu\gam 5\psi^d(q_2)\bar{\psi^e}(q_3)\gam\mu\gam
5\psi^e(q_4)>_0\delta^{(4)}(p_i)\delta^{(4)}(q_i)+O(\Gam_R^3) ,
\enn
where we introduced a short-hand-notation of $\delta^{(4)}(p_i)=\delta^{(3)}(\boldsymbol{p}_1-\boldsymbol{p}_2+\boldsymbol{p}_3-\boldsymbol{p}_4)\delta_{p^0_1,
p^0_2}\delta_{p^0_3, p^0_4}$. All possible first order contributions are given by (Fig. \ref{vertex-1st})
\bnn
&&\limR\f{1}{R}\sum_{a,b,c=1}^{R}<\psi^a_i\bar{\psi}^a_j\psi^a_k\bar{\psi}^a_l\bar{\psi^b}_m(\gam\mu\gam5)_{mn}\psi^b_n\bar{\psi^c}_o(\gam\mu\gam
5)_{op}\psi^c_p>_0\\
&=&\limR\f{1}{R}\sum_{a, b,c=1}^{R}\br[4<\psi^a_i\bar{\psi}^a_j>_0<\psi^a_k\bar{\psi}^a_l>_0<\psi^b_n\bar{\psi^b}_m>_0<\psi^c_p\bar{\psi^c}_o>_0(\gam\mu\gam 5)_{mn}(\gam\mu\gam 5)_{op}-4<\psi^a_i\bar{\psi}^a_j>_0<\psi^a_k\bar{\psi}^a_l>_0\\
&&\times<\psi^b_n\bar{\psi^c}_o>_0<\psi^c_p\bar{\psi^b}_m>_0(\gam\mu\gam 5)_{mn} (\gam\mu\gam 5)_{op}-4<\psi^a_i\bar{\psi}^a_j>_0<\psi^b_n\bar{\psi}^a_l>_0<\psi^a_k\bar{\psi^b}_m>_0<\psi^c_p\bar{\psi^c}_o>_0(\gam\mu\gam 5)_{mn}\\
&&\times(\gam\mu\gam 5)_{op}+4<\psi^a_i\bar{\psi}^a_j>_0<\psi^b_n\bar{\psi}^a_l>_0<\psi^c_p\bar{\psi^b}_m>_0<\psi^a_k\bar{\psi^c}_o>_0(\gam\mu\gam 5)_{mn}(\gam\mu\gam 5)_{op}+4<\psi^b_n\bar{\psi}^a_j>_0<\psi^a_i\bar{\psi}^b_m>_0\\
&&\times<\psi^c_p\bar{\psi^a}_l>_0<\psi^a_k\bar{\psi^c}_o>_0(\gam\mu\gam 5)_{mn} (\gam\mu\gam 5)_{op}\bl]\\
&=&\limR\f{1}{R}\sum_{a,b,c=1}^{R}\br[2G^a_{ij}G^a_{kl}G^b_{nm}G^c_{po}(\gam\mu\gam 5)_{mn}(\gam\mu\gam 5)_{op}\delta_{aa}\delta_{aa}\delta_{bb}\delta_{cc}-2G^a_{ij}G^a_{kl}G^b_{no}G^c_{pm}(\gam\mu\gam 5)_{mn}(\gam\mu\gam 5)_{op}\delta_{aa}\delta_{aa}\delta_{bc}\delta_{cb}\\
&&-4G^a_{ij}G^b_{nl}G^a_{km}G^c_{po}(\gam\mu\gam 5)_{mn}(\gam\mu\gam 5)_{op}\delta_{aa}\delta_{ba}\delta_{ab}\delta_{cc}+4G^a_{ij}G^b_{nl}G^c_{pm}G^a_{ko}(\gam\mu\gam 5)_{mn}(\gam\mu\gam 5)_{op}\delta_{aa}\delta_{ba}\delta_{cb}\delta_{ac}+4G^b_{nj}G^a_{im}\\
&&\times G^c_{pl}G^a_{ko}\delta_{ba}\delta_{ab}\delta_{ca}\delta_{ac}(\gam\mu\gam 5)_{mn}(\gam\mu\gam 5)_{op}\bl]\\
&=&\limR\f{1}{R}\br[2\sum_{a, b, c=1}^{R}[G^a]^2\otimes tr[G^b\gam\mu\gam 5]tr[G^c\gamd\mu\gam 5]\delta_{aa}\delta_{aa}\delta_{bb}\delta_{cc}+2\sum_{a, b, c=1}^{R}[G^a]^2\otimes tr[G^b\gam\mu\gam 5G^c\gamd\mu\gam 5]\delta_{aa}\delta_{aa}\delta_{bc}\delta_{cb}\\
&&-4G^a\otimes G^a\gam\mu\gam 5)G^btr[G^c\gam\mu\gam 5]\delta_{aa}\delta_{ba}\delta_{ab}\delta_{cc}+4G^a\otimes G^a\gam\mu\gam 5G^c\gam\mu\gam 5G^b\delta_{aa}\delta_{ba}\delta_{cb}\delta_{ac}+4G^a\gam\mu\gam 5G^c\otimes G^a\gam\mu\gam 5\\
&&\times G^b\delta_{ba}\delta_{ab}\delta_{ca}\delta_{ac}\bl]
\enn
where $4$ results from identical contributions in the first equality and $-$ comes from the odd number of fermion loops (one loop). The first term is proportional to $R^3$, the second and third, $R^2$, the forth and fifth, $R$. As a result, only the forth and fifth terms survive in the replica limit. But, the forth term is just the product of a bare propagator and a propagator with a Fock self-energy. Therefore, the four-point function and the scattering matrix element are (Fig. \ref{vertex-tree})
\bnn
G^{(1)}(p,p';q)&=&4\l(-\f{\Gam_R}{2}\r)G(p)\gam\mu\gam 5G(p+q)\otimes G(p')\gam\mu\gam 5G(p'-q)\\
M^{(1)}(p,p;q)&=&4\l(-\f{\Gam_R}{2}\r)\gam\mu\gam 5\otimes\gam\mu\gam 5 .
\enn

All possible quantum corrections in the second order are given by
\begin{figure}[t]
\includegraphics[width=1\textwidth]{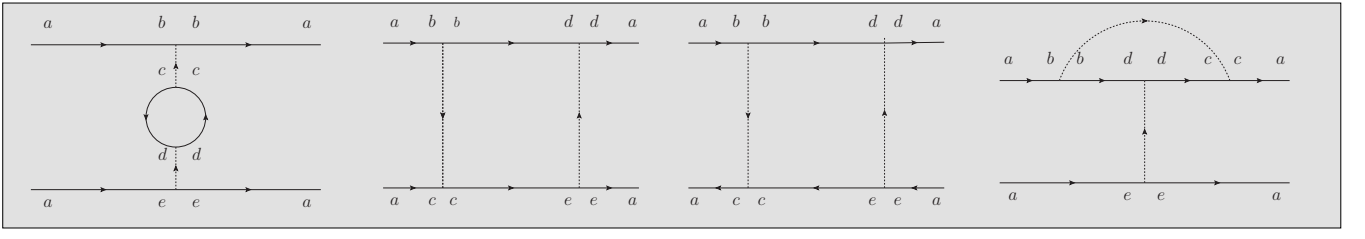}
\caption{Second-order vertex corrections without the replica limit} \label{vertex-2nd}
\end{figure}
\bnn
&&\limR\f{1}{R}\sum_{a,b,c,d,e=1}^{R}\br[<\psi^a_i\bar{\psi}^a_j\psi^a_k\bar{\psi}^a_l\bar{\psi^b}_m(\gam\mu\gam 5)_{mn}\psi^b_n\bar{\psi^c}_o(\gamd\mu\gam 5)_{op}\psi^c_p \bar{\psi^d}_q(\gam\nu\gam 5)_{qr}\psi^d_r\bar{\psi^e}_s(\gamd\nu\gam 5)_{st}\psi^e_t>_0\bl]_{connected}\\
&=&\limR\f{1}{R}\sum_{a, b, c,d,e=1}^{R}\br[16G^a\gam\mu\gam 5G^b\otimes G^a\gamd\nu\gam 5G^etr[G^c\gamd\mu\gam 5G^d\gam\nu\gam 5]\delta_{ab}\delta_{ba}\delta_{cd}\delta_{dc}\delta_{ae}\delta_{ea}+16G^a\gam\mu\gam 5G^b\gam\nu\gam 5G^d\otimes\\
&&G^a\gamd\mu\gam 5G^c\gamd\nu\gam 5G^e\delta_{ab}\delta_{bd}\delta_{da}\delta_{ac}\delta_{ce}\delta_{ea}+16G^a\gam\mu\gam 5G^b\gam\nu\gam 5G^d\otimes G^a\gamd\nu\gam 5G^e\gamd\mu\gam 5G^c\delta_{ab}\delta_{bd}\delta_{da}\delta_{ae}\delta_{ec}\delta_{ca}\\
&&+32G^a\gam\mu\gam 5G^b\gamd\nu\gam 5G^d\gamd\mu\gam 5G^c\otimes G^a\gamd\nu\gam 5G^e\delta_{ab}\delta_{bd}\delta_{dc}\delta_{ca}\delta_{ae}\delta_{ea}\bl] ,
\enn where only diagrams fully connected with the external lines have been taken into account. The first term is proportional to $R^2$ while all other terms are $\sim R$. As a result, the four-point function and the scattering matrix element in the second order are given by (Fig. \ref{p.vertex correction})
\begin{figure}[t]
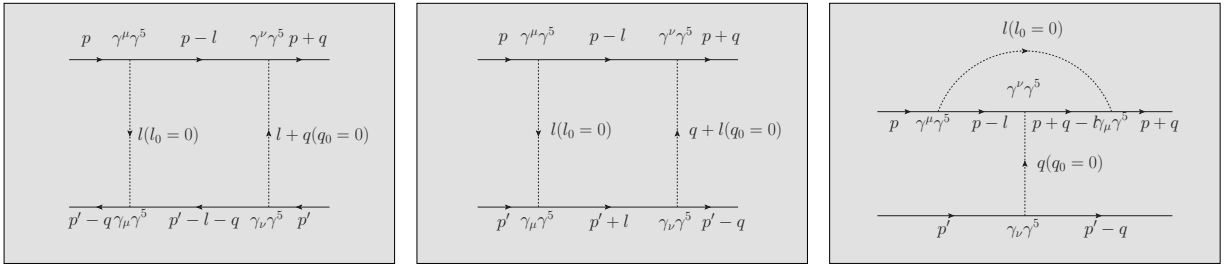

\includegraphics[width=0.3\textwidth]{vertex-ph}
\includegraphics[width=0.3\textwidth]{vertex-pp}
\includegraphics[width=0.3\textwidth]{vertex-ver}
\caption{Second-order vertex corrections in the replica limit} \label{p.vertex correction}
\end{figure}
\bea
G^{(2)}(p,p';q)&=&\l(-\f{\Gam_R}{2}\r)^2\sum_{l}\br[16G(p)\gam\mu\gam 5G(p-l)\gam\nu\gam 5G(p+q)\otimes G(p')\gamd\mu\gam 5G(p'+l)\gamd\nu\gam 5G(p'-q)\non
&&+16G(p)\gam\mu\gam 5G(p-l)\gam\nu\gam 5G(p+q)\otimes G(p')\gamd\nu\gam 5G(p'-l-q)\gamd\mu\gam 5G(p'-q)\non
&&+32G(p)\gam\mu\gam 5G(p-l)\gamd\nu\gam 5G(p+q-l)\gamd\mu\gam 5G(p+q)\otimes G(p')\gamd\nu\gam 5G(p'-q)\bl]\non
M^{(2)}(p,p;q)&=&4A_{pp}+4A_{ph}+8A_{ver}\non
A_{pp}&=&\l(-\f{\Gam_R}{2}\r)^2\sum_{l}\gam\mu\gam 5G(p-l)\gam\nu\gam 5\otimes\gamd\mu\gam 5G(p'+l)\gamd\nu\gam 5\label{e.ver-pp}\\
A_{ph}&=&\l(-\f{\Gam_R}{2}\r)^2\sum_{l}\gam\mu\gam 5G(p-l)\gam\nu\gam 5\otimes G(p')\gamd\nu\gam 5G(p'-l-q)\gamd\mu\gam 5\label{e.ver-ph}\\
A_{ver}&=&\l(-\f{\Gam_R}{2}\r)^2\sum_{l}\gam\mu\gam 5G(p-l)\gamd\nu\gam 5G(p+q-l)\gamd\mu\gam 5\otimes\gamd\nu\gam 5 , \label{e.ver-ver}
\eea
where $pp$ and $ph$ represent ``particle-particle" and ``particle-hole", respectively, and $ver$ means ``vertex".

\subsubsection{Evaluation of relevant Feynman's diagrams}

First, we focus on the particle-hole diagram (the first diagram in Fig. \ref{p.vertex correction}), given by
\bnn
A_{ph}
&=&\l(-\f{\Gam_R}{2}\r)^2\int \f{d^dl}{(2\pi)^d}2\pi\delta (l_0)\gam\mu\gam
5\f{\not{p}-\not{l}-m}{(p-l)^2-m^2}\gam\nu\gam
5\otimes\gamd\mu\gam 5\f{\not{p'}-\not{q}-\not{l}-m}{(p'-q-l)^2-m^2}\gamd\nu\gam 5\\
&=&\f{\Gam_R^2}{4}\int \f{d^{d-1}\boldsymbol{l}}{(2\pi)^{d-1}}\gam\mu \f{p_0\gam 0+p_k\gam k-l_k\gam
k+m}{-p_0^2-(\boldsymbol{p}-\boldsymbol{l})^2-m^2}\gam\nu\otimes\gamd\mu\f{p'_0\gam 0+p'_l\gam l-q_0\gam 0-q_l\gam l-l_l\gam
l+m}{-({p'}_0+q_0)^2-(\boldsymbol{p'}-\boldsymbol{q}-\boldsymbol{l})^2-m^2}\gamd\nu\\
&=&\f{\Gam_R^2}{4}\int \f{d^{d-1}\boldsymbol{l}}{(2\pi)^{d-1}}\f{\gam\mu(p_0\gam 0+p_k\gam k-l_k\gam
l+m)\gam\nu}{(\boldsymbol{l}-\boldsymbol{p})^2+p_0^2+m^2}\otimes\f{\gamd\mu((p'_0-q_0)\gam 0+p'_l\gam l-q_l\gam l-l_l\gam
l+m)\gamd\nu}{(\boldsymbol{l}-\boldsymbol{p'}+\boldsymbol{q})^2+({p'}_0-q_0)^2+m^2}\\
&=&\f{\Gam_R^2}{4}\int \f{d^{d-1}\boldsymbol{l}}{(2\pi)^{d-1}}\f{N}{D}
\enn with
\bnn
D^{-1}&=&\int^1_0
dx[\boldsymbol{l}^2+x(1-x)(\boldsymbol{p}-\boldsymbol{p}'+\boldsymbol{q})^2+xp_0^2+(1-x)({p'}_0-q_0)^2+m^2]^{-2}=\int^1_0dx[\boldsymbol{l}^2+\Delta(p,p'-q)]^{-2}\\
N&=&\gam\mu(-l_k\gam k+(1-x)(p_k-p_k'+q_k)\gam k+p_0\gam 0+m)\gam\nu\otimes\gamd\mu(-l_l\gam
l-x(p_l-p_l'+q_l)\gam l+(p'_0-q_0)\gam
0+m)\gamd\nu\\
&=&l_kl_l\gam\mu\gam k\gam\nu\otimes\gamd\mu\gam l\gamd\nu+\gam\mu((1-x)(p_k-p'_k+q_k)\gam
k+p_0\gam
0+m)\gam\nu\otimes\gamd\mu(-x(p_l-p'_l+q_l)\gam l+(p'_0-q_0)\gam 0+m)\gamd\nu\\
&=&l_kl_l\gam\mu\gam k\gam\nu\otimes\gamd\mu\gam l\gamd\nu+f(p,p'-q) .
\enn Then, we obtain
\bea
A_{ph}
&=&\f{\Gam_R^2}{4}\int^1_0dx\int \f{d^{d-1}\boldsymbol{l}}{(2\pi)^{d-1}}\f{l_kl_l\gam\mu\gam k\gam\nu\otimes\gamd\mu\gam l\gamd\nu+f(p,p'-q)}{[\boldsymbol{l}^2+\Delta]^{-2}}\non
&=&\f{\Gam_R^2}{4}\int^1_0dx\l[\f{-1}{(4\pi)^{d-1}}\f{\gam\mu\gam k\gam\nu\otimes\gamd\mu\gamd
k\gamd\nu}{2}\f{\Gam(1-\f{d-1}{2})}{\Gam(2)}\f{1}{\Delta^{1-\f{d-1}{2}}}+\f{f(p,p'-q)}{(4\pi)^{d-1}}\f{\Gam(2-\f{d-1}{2})}{\Gam(2)}\f{1}{\Delta^{2-\f{d-1}{2}}}\r]\non
&=&\f{\Gam_R^2}{4}\int^1_0dx\l[\f{-1}{(4\pi)^{d-1}}\f{\gam\mu\gam k\gam\nu\otimes\gamd\mu\gamd
k\gamd\nu}{2}\f{\Gam(\f{3-d}{2})}{\Delta^{\f{3-d}{2}}}+\f{f(p,p'-q)}{(4\pi)^{d-1}}\f{\Gam(\f{5-d}{2})}{\Delta^{\f{5-d}{2}}}
\r]\non
&=&\f{\Gam_R^2}{4}\int^1_0dx\l[-\f{\gam\mu\gam k\gam\nu\otimes\gamd\mu\gamd k\gamd\nu}{2(4\pi)}\br(-\f{2}{\epsilon}-\gamma-\log{\Delta(p,p'-q)}+\log{4\pi}+O(\epsilon)\bl)+\f{f(p,p'-q)}{(4\pi)^{\f{2+\epsilon}{2}}}\f{\Gam(\f{2-\epsilon}{2})}{\Delta^{\f{2-\epsilon}{2}}}\r] . \label{r.ver-ph}
\eea

Next, we evaluate the particle-particle diagram (the second diagram in Fig. \ref{p.vertex correction}). It is almost identical with the way for the particle-hole channel how to perform the integral of the particle-particle channel. We find
\bea
A_{pp}
&=&\l(-\f{\Gam_R}{2}\r)^2\int \f{d^dl}{(2\pi)^d}2\pi\delta (l_0)\gam\mu\gam 5\f{\not{p}-\not{l}-m}{(p-l)^2-m^2}\gam\nu\gam 5\otimes\gamd\mu\gam 5\f{\not{p'}+\not{l}-m}{(p'+l)^2-m^2}\gamd\nu\gam 5\non
&=&A_{ph}(p'-q\rightarrow p', l^2\rightarrow -l^2)\non
&=&\f{\Gam_R^2}{4}\int^1_0dx\l[+\f{\gam\mu\gam k\gam\nu\otimes\gamd\mu\gamd k\gamd\nu}{2(4\pi)}\br(-\f{2}{\epsilon}-\gamma-\log{\Delta(p,p')}+\log{4\pi}+O(\epsilon)\bl)+\f{f(p,p')}{(4\pi)^{\f{2+\epsilon}{2}}}\f{\Gam(\f{2-\epsilon}{2})}{\Delta(p,p')^{\f{2-\epsilon}{2}}}\r] . \label{r.ver-pp}
\eea

Lastly, we evaluate the third diagram in Fig. \ref{p.vertex correction}, given by
\bnn
A_{ver}
&=&\l(-\f{\Gam_R}{2}\r)^2\int \f{d^dl}{(2\pi)^d}2\pi\delta (l_0)\gam\mu\gam 5\f{\not{p}-\not{l}-m}{(p-l)^2-m^2}\gam\nu\gam 5\f{\not{p}-\not{l}+\not{q}-m}{(p-l+q)^2-m^2}\gamd\mu\gam 5\otimes\gamd\nu\gam 5\\
&=&\f{\Gam_R^2}{4}\int \f{d^{d-1}\boldsymbol{l}}{(2\pi)^{d-1}}\gam\mu \f{p_0\gam 0+p_k\gam k-l_k\gam k+m}{-p_0^2-(\boldsymbol{p}-\boldsymbol{l})^2-m^2}\gam\nu\f{(p_0+q_0)\gam 0+(p_l+q_l)\gam l-l_l\gam l-m}{-(p_0+q_0)^2-(\boldsymbol{p}+\boldsymbol{q}-\boldsymbol{l})^2-m^2}\gamd\mu\gam 5\otimes\gamd\nu\gam 5\\
&=&\f{\Gam_R^2}{4}\int \f{d^{d-1}\boldsymbol{l}}{(2\pi)^{d-1}}\f{\gam\mu(-l_k\gam k+p_0\gam 0+p_k\gam k+m)\gam\nu((p_0+q_0)\gam 0+(p_l+q_l)\gam l-l_l\gam l-m)\gamd\mu\gam 5\otimes\gamd\nu\gam 5}{((\boldsymbol{l}-\boldsymbol{p})^2+p_0^2+m^2)((\boldsymbol{l}-\boldsymbol{p}-\boldsymbol{q})^2+(p_0+q_0)^2+m^2)}\\
&=&\f{\Gam_R^2}{4}\int \f{d^{d-1}\boldsymbol{l}}{(2\pi)^{d-1}}\f{N}{D}
\enn with
\bnn
D^{-1}&=&\int^1_0
dx[\boldsymbol{l}^2+x(1-x)\boldsymbol{q}^2+xq_0^2+2xp_0q_0+p_0^2+m^2]^{-2}=\int^1_0dx[\boldsymbol{l}^2+\Delta'(p,q)]^{-2}\\
N&=&\gam\mu(-l_k\gam k-xq_k\gam k+p_0\gam 0+m)\gam\nu(-l_l\gam l-(1-x)q_l\gam l+(p_0+q_0)\gam 0-m)\gamd\mu\gam 5\otimes\gamd\nu\gam 5\\
&=&l_kl_l\gam\mu\gam k\gam\nu\gam l\gamd\mu\gam 5\otimes\gamd\nu\gam 5+\gam\mu(-xq_k\gam k+p_0\gam 0+m)\gam\nu((1-x)q_l\gam l+(p_0+q_0)\gam 0-m)\gamd\mu\gam 5\otimes\gamd\nu\gam 5\\
&=&l_kl_l(-2\gam l\gam\nu\gam k+(4-d)\gam k\gam\nu\gam l)\gam 5\otimes\gamd\nu\gam 5+g(p,q) .
\enn Following the similar procedure as the above, we obtain
\bea
A_{ver}
&=&\f{\Gam_R^2}{4}\int^1_0dx\int \f{d^{d-1}\boldsymbol{l}}{(2\pi)^{d-1}}\f{l_kl_l(-2\gam l\gam\nu\gam k+(4-d)\gam k\gam\nu\gam l)\gam 5\otimes\gamd\nu\gam 5+g(p,q)}{[\boldsymbol{l}^2+\Delta'(p,q)]^2}\non
&=&\f{\Gam_R^2}{4}\int^1_0dx\l[\f{1}{(4\pi)^{\f{d-1}{2}}}\f{(-2\gam l\gam\nu\gam k+(4-d)\gam k\gam\nu\gam l)\gam 5\otimes\gamd\nu\gam
5}{2}\f{\Gam(\f{3-d}{2})}{\Gam(2)}\f{1}{\Delta'(p,q)^{\f{3-d}{2}}}+\f{g(p,q)}{(4\pi)^{d-1}}\f{\Gam(\f{5-d}{2})}{\Gam(2)}\f{1}{\Delta'(p,q)^{\f{5-d}{2}}}\r]\non
&=&\f{\Gam_R^2}{4}\int^1_0dx\l[\f{(2-d)}{(4\pi)^{\f{d-1}{2}}}\f{\gam l\gam\nu\gamd l\gam 5\otimes\gamd\nu\gam
5}{2}\f{\Gam(\f{3-d}{2})}{\Gam(2)}\f{1}{\Delta'(p,q)^{\f{3-d}{2}}}+\f{g(p,q)}{(4\pi)^{d-1}}\f{\Gam(\f{5-d}{2})}{\Gam(2)}\f{1}{\Delta'(p,q)^{\f{5-d}{2}}}\r]\non
&=&\f{\Gam_R^2}{4}\int^1_0dx\l[\f{\epsilon(\epsilon+1)}{4\pi}\br(-\f{2}{\epsilon}-\gamma-\log{\Delta'(p,q)}+\log{4\pi}+O(\epsilon)\bl)\gam\nu\gam 5\otimes\gamd\nu\gam 5+\f{g(p,q)}{(4\pi)^{\f{2+\epsilon}{2}}}\f{\Gam(\f{2-\epsilon}{2})}{\Delta'(p,q)^{\f{2-\epsilon}{2}}}\r] . \label{r.ver-ver}
\eea

\section{Evaluation of renormalization constants}

Combining Eq. (\ref{r.self-1st}), Eq. (\ref{r.self-2nd-rainbow}), and Eq. (\ref{r.self-2nd-cross}) in the following way
\bnn
&&2\times\Sigma^{(1)}+8\times\Sigma^{(2),r}+8\times\Sigma^{(2),c}+\l[\delta_\psi^\omega p_0\gam 0+\delta_\psi^{\boldsymbol{k}} p_k\gam k+\delta_m m\r]\\
&=&2\times\f{-\Gam_R}{4\pi}\f{1}{\epsilon}(-2p_0\gam 0+4m)+8\times\f{\Gam_R^2}{16\pi^2}\f{1}{\epsilon}(3p_0\gam 0+10m)+8\times\f{\Gam_R^2}{128\pi^2}\f{1}{\epsilon}(34p_0\gam 0+8p_m\gam m+2m)\\
&&+\delta_\psi^\omega p_0\gam0+\delta_\psi^{\boldsymbol{k}} p_k\gam k+\delta_m m+O(1)\\
&=&p_0\gam 0\l(\f{\Gam_R}{\pi}\f{1}{\epsilon}+\f{29\Gam_R^2}{8\pi^2}\f{1}{\epsilon}+\delta_\psi^\omega\r)+p_m\gam m\l(\f{\Gam_R^2}{2\pi^2}\f{1}{\epsilon}+\delta_\psi^{\boldsymbol{k}}\r)+m\l(-\f{2\Gam_R}{\pi}\f{1}{\epsilon}+\f{41\Gam_R^2}{8\pi^2}\f{1}{\epsilon}\r) ,
\enn
we find counter terms
\be \delta_{\psi}^{\omega}=-\f{\Gam_R}{\pi}\f{1}{\epsilon}-\f{29\Gam_R^2}{8\pi^2}\f{1}{\epsilon}, ~~~~~ \delta_{\psi}^{\boldsymbol{k}}=-\f{\Gam_R^2}{2\pi^2}\f{1}{\epsilon},  ~~~~~ \delta_{m}=\f{2\Gam_R}{\pi}\f{1}{\epsilon}-\f{41\Gam_R^2}{8\pi^2}\f{1}{\epsilon} , \no\ee
which give rise to renormalization constants of
\be\left\{\begin{matrix}
Z_{\psi}^{\omega}=1-\f{\Gam_R}{\pi}\log{\mu}-\f{29\Gam_R^2}{8\pi^2}\log{\mu}\simeq\exp{\l[-\f{\Gam_R}{\pi}\log{\mu}-\f{29\Gam_R^2}{8\pi^2}\log{\mu}\r]}\\
Z_{\psi}^{\boldsymbol{k}}=1-\f{\Gam_R^2}{2\pi^2}\log{\mu}\simeq\exp{\l[-\f{\Gam_R^2}{2\pi^2}\log{\mu}\r]}\\
Z_{m}=1+\f{2\Gam_R}{\pi}\log{\mu}-\f{41\Gam_R^2}{8\pi^2}\log{\mu}\simeq\exp{\l[\f{2\Gam_R}{\pi}\log{\mu}-\f{41\Gam_R^2}{8\pi^2}\r]}\label{r.Rfactors}
\end{matrix}\right\} , \ee
where $\f{1}{\epsilon}$ is replaced with a cutoff scale, $\log{\mu}$.

Similarly, one can find the renormalization constant for vertex corrections. It turns out that the particle-hole contribution Eq. (\ref{r.ver-ph}) cancels the particle-particle correction Eq. (\ref{r.ver-pp}) while the vertex part Eq. (\ref{r.ver-ver}) is finite, given by \be 4\times A_{ph}+4\times A_{pp}+8\times A_{ver}+\delta_\Gam\f{\Gam_R}{2}(\gam\nu\gam 5\otimes\gamd\nu\gam 5)=O(1)+\delta_\Gam\f{\Gam_R}{2}(\gam\nu\gam 5\otimes\gamd\nu\gam 5) . \no\ee As a result, we find $Z_{\Gam}=1$ up to the $\Gam_{R}^2$ order.

\section{Derivation of renormalization group equations}

Recall the relation between the bare and renormalized coupling constant: $\Gam_B=\mu^{3-d}(Z_\psi^\omega)^{-2}Z_\Gam\Gam_R$. It is straightforward to find the renormalization group equation for the variance parameter
\bea
&& \f{d\log{\Gam_R}}{d\log{\mu}}=d-3+2\f{d\log{Z_{\psi}^{\omega}}}{d\log{\mu}}-\f{d\log{Z_\Gam}}{d\log{\mu}} . \label{e.RGE-gam}
\eea
Similarly, $m_B=\mu (Z_\psi^\omega)^{-1}Z_mm_R$ results in
\bea
&& \f{d\log{m_R}}{d\log{\mu}}=-1+\f{d\log{Z_{\psi}^{\omega}}}{d\log{\mu}}-\f{d\log{Z_m}}{d\log{\mu}} . \label{e.RGE-m}
\eea

Substituting Eq. (\ref{r.Rfactors}) and $Z_\Gam=1$ into Eq. (\ref{e.RGE-gam}) and Eq. (\ref{e.RGE-m}), we obtain the renormalization group equations for the variance and mass parameter
\bea
\f{d\log{\Gam_R}}{d\log{\mu}}&=&1-\f{2\Gam_R}{\pi}-\f{29\Gam_R^2}{4\pi^2}\label{r.RGE-gam}\\
\f{d\log{m_R}}{d\log{\mu}}&=&-1-\f{3\Gam_R}{\pi}+\f{3\Gam_R^2}{2\pi^2} .\label{r.RGE-m}
\eea
\begin{figure}[t]
\includegraphics[width=0.3\textwidth]{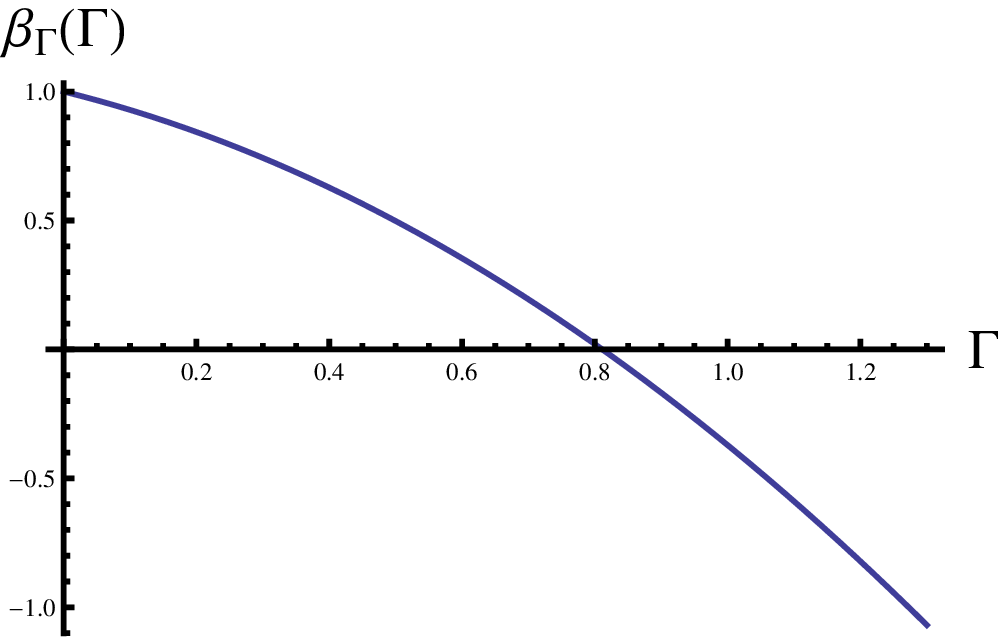}
\includegraphics[width=0.3\textwidth]{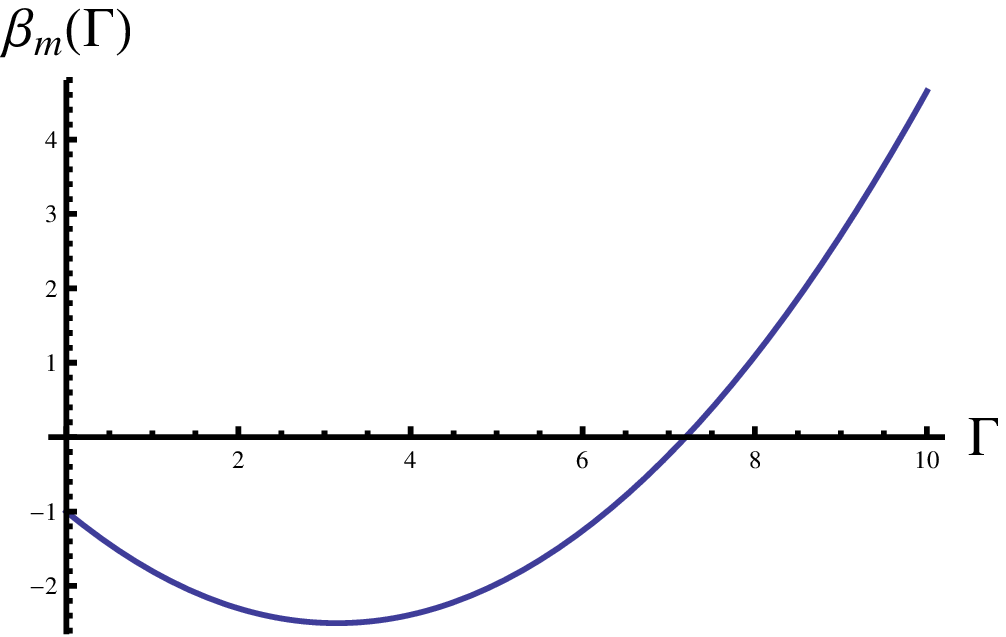}
\caption{Beta functions} \label{p.betafunction}
\end{figure}


\begin{thebibliography}{9}
\bibitem{DMFT_RMP} Georges, A., Kotliar, G., Krauth, W. $\&$ Rozenberg, M. J. Dynamical mean-field theory of strongly correlated fermion systems and the limit of infinite dimensions. Rev. Mod. Phys. {\bf 68}, 13 (1996).
\bibitem{Lee_Nagaosa_Wen_RMP} Lee, P. A., Nagaosa, N. $\&$ Wen. X.-G. Doping a Mott insulator: Physics of high-temperature superconductivity. Rev. Mod. Phys. {\bf 78}, 17 (2006).
\bibitem{HFQCP_RMP} L$\ddot{o}$hneysen, H. v., Rosch A., Vojta, M., $\&$ W$\ddot{o}$lfle, P. Fermi-liquid instabilities at magnetic quantum phase transitions. Rev. Mod. Phys. {\bf 79}, 1015 ( 2007).
\bibitem{Doped_Si_Review} L$\ddot{o}$hneysen, H. v. Electron-electron interactions and the metal-insulator transition in heavily doped silicon. Ann. Phys. (Berlin) {\bf 523}, 599 (2011).
\bibitem{Disorder_Review} Lee, P. A. $\&$ Ramakrishnan, T. V. Disordered electronic systems. Rev. Mod. Phys. {\bf 57}, 287 (1985).
\bibitem{Disorder_Interaction_Review} Dobrosavljevic, V. Introduction to Metal-Insulator Transitions. arXiv:1112.6166, "Conductor Insulator Quantum Phase Transitions", edited by V. Dobrosavljevic, N. Trivedi, and J.M. Valles Jr., Oxford University Press, 2012, ISBN 9780199592593.
\bibitem{DMS_Review} Dietl, T. $\&$ and Ohno, H. Dilute ferromagnetic semiconductors: Physics and spintronic structures. Rev. Mod. Phys. {\bf 86}, 187 (2014).
\bibitem{TI_Review_I} Hasan, M. Z. $\&$ Kane, C. L. Colloquium: Topological insulators. Rev. Mod. Phys. {\bf 82}, 3045 (2010).
\bibitem{TI_Review_II} Qi, X.-L. $\&$ Zhang, S.-C. Topological insulators and superconductors. Rev. Mod. Phys. {\bf 83}, 1057 (2011).
\bibitem{Weyl_Metal_I} Haldane, F. D. M. Berry Curvature on the Fermi Surface: Anomalous Hall Effect as a Topological Fermi-Liquid Property. Phys. Rev. Lett. {\bf 93}, 206602 (2004).
\bibitem{Weyl_Metal_II} Murakami, S. Phase transition between the quantum spin Hall and insulator phases in 3D: emergence of a topological gapless phase. New J. Phys. {\bf 9}, 356 (2007).
\bibitem{Weyl_Metal_III} Burkov, A. A. $\&$ Balents, L. Weyl Semimetal in a Topological Insulator Multilayer. Phys. Rev. Lett. {\bf 107}, 127205 (2011).
\bibitem{Weyl_Metal_IV} Nielsen, H. B. $\&$ Ninomiya, M. The Adler-Bell-Jackiw anomaly and Weyl fermions in a crystal. Phys. Lett. {\bf 130B}, 389 (1983).
\bibitem{Kim_Kim_Sasaki_FMTI} Kim, H.-J., Kim, K.-S., Wang, J.-F., Kulbachinskii, V. A., Ogawa, K., Sasaki, M., Ohnishi, A., Kitaura, M., Wu, Y.-Y., Li, L., Yamamoto, I., Azuma, J., Kamada, M. $\&$ Dobrosavljevic, V. Topological Phase Transitions Driven by Magnetic Phase Transitions in Fe$_x$Bi$_2$Te$_3$ ($0 \leq x \leq 0.1$) Single Crystals. Phys. Rev. Lett. {\bf 110}, 136601 (2013).
\bibitem{Axion_EM1} Wilczek, F. Two applications of axion electrodynamics. Phys. Rev. Lett. {\bf 58}, 1799 (1987).
\bibitem{Axion_EM2} Kim, H.-J., Kim, K.-S., Wang, J.-F., Sasaki, M., Satoh, N., Ohnishi, A., Kitaura, M., Yang, M., $\&$ Li, L. Dirac versus Weyl Fermions in Topological Insulators: Adler-Bell-Jackiw Anomaly in Transport Phenomena. Phys. Rev. Lett. {\bf 111}, 246603 (2013).
\bibitem{Axion_EM3} Kim, K.-S., Kim, H.-J. $\&$ Sasaki, M. Boltzmann equation approach to anomalous transport in a Weyl metal. Phys. Rev. B {\bf 89}, 195137 (2014).
\bibitem{RKKY} Ruderman, M. A. $\&$ Kittel, C. Indirect Exchange Coupling of Nuclear Magnetic Moments by Conduction Electrons. Phys. Rev. {\bf 96}, 99 (1954); Kasuya, T. A Theory of Metallic Ferro- and Antiferromagnetism on Zener's Model. Prog. Theor. Phys. {\bf 16}, 45 (1956); Yosida, K. Magnetic Properties of Cu-Mn Alloys. Phys. Rev. {\bf 106}, 893 (1957).
\bibitem{TI_BI_QCP} Fu, L. $\&$ Kane, C. L. Topological insulators with inversion symmetry. Phys. Rev. B {\bf 76}, 045302 (2007).
\bibitem{Percolation} Kirkpatrick, S. Percolation and Conduction. Rev. Mod. Phys. {\bf 45}, 574 (1973).
\bibitem{Fu_Kane_NLsM} Fu, L. $\&$ Kane, C. L. Topology, Delocalization via Average Symmetry and the Symplectic Anderson Transition. Phys. Rev. Lett. {\bf 109}, 246605 (2012).
\bibitem{Spin_Textbook} Auerbach, A. {\it Interacting Electrons and Quantum magnetism} (Springer-Verlag, New York, 1994).
\bibitem{Kettemann} Kettemann, S., Mucciolo, E. R., Varga, I., $\&$ Slevin, K. Kondo-Anderson transitions. Phys. Rev. B {\bf 85}, 115112 (2012).
\bibitem{Quantum_Griffiths} Miranda, E. $\&$ Dobrosavljevic, V. Disorder-Driven Non-Fermi Liquid Behavior of Correlated Electrons. Rep. Prog. Phys. {\bf 68}, 2337 (2005).
\bibitem{Anderson_Mott} Byczuk, K., Hofstetter, W., $\&$ Vollhardt, D. Anderson localization vs. Mott-Hubbard metal-insulator transition in disordered, interacting lattice fermion systems. Int. J. Mod. Phys. B {\bf 24}, 1727 (2010).
\end{thebibliography}
\end{document}